\documentclass[12pt]{article}
\newlength{\abstractwidth}
\usepackage{epsf}
\usepackage{color}

\renewcommand{\thefootnote}{\fnsymbol{footnote}}
\renewcommand{\thanks}[1]{\footnote{#1}}
\newcommand{\starttext}{
\setcounter{footnote}{0}
\renewcommand{\thefootnote}{\arabic{footnote}}}

\newcommand{\bea}{\begin{eqnarray}}
\newcommand{\eea}{\end{eqnarray}}
\newcommand{\ee}{\end{equation}}
\newcommand{\be}{\begin{equation}}

\def\cC{{\cal C}}
\def\cD{{\cal D}}

\def\cG{{\cal G}}
\def\cH{{\cal H}}

\def\cM{{\cal M}}
\def\cN{{\cal N}}
\def\cO{{\cal O}}

\def\cS{{\cal S}}

\def\cV{{\cal V}}

\def\cGb{{\cal G} _{\bar 0}}
\def\cGf{{\cal G} _{\bar 1}}

\def\bC{{\bf C}}

\def\bH{{\bf H}}
\def\bR{{\bf R}}
\def\bZ{{\bf Z}}

\def\tr{{\rm tr}}
\def\str{{\rm str}}

\def\half{ {1\over 2}}

\def\tim{\! \times \! }

\def\a{\alpha}
\def\b{\beta}

\def\ep{\varepsilon}

\def\g{\gamma}
\def\o{\omega}

\def\G{\Gamma}

\def\g{\gamma}
\def\s{\sigma}

\def\no{\nonumber}
\def\sm{\smallskip}

\def\sa{superalgebra}

\begin{document}
\starttext
\setcounter{footnote}{0}

\begin{flushright}
UCLA/08/TEP/17 \\
LAPTH-1268/08 \\
8 November 2008
 \end{flushright}

\bigskip

\begin{center}

{\Large \bf Half-BPS Supergravity  Solutions  and
Superalgebras}\footnote{\noindent This work was supported in
part by NSF grants PHY-04-56200 and PHY-07-57702.}

\vskip .5in

{\large  Eric D'Hoker,  John Estes\footnote{Present address:
Centre de Physique Th\'eorique, Ecole Polytechnique,
F-91128 Palaiseau,  France.},
Michael Gutperle, Darya Krym}

\vskip .2in

{ \sl Department of Physics and Astronomy }\\
{\sl University of California, Los Angeles, CA 90095, USA}

\vskip .5in

{\large Paul Sorba}

\vskip .2in

{\sl LAPTH,}\footnote{Laboratoire de Physique Th\'eorique d'Annecy-le-Vieux, UMR 5108.

\medskip

\noindent
{\sl E-mail addresses}: 
dhoker@physics.ucla.edu; johnaldonestes@gmail.com; gutperle@physics.ucla.edu; \\
dk320@physics.ucla.edu; sorba@lapp.in2p3.fr.} 
{\sl Universit\'e de Savoie, CNRS;}\\
{\sl 9, Chemin de Bellevue, BP110, F-74941 Annecy-le-Vieux, Cedex, France.}

\end{center}

\vskip .2in

\begin{abstract}

\vskip 0.1in

We establish a correspondence between
certain Lie superalgebras with 16 fermionic generators, and half-BPS solutions
to  supergravities with 32 supersymmetries.  Three cases are discussed.
For Type IIB supergravity, we relate semi-simple Lie superalgebras $\cH$
with 16 fermionic generators which are subalgebras of $PSU(2,2|4)$,
to families of half-BPS solutions which are invariant under $\cH$, and locally
asymptotic to $AdS_5 \times S^5$. Similarly, for M-theory, we relate
semi-simple Lie superalgebras $\cH$ with 16 fermionic generators which are
subalgebras of $OSp(8^*|4)$ or $OSp(8|4,\bR )$ to families of half-BPS solutions
which are invariant under $\cH$, and locally asymptotic to $AdS_7\times S^4$
or $AdS_4\times S^7$ respectively. Possible enhancements to more than
16 supersymmetries, such as 24, are also analyzed.
The classification of semi-simple
subalgebras of $PSU(2,2|4)$, $OSp(8^*|4)$, and $OSp(8|4,\bR )$ derived
in this paper, leads us to conjecture the existence of various new families of
half-BPS solutions to Type IIB supergravity and M-theory.

\end{abstract}

\baselineskip=16pt
\setcounter{equation}{0}
\setcounter{footnote}{0}


\newpage

\section{Introduction}
\setcounter{equation}{0}

Supersymmetry places powerful constraints on the existence of solutions to
supergravity. Invariance of a solution under supersymmetry is accompanied
by invariance under bosonic symmetries, such as isometries, which result from
the composition of supersymmetry transformations. The isometries strongly
constrain the space-time manifold, metric, and other fields of the solution. Together,
supersymmetries and bosonic symmetries close into a Lie superalgebra,
under which the solution is said to be invariant.

\sm

In 10-dimensional Type IIB supergravity and in 11-dimensional M-theory,
the maximal number of supersymmetries preserved by any solution is 32.
The only solutions with maximal supersymmetry correspond to the  following
space-time manifolds and  Lie superalgebras,
\bea
\hbox{Type~IIB} \qquad &\qquad AdS_5 \times S^5 \qquad& PSU(2,2|4)
\no \\
\hbox{M-theory} \qquad &\qquad AdS_7 \times S^4 \qquad& OSp(8^*|4)
\no \\
\hbox{M-theory} \qquad &\qquad AdS_4 \times S^7 \qquad& OSp(8|4,\bR)
\eea
and their respective flat Minkowski space-time and plane-wave limits
\cite{Freund:1980xh,Romans:1984an,FigueroaO'Farrill:2002ft}.
Solutions invariant under 28 or more supersymmetries have
been shown to automatically possess maximal supersymmetry \cite{Gran:2007eu}.
With fewer than 28 supersymmetries, many solutions have been constructed explicitly,
but no general classification is available.

\sm

The goal of the present paper is to classify, in Type IIB supergravity and M-theory,
solutions which are invariant under at least 16 supersymmetries (so-called
{\sl half-BPS solutions}), and are subject to certain asymptotic conditions,
to be spelled out below. The ultimate goal will be to obtain these solutions
in exact form, but this will be left for future work.

\sm

Various half-BPS solutions have long played a prominent role
in string theory already, including the F1/D1, D3, NS5/D5, and D7 branes in Type IIB,
and the M2 and M5 branes in M-theory. Certain intersections of branes
preserve just 8 supersymmetries (so-called {\sl quarter-BPS solutions}),
but these may be enhanced to 16 supersymmetries in the near-horizon limit,
thereby yielding further half-BPS solutions. The AdS/CFT correspondence
maps these half-BPS solutions onto half-BPS chiral operators in the dual
conformal field theory.

\sm

Over the past few years, a wealth of half-BPS solutions have been derived in exact
form. They include the exact solutions of \cite{Lin:2004nb}  which are AdS/CFT dual to
{\sl local half-BPS gauge invariant operators} in Type IIB and in M-theory
\cite{Berenstein:2004kk}.
In Type IIB, they include the exact solutions of \cite{DHoker:2007xy,DHoker:2007xz} and
of \cite{DHoker:2007fq}, which are AdS/CFT dual respectively to a half-BPS
{\sl planar interface} and to a half-BPS {\sl Wilson line} in $\cN=4$ super
Yang-Mills theory (SYM). In M-theory, they include the exact solutions of \cite{DHoker:2008wc},
which are dual to a half-BPS {\sl line-interface} and a half-BPs {\sl boundary operator} respectively in the associated 2+1- and 5+1-dimensional CFTs.
In all these cases, one actually has families of solutions described by non-trivial
moduli spaces, which may be related to the near-horizon limits of quarter-BPS localized
brane intersections.

\sm

A general program aimed at classifying the local form of supersymmetric 
solutions, in terms of Killing spinor bilinears and $G$-structures,
was proposed in \cite{Gauntlett:2002sc,Gauntlett:2002fz} (for a review,
see \cite{Gauntlett:2005bn}).
These methods have since been widely applied to supersymmetric AdS-type 
solutions, beginning in \cite{Gauntlett:2004zh}, and in subsequent papers, many of  
which will be mentioned later.

\subsection{The half-BPS solution -- superalgebra correspondence}

Given the special importance of half-BPS solutions to string theory and to the AdS/CFT
correspondence, it will clearly be of value to obtain their complete classification.
In this paper, we shall propose a solution to this problem in terms of Lie superalgebras.

\sm

Specifically, we shall exhibit a correspondence between a Lie superalgebra
$\cH$ which is a subalgebra with 16 fermionic generators of $PSU(2,2|4)$,
and a family of half-BPS solutions to Type IIB supergravity whose asymptotics at
space-time infinity approaches $AdS_5 \times S^5$. We exhibit a similar
correspondence between a subalgebra $\cH$ with 16 fermionic generators of
$OSp(8^*|4)$ or $OSp(8|4,\bR)$ and a family of half-BPS solutions to M-theory
whose asymptotics at space-time infinity approaches respectively
$AdS_7 \times S^4$ or $AdS_4 \times S^7$. For simplicity,  we shall
restrict to subalgebras $\cH$ which are {\sl basic Lie superalgebras or direct sums
thereof}.\footnote{Basic Lie superalgebras will be defined in Section 4; they are the
simple Lie superalgebras closest in properties to the ordinary simple Lie algebras.
It would be interesting to study also \sa s $\cH$ which are non-semi-simple
(such as those arising from the contraction of semi-simple
\sa s in the PP-wave problem), or which include the Lie \sa\ $PSU(1|1)$,
the ``strange" classical \sa s $P(n)$ and $Q(n)$, or those of ``Cartan type",
$W(n)$, $S(n)$, $\tilde S(n)$, and $H(n)$, but this will not be done here.}

\sm

More precisely, we shall argue that a half-BPS solution of Type IIB supergravity
which is asymptotic to $AdS_5 \times S^5$ must be invariant under a superalgebra
with 16 fermionic generators $\cH$ which is a subalgebra of $PSU(2,2|4)$
(and analogously for  M-theory). We shall show that there exist a finite number of
such subalgebras $\cH$ so that, as a result, this correspondence
provides a classification of the half-BPS solutions with the above asymptotics.
All cases of known exact solutions of Type IIB supergravity and M-theory
with respectively $AdS_5 \times S^5$ and $AdS_7 \times S^4$ or $AdS_4 \times S^7$
asymptotic behavior indeed obey this correspondence.

\sm

A classification of all basic Lie superalgebras $\cH$ with 16 fermionic
generators contained in $PSU(2,2|4)$, $OSp(8^*|4)$, and $OSp(8|4,\bR )$
did not appear to be available prior to this work, and will thus be derived here.
In particular, we shall show that none of the exceptional Lie superalgebras,
$F(4)$, $G(3)$, or $D(2,1;c)$ with $c \not= 1,-2,-1/2$, or any of their real forms,
is a subalgebra of $PSU(2,2|4)$, $OSp(8^*|4)$ or $OSp(8|4,\bR)$.

\sm

As a result, in Type IIB supergravity, no solutions exist which are asymptotic to
$AdS_5 \times S^5$ and invariant under a basic  Lie superalgebra,
or direct sum of basic Lie superalgebras,  $\cH$, which is not in the
above classification (e.g. when $\cH$ is an exceptional superalgebra).
The analogous result holds for M-theory.
The {\sl converse of the correspondence}, namely whether every regular solution to
Type IIB  supergravity  invariant under a basic Lie superalgebra $\cH$
with 16 fermionic generators contained in $PSU(2,2|4)$  must be asymptotic to
$AdS_5 \times S^5$ (and analogously  for M-theory), appears not to hold.

\sm

All known cases of exact solutions actually consist of families, with solutions
depending on a number of free parameters,  referred to as moduli.
To each basic Lie superalgebra, or direct sum of basic Lie superalgebras,
$\cH$, with 16 fermionic generators, which is a subalgebra  of $PSU(2,2|4)$,
$OSp(8^*|4)$, or $OSp(8|4,\bR )$, we associate a moduli space $\cM_\cH$ of
regular solutions with the above asymptotics for each respective supergravity.
The solution with 32 supersymmetries always trivially belongs  to $\cM_\cH$.
For all the  non-trivial known cases,
\cite{Lin:2004nb,DHoker:2007xz,DHoker:2007xy,DHoker:2007fq,DHoker:2008wc},
the moduli spaces display non-trivial topology and geometry, and it becomes
an interesting mathematical question to characterize $\cM_\cH$ for each $\cH$.
In particular, the connectedness properties of  $\cM_\cH$ will reveal to what degree
solutions are continuously  connected to the maximally symmetric solution.

\sm

These moduli spaces may contain subvarieties where supersymmetry
is enhanced, and the number of supersymmetry generators is larger than 16,
but smaller than 28. We shall also classify these possible solutions via
their invariance superalgebras.

\sm

For the cases of superalgebras $\cH$ where no non-trivial half-BPS solutions
are known yet,  we shall compare arguments based on intersecting and probe
branes, the CFT duals, superalgebra theory, and the structure of
the Ansatz for supergravity fields to argue either for the existence or for the
absence of new non-trivial half-BPS solutions. The list of all cases produced by our
proposed correspondence between Lie superalgebras and half-BPS solutions
for Type IIB will be given in Table \ref{table10}, and for M-theory in Tables
\ref{table11} and \ref{table12}. 

\subsection{Half-BPS solutions related to D7 branes}

Cases related to D7 branes in Type IIB merit to be highlighted here.
On the one hand, the existence of near-horizon limits involving
intersections of D7 branes with D3 branes, or of D7 branes viewed as probes in
the $AdS_5 \times S^5$ background, suggests that fully back-reacted half-BPS 
solutions can exist. On the other hand, D7 branes produce flavor multiplets 
in the dual CFT, which generate a non-vanishing $\beta$-function, and
thus vitiate solutions with $AdS_3$ or $AdS_5$ factors on the gravity side.
Arguments have been presented in \cite{Kirsch:2005uy} and in 
\cite{Buchbinder:2007ar,Harvey:2008zz} (following earlier work on D7 branes
in \cite{Aharony:1998xz,Grana:2001xn}), that no fully back-reacted near-horizon limit 
solutions corresponding to D7 branes should exist. 

\sm

Our analysis will reveal the existence of two cases (namely II$^*$ and III$^*$ of Table 
\ref{table10}) corresponding to the superalgebras $SU(1,1|4) \oplus SU(1,1)$ 
and $SU(2,2|2) \oplus SU(2)$ which are subalgebras of $PSU(2,2|4)$, and 
whose global symmetries and  space-time structure precisely match those of the D7 
probe or D7/D3 intersecting brane analysis. 
These cases correspond respectively to D7/D3 intersections of dimensions 2 or 4. 
Closer analysis of the superalgebra structure reveals, however,  
further layers of subtlety (see section 5.4).  Each subalgebra, $SU(1,1|4) \oplus SU(1,1)$ 
or $SU(2,2|2) \oplus SU(2)$,  possesses a {\sl purely bosonic invariant subalgebra} 
(respectively $SU(1,1)$ and $SO(3)$), which is not needed for the half-BPS
condition, and thus not protected by supersymmetry. Arguments, to be presented 
in subsection 5.4, will indicate that these extra bosonic invariant subalgebras
are incompatible with asymptotic $AdS_5 \times S^5$ behavior (though compatible
with {\sl local} asymptotic behavior). Removing these bosonic factors
leads to corresponding ``general  cases", namely II and III of Table \ref{table10},
for which superalgebra arguments suggest the existence of solutions.
The general cases II and III now have, however,  insufficient symmetry to
support fully back-reacted half-BPS solutions for the near-horizon limit of D7/D3 
branes.

\subsection{Families of half-BPS solutions in M-theory}

The correspondence may be further illustrated by the M-theory solutions
of \cite{Boonstra:1998yu} (see also \cite{Gauntlett:1998kc,deBoer:1999rh})
and the M-theory solutions of \cite{DHoker:2008wc}, which both
form families of solutions  invariant under the exceptional Lie superalgebra
$\cH= D(2,1;c;0) \oplus D(2,1;c;0)$. Here, $c\not=0$ is a free real parameter,
and the last entry ``0" in $D(2,1;c;0)$ refers to the real form of $D(2,1;c)$
whose maximal bosonic subalgebra is $SO(1,2) \oplus SO(4)$.
For $c=-2$ or $c=-1/2$, $\cH$ is a subalgebra of $OSp(8^*|4)$ and
the exact solutions of \cite{DHoker:2008wc} are locally asymptotic to $AdS_7 \times S^4$.
Similarly, for $c=1$, $\cH$ is a subalgebra of $OSp(8|4,\bR)$ and the associated
exact solutions of \cite{DHoker:2008wc} are locally asymptotic to $AdS_4 \times S^7$.
Finally, for $c\not= 1,-2,-1/2$,  $\cH$ is a subalgebra of neither  $OSp(8^*|4)$
nor $OSp(8|4,\bR)$, and it has been argued in \cite{DHoker:2008wc} that the 
local asymptotics
of the solutions can be neither $AdS_7 \times S^4$ nor  $AdS_4 \times S^7$.
The solutions of \cite{Boonstra:1998yu,Gauntlett:1998kc} have a space-time manifold
of the form $AdS_3 \times S^3 \times S^3 \times E_2$ for all values of $c$.
Here, the product of the space-time factors is genuinely direct, so that no warping 
occurs over the
flat surface $E_2$ with boundary. In particular, for $c=1,-2,-1/2$, there exist solutions
asymptotic to $AdS_3 \times S^3 \times S^3 \times E_2$, as well as
to $AdS_4 \times S^7$ or $AdS_7 \times S^4$,
which indicates that the converse of the correspondence does not hold.

\subsection{Organization}

The remainder of this paper is organized as follows. In Sections 2 and 3, we review
known constructions of half-BPS configurations in respectively Type IIB and M-theory
from the points of view of their CFT duals, probe and intersecting branes, and exact
supergravity solutions. In Section 4, we produce a classification of all semi-simple
basic Lie \sa s which are subalgebras of $PSU(2,2|4)$, $OSp(8^*|4)$, or $OSp(8|4,\bR )$.
In Section 5, we apply the results of Section 4 towards the classification of old and new
half-BPS solutions. In Section 6, conformal superalgebras \cite{VanProeyen:1986me}
with 16 supersymmetries in various dimensions are matched with the superalgebras
$\cH$ which arise in  our classification. In Section 7, a discussion is presented
of when supersymmetry of half-BPS solutions may be enhanced in a non-trivial way,
and how the M-theory solutions of \cite{DHoker:2008wc} illustrate the correspondence.
In Appendix A, relevant facts of ordinary Lie algebras are reviewed. In Appendix B,
subalgebra relations between non-exceptional basic Lie superalgebras are proven.
In Appendix C, it is demonstrated that none of the exceptional Lie superalgebras is a
subalgebra of  $PSU(2,2|4)$, $OSp(8^*|4)$, or $OSp(8|4,\bR)$.
Finally, in Appendix D, we derive the superalgebra of the M-theory solution of
\cite{DHoker:2008wc}.
Throughout, our analysis will be carried out at the level of Lie superalgebras
rather than supergroups, thus leaving possible global issues for later investigation.

\newpage

\section{Half-BPS configurations in Type IIB}
\label{sectwo}
 \setcounter{equation}{0}

Before initiating the classification of subalgebras of $PSU(2,2|4)$, $OSp(8^*|4)$,
and $OSp(8|4,\bR )$ in Section 4, and its application to half-BPS solutions in
Section 5, we begin by assembling evidence for the existence of various families
of half-BPS  solutions, by drawing together arguments from the
AdS/CFT correspondence, from intersecting and probe brane dynamics,
and from existing families of exact solutions.

\subsection{The AdS/CFT correspondence for Type IIB}

The AdS/CFT correspondence provides a map between Type IIB superstring
theory on $AdS_5 \times S^5$ and 4-dimensional $\cN=4$ SYM
with gauge group $SU(N)$ in its conformal phase \cite{Maldacena:1997re}.
The global symmetries, including the supersymmetries, on the AdS side match
those on the CFT side, and are described by the \sa\ $PSU(2,2|4)$,
whose maximal bosonic subalgebra is $SU(2,2) \oplus SU(4) = SO(2,4) \oplus SO(6)$.
On the CFT side, $SO(2,4)$ is the 4-dimensional conformal algebra,
and $SU(4)$ is the R-symmetry.
On the AdS side,  $SO(2,4) \oplus SO(6)$ is the isometry algebra of
$AdS_{5}\times S^{5}$ space-time, and the fermionic generators are
associated with 32 Killing spinors of this maximally supersymmetric solution.
Furthermore, there exists an explicit map between correlators of gauge
invariant operators in $\cN=4$ SYM theory (in the `t Hooft limit, and
large `t Hooft coupling) and solutions to Type IIB supergravity subject to suitable
boundary conditions \cite{Gubser:1998bc,Witten:1998qj} (for reviews, see
\cite{Aharony:1999ti}, and \cite{D'Hoker:2002aw}).

\sm

The CFT vacuum has maximal symmetry, given by the superalgebra $PSU(2,2|4)$,
and corresponds to the maximally symmetric $AdS_5 \times S^5$ solution on the AdS side. Inserting an operator $\cO$ in the CFT vacuum produces a new state whose
symmetry is reduced to the subalgebra $\cH$ of $PSU(2,2|4)$ that leaves $\cO$ invariant.
The insertion of $\cO$ on the CFT side causes the Type IIB solution to deform
away from $AdS_5 \times S^5$ on the AdS side.
If the operator $\cO$ is  ``sufficiently local", then there will exist an asymptotic bulk
region in $\bR^4$, away from the support of $\cO$, where locally the state of the CFT
is close to the vacuum state. The asymptotics at space-time infinity of the AdS
dual supergravity solution then coincides locally with $AdS_5 \times S^5$.
Here, ``sufficiently local" operators include, for example, those whose
support in $\bR^4$ is at least co-dimension 1, which is the case of local operators,
Wilson loops, interface operators, and surface operators.

\sm

This AdS/CFT argumentation shows that any half-BPS solution which is AdS/CFT
dual to a ``sufficiently local" operator in the CFT will obey the local asymptotics
of  $AdS_5 \times S^5$, and will be invariant under a  subalgebra $\cH$ with 16
fermionic generators of $PSU(2,2|4)$.

\subsection{Half-BPS operators in ${\cal N}=4$ SYM}
\label{sectwoone}

In this subsection we shall review half-BPS gauge invariant operators in
${\cal N}=4$ SYM. It is customary to distinguish between  local and non-local operators.
By the state operator mapping, local operators are associated with
superconformal chiral primary states of the CFT. Examples of non-local operators
include the Wilson loop, the `t Hooft loop, surface operators, and interface/defect
operators. Non-local operators are useful as probes and often provide order
parameters which can distinguish the different phases of the theory.

\sm

$\bullet$ A local half-BPS operator forms a multiplet associated with a chiral primary state,
which can be constructed as follows. The scalar fields of $\cN=4 $ SYM
will be denoted $\Phi ^i$, with $i=1,\cdots, 6$, and the gauge field by $A_\mu$,
both of which transform under the adjoint representation of the $SU(N)$ gauge group.
One considers the theory on $\bR \times S^{3}$, and chooses a complex combination
$Z=\Phi^{1}+i \Phi^{2}$ of the scalars. The scalar field $Z$ carries a charge $J=1$
under the $U(1)$ subalgebra of the $SU(4)$ R-symmetry corresponding to rotations
of scalars in the $1-2$ plane.  The chiral primary state is given by a multi-trace operator,
\bea
\label{halfBPSloca}
{\cal O}= \prod_{i} \tr \left ( Z^{n_{i}} \right )
\hskip 1in J = \sum _i n_i
\eea
The conformal dimension  $\Delta$ saturates the BPS bound, and we have  $\Delta=J$.
The operator $\cO$ is invariant under time translations and $SO(4 )$ rotations of $S^{3}$.
The particular linear combination chosen to define $Z$ in terms of $\Phi^i$
breaks the $SU(4)$ R-symmetry to its $SO(4)$ subalgebra.
In total, the operator (\ref{halfBPSloca}) is invariant under 
$  \bR \oplus SO(4) \oplus SO(4)$
and preserves 16 supersymmetries, enlarging the symmetry to a
$ \bR \oplus PSU(2|2)\oplus PSU(2|2)$ superalgebra.

\sm

$\bullet$ A supersymmetric Wilson loop is a non-local operator associated with
a  curve  ${\cal C}$ in $\bR^{1,3}\times \bR^{6}$ parameterized
by $\{x^{\mu}(\tau), y^{i}(\tau )\}$, where $\mu=0,1,2,3$ and $i=1,\cdots ,6$,
\bea
W(R,{\cal C})= \tr_{R} \left \{ P \exp  i \oint_\cC
\left ( A_{\mu} d  x^{\mu} + \Phi^{i} d  y^{i} \right ) \right \}
\eea
Here, $P$ stands for path ordering the exponential, and the trace $\tr _R$ is to be
evaluated in a representation $R$ of the gauge group $SU(N)$.
It has been shown in \cite{Drukker:1999zq,Bianchi:2002gz} that a Wilson loop
preserves 16 supercharges if the curve ${\cal C}$ is a straight line, given by
$x^{0}=\tau$, $x^{1} =x^{2}=x^{3}=0$, and $y^{i} = n^{i} \tau$ where $n^{i}$ is a
vector of unit norm in $\bR^{6}$.  Such a Wilson loop breaks the bosonic
$SO(2,4) \oplus SO(6)$ symmetry to $SO(2,1) \oplus SO(3) \oplus SO(5)$.
The \sa\ preserved by the Wilson loop is $OSp(4^*|4)$.

\sm

$\bullet$ A surface operator is associated with a 2-dimensional surface ${\cal S}$
in $\bR^{1,3}$.
In ${\cal N}=4$ SYM it may be defined \cite{Gukov:2006jk,Gukov:2008sn}
as a disorder operator which prescribes a codimension 2 singularity for the YM
fields along the surface ${\cal S}$. The surface operator is therefore a
higher-dimensional analog of the `t Hooft loop in the sense that neither can
be expressed as a path integral insertion using the local fields that appear in
the Lagrangian.
A  surface operator is half-BPS if ${\cal S}=\bR^{1,1}$ and the singularity along
${\cal S}$ selects a single complex scalar field $Z$.  The bosonic symmetries
are the conformal $SO(2,2)$ symmetry of the surface world-volume, and the
$SO(4)$ R-symmetry unbroken by the choice of $Z$. The superalgebra preserved
by a surface operator is $PSU(1,1|2) \oplus PSU(1,1|2) \oplus U(1)$,
see \cite{Drukker:2008wr,Gomis:2007fi}.

\sm

$\bullet$ A defect/interface operator is associated with a 2+1-dimensional
space ${\cal D}$  in $\bR^{1,3}$. Boundary conditions for the SYM fields,
and possible worldvolume actions  which preserve 16  supersymmetries were
analyzed in \cite{DHoker:2006uv,Gaiotto:2008ak,Gaiotto:2008sa,Gaiotto:2008sd}.
It was shown that if $\cD$ is chosen to be ${\cal D}=\bR^{1,2}$ and
the  $SU(4)$ R-symmetry is broken to $SU(2)\oplus SU(2)$, the conformal algebra
$SO(2,3)$,  as well as 16 supersymmetries, are unbroken by the defect/interface.
This half-BPS planar defect/interface preserves the \sa\ $OSp(4|4,\bR)$.

\subsection{Intersecting branes in flat space-time}

A stack of parallel D3-branes in flat space-time preserves 16 of the 32 
supersymmetries.
More generally, the unbroken supersymmetries of a Dp-brane in flat 
space are given by
\bea
\label{susycon}
Q=Q_L + \prod_{i} \Gamma^iQ_R
\eea
where the index $i$ runs over all space-time direction transverse to the
Dp-brane world-volume.

\begin{table}[htdp]
\begin{center}
\begin{tabular}{|c|c|c|c|c|} \hline
Case  &  branes & dim intersec  & $\# _{ND}$  & bosonic symmetry
\\ \hline \hline
  1 & D3/D1  &1& 4 & $ \bR \oplus SO(3) \oplus  SO(5)$
\\  \hline
2 &  D3/D5 &1 & 8 &$ \bR \oplus SO(3) \oplus  SO(5)$
\\  \hline
3 & D3/D3 & 2& 4&  $ ISO(1,1) \oplus  SO(4) \oplus  SO(2) $
\\  \hline
4 & D3/D7 & 2& 8&  $ ISO(1,1) \oplus  SO(6) \oplus  SO(2) $
\\  \hline
5&D3/D5& 3& 4&$ ISO(2,1) \oplus  SO(3) \oplus  SO(3) $
\\  \hline
6&D3/D7& 4& 4 & ~ $ ISO(3,1) \oplus  SO(4) \oplus  SO(2) $ ~
\\ \hline \hline
\end{tabular}
\end{center}
\caption{Intersecting D3/Dp branes in Type IIB preserving 8 supersymmetries}
\label{table1}
\end{table}

\sm

The intersection of a Dp and a Dp' brane preserves 8 of the 32 supersymmetries
if the number $\#_{ND}$ of Neumann-Dirichlet directions  (i.e. the directions which are transverse to one and longitudinal to the other  brane, but not both) is either 4 or 8.
In Table \ref{table1}, we list all supersymmetric intersections of a D3 brane with
a Dp brane which preserve 8 supersymmetries.   The cases are ordered by
increasing dimensionality  of the  intersection.\footnote{Although the bosonic symmetry
is allowed to have an extra $SO(2)$, in addition to the bosonic symmetries
listed, in case 3 of Tables \ref{table1}, \ref{table2}, \ref{table3}, and \ref{table4},
and in case 4 of Table \ref{table5},  the generic solution has no such factor \cite{Drukker:2008wr,Constable:2002xt}. 
We thank Jaume Gomis for stressing this result to us.}

\sm

In cases 1, 2, and 5 it is known that the near-horizon limit
will enhance the number of unbroken supersymmetries from 8 to 16.
In addition, the Poincar\'e symmetry $ISO(1, d-1)$ will be enhanced to the conformal
symmetry $SO(2,d)$.  In the absence of fully localized supergravity solutions for
the intersecting branes in cases 3, 4, and 6, it remains an open question whether
the Poincar\'e invariance and supersymmetry will be enhanced in those cases,
particularly the ones involving D7 branes (see also the subsequent subsection).

\subsection{Probe branes in $AdS_{5}\times S^{5}$}

A different approach to investigating half-BPS brane configurations
starts from the  $AdS_{5}\times S^5$ near-horizon
limit of D3 branes and considers inserting supersymmetric probe branes.
In the probe approximation,  the strength of the back-reaction of the probe
brane onto the geometry is neglected.  The back-reaction of the probe brane is
governed by $g_{s} N_{probe}$, where $g_{s}$ is the string coupling constant
and $N_{probe}$ is the number of probe branes. In the `t Hooft limit, where
$g_{s}\to 0, N\to \infty$ with $\lambda = g_{s}N$ held fixed, the back-reaction
is negligible for finite $N_{probe}$. On the field theory side this is reflected by
the fact that the world-volume theory on  the probe is in general not conformal
and that the $\beta$-function is proportional to $g_{YM} ^3 N_{probe}$ which
vanishes in the `t Hooft limit for finite $N_{probe}$.

\sm

Consistency of the probe brane requires that it solve the  equations of motion
for the embedding of the probe into $AdS_5 \times S^5$, derived from the
Dirac-Born-Infeld  world-volume action for the probe brane. The embedding
of the probe $p$-brane is parameterized by a  $p+1$-dimensional submanifold
$M$ of $AdS_{5}\times S^5$. To describe  this more explicitly, we represent
$AdS_{5}$ as a hyperboloid in $\bR^{2,4}$, and $S^5$ as a sphere in $\bR^{6}$,
both with the same radius $R$,
\bea
 X_{0}^{2}+X_{5}^{2}- X_{1}^{2}-X_{2}^{2}-X_{3}^{2}-X_{4}^{3} & = & R^2
\no \\
Y_{1}^{2}+Y_{2}^{2}+\cdots+ Y_{6}^{2} & = &  R^2
\eea
We are interested in cases where the manifold $M$ is a product of an
$AdS_{n}$  submanifold in $AdS_{5}$   parametrized by
\bea
X_{0}^{2}+X_{5}^{2}- X_{1}^{2}-X_{n-1} ^2 & = & R^{2}\cosh^{2}\rho
\no \\
X_{n}^{2} +\cdots +X_{4}^{2} & = & R^{2}\sinh^{2}\rho
\eea
and an $S^m$ submanifold in $S^5$ parameterized by
\bea
Y_{1}^{2}+Y_{2}^{2}+\cdots Y_{m+1} ^2 & = & R^{2}\cos^{2}\alpha
\no \\
Y_{m+2}^{2} +\cdots +Y_{6}^{2} & = & R^{2}\sin^{2}\alpha
\eea
The Dirac-Born-Infeld equations of motion for the probe brane then determine
the position $\rho$ of the $AdS_{n}$ inside $AdS_5$, and the position $\alpha$
of the $S^m$ inside the $S^5$. Note that the spheres $S^m$ have slipping modes,
which tend to decrease their volume. Although one might  think
that such modes indicate an instability, they actually turn out to be above the
Breitenlohner-Freedman bound \cite{Breitenlohner:1982bm}, and hence do
not introduce instabilities \cite{Skenderis:2002vf,Karch:2000ct, Karch:2000gx}.

\sm

In Table 2 below, the possible supersymmetric Dp branes in $AdS_5 \times S^5$
which are related to the brane intersections in flat space-time of Table 1 are
listed in order of increasing world-volume dimension.

\begin{table}[htdp]
\begin{center}
\begin{tabular}{|c|c|c|c|} \hline
Case  & probe & worldvolume     & bosonic symmetry
\\ \hline \hline
1 & D1   &$ AdS_{2}$   & ~$ SO(2,1) \oplus SO(3) \oplus  SO(5) $ ~
\\  \hline
2 & D5   &$ AdS_{2} \times S^4$   & $ SO(2,1) \oplus  SO(3) \oplus  SO(5) $
\\  \hline
3 & D3   &$ AdS_{3} \times S^1$    & $ SO(2,2) \oplus  SO(4) \oplus  SO(2) $
\\  \hline
4 & D7   &$ AdS_{3} \times S^5$ & $ SO(2,2) \oplus  SO(6) \oplus  SO(2) $
\\  \hline
5 & D5   &$ AdS_{4} \times S^2$    & $ SO(2,3) \oplus  SO(3) \oplus  SO(3) $
\\  \hline
6 & D7   &$ AdS_{5} \times S^3$    & $ SO(2,4) \oplus  SO(4) \oplus  SO(2) $
\\  \hline \hline
 \end{tabular}
\end{center}
\caption{Probe Dp branes  in $AdS_5\times S^5$ preserving 16 supersymmetries}
\label{table2}
\end{table}

The supersymmetry of the probe branes is established by checking the existence
of a $\kappa$-symmetry projector for the Dp-brane world-volume action
\cite{Cederwall:1996ri,Aganagic:1996pe,Bergshoeff:1996tu}.
The projector may be used to gauge fix the  world-volume  action and to exhibit
equal numbers of bosonic and fermionic degrees of freedom.
It was shown in \cite{Skenderis:2002vf}
that the probe branes  in Table \ref{table2} are all supersymmetric.

\sm

Probe branes allow us to introduce additional structure into the AdS/CFT
correspondence.  In the probe approximation, the AdS/CFT duality  is expected
to ``act twice" \cite{Karch:2000ct, Karch:2000gx}, with the open strings on the
probe branes being dual to the localized degrees of freedom on the intersection. 
Cases 1 and 2 have been discussed in \cite{Gomis:2006sb}. Note that, in  case 1, 
the D1-brane can be replaced  by a ``giant'' D3-brane with the same symmetries 
and  magnetic flux on its $AdS_{2}\times S^{2}$ worldvolume. Case 3 has been 
discussed in  \cite{Constable:2002xt}, where it was shown  that a  supersymmetric 
embedding   preserves only one of the two $SO(2)$ symmetries.
Case 4 has been discussed in the probe approximation in \cite{Buchbinder:2007ar}, 
and with the inclusion of the fully back-reacted geometry in \cite{Harvey:2008zz}.
Case 5 was argued
to correspond to a holographic description of a defect conformal field theory
\cite{Karch:2000ct, Karch:2000gx,DeWolfe:2001pq,Erdmenger:2002ex}.
Case 6 was used to introduce flavor into  AdS/CFT \cite{Karch:2002sh}.

\sm

It is a priori not determined whether a probe solution can be extended to an
exact supergravity solution with the same symmetries. The question  whether
including the gravitational back-reaction of the probe  preserves all or destroys
some of the symmetries of the configuration has to be decided on a case by
case basis.

\subsection{Structure of known exact supergravity solutions}
\label{sectwofour}

The presently known exact supergravity solutions which preserve 16  supersymmetries
share the following structure. We will focus on solutions which  asymptotically
approach $AdS_{5}\times S^{5}$. The Ansatz has the structure of a fibration
of several metric factors over a base manifold $\Sigma$,
\bea
\label{metans}
ds^{2} = f_{1}^{2} ds_{M_{1}}^{2}+
f_{2}^{2}ds_{M_{2}}^{2}+\cdots +
f_{n}^{2}ds_{M_{n}}^{2}+ ds_{\Sigma}^{2}
\eea
The metric factors $f_{i}$ and $ds_{\Sigma}^{2}$, as well as the axion and dilaton
fields are functions of  $\Sigma$ only. The 3- and 5-form field strengths also depend
on $\Sigma $; in view of their invariance under bosonic symmetries,
they depend on $M_i$ only through the invariant volume forms on $M_i$.

\sm

The spaces $M_{i}$ are  chosen to realize the bosonic
symmetries of the half-BPS configuration which often leads to a product of
lower dimensional $AdS$ and sphere factors.\footnote{For some of the new 
solutions, which we shall advocate
in Section 5, the isometry algebra will have components of the form
$SU(3)$, $SU(1,2)$ and $SU(1,3)$, with corresponding space-time manifold
factors $M_i$ of the form $CP_2$, $CH_2$, and $CH_3$. This generalization
of (\ref{metans}) will be discussed systematically in section 5.2.}
  In view of the overall Minkowski
signature of the full space-time metric $ds^2$, no more than a single Minkowski
signature $AdS$ factor should occur. In some solutions, an extra factor of
the real line $\bR$ will be part of the space-time manifold. Thus, we may
have,\bea
&&M_{i}= AdS_{d+1}  \quad  {\rm isometry:} \;\; SO(2,d)
\no\\
&&M_i= S^{d} \quad\quad \quad  {\rm isometry:} \;\; SO(d+1)
\no \\
&& M_i = \bR \hskip 0.52in {\rm isometry:} \;\; \bR
\eea
On the CFT side, the $SO(2,d)$ and $SO(d+1)$  isometries are associated
respectively with the  conformal symmetry, and the R-symmetry of the
corresponding half-BPS operator.

\sm

A purely bosonic field configuration will preserve 16 supersymmetries provided
there exist 16 linearly independent spinors $\ep$ for which the supersymmetry
transformations of the fermion fields, namely the gravitino and the dilatino, vanish.
Using the supersymmetry variations of these fermion fields \cite{Schwarz:1983qr},
the condition leads to the following BPS equations,
 \bea
\label{BPS}
0 &=& i (\G \cdot P)   \ep^*  -{i\over 24} (\G \cdot G) \ep
\label{susy1} \no  \\
0 &=& D _M  \ep
+ {i\over 480}(\G \cdot F )  \Gamma_M  \ep
-{1\over 96}\left ( \Gamma_M (\G \cdot G)
+ 2 (\G \cdot G) \G^M \right )  \ep^*
\eea
where $P$, $G$, and $F$ stand for the Type IIB dilaton/axion, 3-form, and
5-form field strengths respectively. The unbroken supersymmetries are
constructed using the invariant or {\sl Killing spinors} on the spaces $M_{i}$.
We will briefly review the known solutions associated with half-BPS operators
in the next subsection. A unifying feature of all solutions is that the extremely
complicated system of BPS equations arising from (\ref{BPS}) can be solved
exactly in terms of harmonic functions subject to certain prescribed boundary conditions.

\subsection{Known families of half-BPS solutions}

In this subsection, we shall summarize the known families of half-BPS solutions.
In each case, the solutions are locally asymptotic to $AdS_5 \times S^5$, and
invariant under a subalgebra with 16 fermionic generators of $PSU(2,2|4)$.
(The embeddings of Lie superalgebras will be discussed in detail
in Section 4, and Appendices B and C.)
\sm

$\bullet$ ~
The supergravity solutions of \cite{Lin:2004nb} correspond to the
AdS dual of localized half-BPS operators on the CFT side.
The solutions have $\bR \oplus  SO(4) \oplus SO(4)$ isometry and
hence  $M_{1}= S^{3}$, $M_{2}=S^{3}$, $M_{3} = \bR$. The only non-trivial fields
are the metric and the five form $F$, while the dilaton is constant, and the
axion, and 3-form antisymmetric field strengths vanish.  The 3-dimensional
base space $\Sigma$ is constrained further by Killing spinors. The most general
half-BPS solution  with these symmetries is given by a 3-dimensional
harmonic function $h$ defined on a 3-dimensional half space $(x_{1},x_{2},y)$
where $y>0$. The boundary condition $h=\pm {1\over2}$ at $y=0$  completely
determines the regular solutions. The symmetry preserved by these solutions
is the superalgebra $\bR \oplus PSU(2|2) \oplus PSU(2|2) $.

\sm

$\bullet$ ~
The full supergravity solution corresponding to half-BPS Wilson loops was
constructed exactly in  \cite{DHoker:2007fq}. (For an earlier derivation of 
the BPS equations and discussion of suitable boundary conditions,
see \cite{Yamaguchi:2006te,Lunin:2006xr}).
The isometries are $SO(2,1) \oplus SO(3) \oplus SO(5)$ and hence we have
$M_{1}=AdS_{2}, M_{2}=S^{2}$, and $M_{3}=S^{4}$. The 3-form field strength
$G$ is, in general, non-vanishing.
The base space $\Sigma$ is a 2-dimensional hyperelliptic Riemann surface with
a boundary and the solution is completely determined by two harmonic functions
with alternating Neumann and vanishing Dirichlet conditions on segments of the boundary.
These solutions were analyzed further in \cite{Gomis:2008qa,Okuda:2008px}.
The symmetry preserved by these solutions  is the superalgebra $OSp(4^*|4)$.

\sm

$\bullet$ ~
The supergravity solutions dual to half-BPS surface operators were constructed in
\cite{Gomis:2007fi,Drukker:2008wr} as double analytic continuations of the
solutions of  \cite{Lin:2004nb}. The isometries are $SO(2,2) \oplus SO(4) \oplus U(1)$,
and hence $M_{1} =AdS_{3}$, $M_{2}=S^3$, and $M_3=\bR$. The base space
$\Sigma$ is 3-dimensional in view of the $U(1)$ isometry.
The symmetry preserved by these solutions
is the superalgebra $PSU(1,1|2) \oplus PSU(1,1|2) \oplus U(1)$.

\sm

$\bullet$ ~
The supergravity  solutions  corresponding to the planar half-BPS defect/interface
were obtained exactly  in \cite{DHoker:2007xy,DHoker:2007xz}. (For an earlier 
derivation of the BPS equations and of suitable boundary conditions, see 
\cite{Gomis:2006cu}.)
These solutions  have  isometry algebra $SO(2,3) \oplus SO(3) \oplus SO(3)$, 
and hence we have
$M_{1}=AdS_{4}, M_{2}= S^{2}$ and $M_{3}=S^{2}$. The 3-form field strength $G$ is,
in general, non-vanishing. As in the case of the supergravity solutions
dual to the Wilson loop, the base space $\Sigma$ is a  hyperelliptic Riemann
surface with boundary and the solution is determined by two harmonic functions
with alternating Neumann and Dirichlet boundary conditions, although the
detailed regularity conditions are different. The symmetry preserved by the
defect/interface solutions  is the superalgebra $OSp(4|4,\bR)$.

\sm

Whenever the CFT operator is not strictly speaking local, the corresponding
AdS dual supergravity solutions are {\sl not globally asymptotic} to
$AdS_5 \times S^5$, as the boundary of space-time is deformed at the
imprint of the operator on the boundary. This is the case for a Wilson line
where the dual string ends at the boundary of $AdS_5 \times S^5$,
as well as for an interface  where the boundary consists of multiple asymptotic
regions of $AdS_5 \times S^5$ glued together along the interface. In both cases,
the solutions are asymptotically $AdS_5 \times S^5$ away from the support of the operator.

\newpage

\section{Half-BPS configuration in M-theory}
\setcounter{equation}{0}

In this section we shall generalize the analysis of  Section 2 to M-theory.
Eleven-dimensional supergravity admits two maximally supersymmetric
Freund-Rubin solutions \cite{Freund:1980xh}.
One is $AdS_7\times S^4$, which arises in the near-horizon limit of a stack of parallel
$M5$ branes; the other is $AdS_4\times S^7$,  which arises in the near-horizon limit
of a stack of parallel $M2$ branes \cite{Maldacena:1997re}.

\subsection{The AdS/CFT correspondence for M-theory}

Two cases may be distinguished according to the two Freund-Rubin solutions.
The first case provides a correspondence between 11-dimensional supergravity
on $AdS_7 \times S^4$ and the 5+1-dimensional CFT with 32 supersymmetries,
which we shall denote by CFT$_6$. The global symmetries here are encoded in
the \sa\  $OSp(8^*|4)$, whose maximal bosonic subalgebra is
$SO(8^*) \oplus Sp(4) = SO(2,6) \oplus SO(5)$. On the CFT side,
$SO(8^*) = SO(2,6)$ is the 5+1-dimensional conformal algebra,
while $SO(5)$ is the R-symmetry. On the AdS side, $SO(2,6) \oplus SO(5)$ is the
isometry of $AdS_7 \times S^4$. The second case provides
a correspondence between 11-dimensional supergravity on $AdS_4 \times S^7$ and
the 2+1-dimensional CFT with 32 supersymmetries, which we shall denote by CFT$_3$.
The global symmetries are encoded in the \sa\ $OSp(8|4,\bR )$, whose maximal
bosonic subalgebra is $SO(8) \oplus Sp(4,\bR)$.
On the CFT side, $Sp(4,\bR) = SO(2,3)$ is the 2+1-dimensional conformal algebra,
while $SO(8)$ is the R-symmetry. On the AdS side, $SO(8) \oplus SO(2,3)$ is the
isometry of $AdS_4 \times S^7$.

\sm

Although CFT$_6$ and CFT$_3$ are not as well understood as their 3+1-dimensional
counterpart, many helpful facts analogous to the results derived in $\cN=4$ SYM
remain true.  The unique CFT$_6$ vacuum has maximal symmetry, given by the
superalgebra $OSp(8^*|4)$, while the unique CFT$_3$ vacuum has symmetry
$OSp(8|4,\bR)$. Inserting an operator $\cO$ in the CFT vacuum produces a new
state whose symmetry is reduced to the subalgebra $\cH$ of $OSp(8^*|4)$ or
$OSp(8|4,\bR)$ that leaves $\cO$ invariant. The insertion of $\cO$ on the CFT side causes the
M-theory solution to deform away from $AdS_7 \times S^4$ or $AdS_4 \times S^7$.
If the operator $\cO$ is  ``sufficiently local", then there will exist an asymptotic bulk
region in $\bR^6$ or $\bR^3$, away from the support of $\cO$, where locally the
state of the CFT is close to the vacuum state. The asymptotics at space-time infinity
of the $AdS$ dual supergravity solution then coincide locally with $AdS_7 \times S^4$
or $AdS_4 \times S^7$.

\sm

This AdS/CFT argumentation shows that any half-BPS solution which is AdS/CFT
dual to a ``sufficiently local" operator in the CFT will obey the local asymptotics
of $AdS_7 \times S^4$ or  $AdS_4 \times S^7$, and will be invariant under
a  subalgebra $\cH$ with 16 fermionic generators of respectively
$OSp(8^*|4)$ or $OSp(8|4,\bR)$.

\subsection{Half-BPS operators in M-theory}
\label{secthreeone}

The CFT$_6$ may be viewed as the decoupling limit of the $(0,2)$ worldvolume
theory of a stack of parallel M5 branes, while the CFT$_3$ may be defined as the
IR fixed point of a 2+1-dimensional supersymmetric $SU(N)$ SYM.
In both cases, the definition of the CFT does not provide a direct Lagrangian
description suitable to a weak coupling expansion of the quantum theory. This
situation makes the analysis of the CFT more complicated than in the case
of 3+1-dimensional $\cN=4$ SYM. (For recent progress on a Lagrangian formulation
for multiple $M2$-branes,  see e.g. \cite{Bagger:2007vi,Bagger:2007jr,Gustavsson:2007vu,Aharony:2008ug}).

\sm

The local half-BPS operators of CFT$_6$ were
classified in \cite{Aharony:1997an}. For  CFT$_6$ on $\bR \times S^{5}$,
the chiral primary operators are time-translation invariant, do not depend
on $S^{5}$, and preserve an $SO(3)$ subgroup of  R-symmetry of the
$(0,2)$ supersymmetry. The resulting bosonic symmetry associated
with localized half-BPS operators is thus $\bR \oplus SO(6) \oplus SO(3) $.
As far as we know, a systematic analysis of non-local half-BPS operators in
M-theory along the lines of the previous section has not been carried out.

\subsection{Intersecting branes}

The rules for supersymmetric intersections of M-branes are most easily derived by
descending to 10-dimensional Type IIA superstring theory via compactification on a circle.
The supersymmetric intersections of  M5 and M2 branes preserving 8 of the 32 supersymmetries are listed in Table \ref{table3}. A near-horizon limit may lead to an
enhancement of the Poincar\'e
and supersymmetry to superconformal symmetry on the intersection.

\begin{table}[htdp]
\begin{center}
\begin{tabular}{|c|c|c|c|} \hline
Case  &  branes & dim intersec    & bosonic symmetry
\\ \hline \hline
1 & M5/M5 &2   & $ ISO(1,1) \oplus SO(4)\oplus SO(4)$
\\  \hline
2 & M5/M2 &2 & $ ISO(1,1) \oplus SO(4) \oplus SO(4)$
 \\  \hline
3 & M5/M5 &4  & $ ISO(3,1) \oplus SO(3) \oplus SO(2) $
 \\  \hline
 4 & M2/M2 &1 & ~ $ \bR \oplus SO(6) \oplus SO(2) \oplus SO(2)$ ~
\\  \hline
\end{tabular}
\end{center}
\caption{Intersecting M2 and M5 branes with 8 supersymmetries}
\label{table3}
\end{table}

\subsection{Probe branes in $AdS_7\times S^4$ and $AdS_4\times S^7$ }

Supersymmetric probe branes can be considered in the $AdS_7\times S^4$
near-horizon limit of $M5$ branes and in the $AdS_4\times S^7$ 
near-horizon limit of $M2$ branes.  Just as in the case of $Dp$ branes discussed
in the previous section, one finds  stable probe branes with world-volumes
which are products of a lower dimensional $AdS$ space embedded into
$AdS_7/AdS_4$ and a lower dimensional sphere embedded into $S^4/S^7$
respectively. The supersymmetry of the
embedding follows from the existence of a $\kappa$-symmetry projector
as in the Type IIB case. The supersymmetry of the probe brane was
analyzed in \cite{Kim:2002tj,Lunin:2007ab}.

\sm

Probe branes preserving 16 supersymmetries in $AdS_7\times S^4$, and in
$AdS_4 \times S^7$ backgrounds are listed respectively in Table \ref{table4},
and Table \ref{table5}.

\begin{table}[htdp]
\begin{center}
\begin{tabular}{|c|c|c|c|} \hline
Case  & probe & worldvolume $M$    & bosonic symmetry
\\  \hline \hline
1& M5   &$ AdS_{3} \times S_{3}$  & $ SO(2,2) \oplus SO(4) \oplus SO(4) $
\\  \hline
2 & M2   &$ AdS_{3} $   & $ SO(2,2) \oplus SO(4) \oplus SO(4) $
\\  \hline
3 & M5   &$ AdS_{5} \times S^1$   & ~$ SO(2,4) \oplus SO(3) \oplus SO(2)  $~
\\  \hline
 \end{tabular}
\end{center}
\caption{Probe branes in $AdS_7\times S^4$ preserving 16 supersymmetries}
\label{table4}
\end{table}

\begin{table}[htdp]
\begin{center}
\begin{tabular}{|c|c|c|c|} \hline
Case  & probe & worldvolume $M$    & bosonic symmetry
\\ \hline \hline
4& M2   &$ AdS_{2} \times S^1$ & $ SO(2,1) \oplus SO(6) \oplus SO(2)  $
\\  \hline
5 & M5   &$ AdS_{3} \times S^3  $	& ~ $ SO(2,2) \oplus SO(4) \oplus SO(4) $~
 \\  \hline
 \end{tabular}
\end{center}
\caption{Probe branes in $AdS_4 \times S^7$ preserving 16 supersymmetries}
\label{table5}
\end{table}

Case 2 of Table \ref{table4}, and case 5 of Table \ref{table5} may be viewed
as different near-horizon limits of the M2/M5 intersecting brane configurations,
and thus both correspond to case 2 of Table~\ref{table3}. The other cases
of Tables \ref{table4} and  \ref{table5} match the remaining cases in Table \ref{table3}.
Note that, for cases 3 and  4, a supersymmetric embedding only preserves a single 
$SO(2)$ symmetry. 

\subsection{Exact supergravity solutions}

Supergravity solutions to 11-dimensional supergravity are constructed following
the same methods as were discussed in Section \ref{sectwofour} for Type IIB supergravity.
A purely bosonic field configuration will preserve 16 supersymmetries, provided
there exist 16 linearly independent spinors $\ep$ for which the supersymmetry
variation of the gravitino field vanishes. Using the supersymmetry transformations
of \cite{Cremmer:1978km}, the condition leads to the following BPS equation,
\bea
\label{gravim}
\nabla_M \ep +{1\over 288} \Big(\Gamma_M{}^{NPQR}
- 8 \delta_M{}^N \Gamma^{PQR} \Big) F_{NPQR} \, \ep=0
\eea
The metric is given by a fibration  (\ref{metans}) over a base space $\Sigma$,
and products of space $M_{i}$ respecting the expected symmetry of the solution.
For M-theory,  the total dimension of $M_{i}$'s and $\Sigma$ is eleven.
The 4-form field strength $F_{MNPQ}$ can depend on $\Sigma$ and on the invariant
volume forms of $M_{i}$.

\subsection{Known families of half-BPS solutions}

In this section, we shall  summarize the known or partially known half-BPS
solutions to 11-dimensional supergravity.

\sm

$\bullet$ ~
Half-BPS solutions dual to local half-BPS operators were constructed in \cite{Lin:2004nb}.
They have $SO(6) \oplus SO(3) \oplus \bR$ isometry and thus we have
$M_{1}=S^{5}, \, M_{2}=S^{2}, \, M_3 = \bR$.  The remaining base manifold
$\Sigma$ is 3-dimensional. In contrast to the Type IIB case, obtaining  solutions here
requires solving of a 3-dimensional Toda-like field equation with certain boundary
conditions imposed for regularity. This equation has not been solved
explicitly, so that solutions are only partially known. The solutions are asymptotically
$AdS_4 \times S^7$. They are invariant under the $SU(4|2)$
subalgebra of $OSp(8|4,\bR)$, which in turn leaves $AdS_4 \times S^7$
invariant.

\sm

$\bullet$ ~
Half BPS-solutions dual to a 1+1-dimensional planar defect/interface
in CFT$_3$,  and to a  2-dimensional  surface operator in CFT$_6$
were constructed exactly in \cite{DHoker:2008wc}, at the local  level. 
(For an earlier derivation of the BPS equations, and a 
partial solution, see \cite{Yamaguchi:2006te,Lunin:2007ab}.) Regularity conditions,
and global solutions are obtained in \cite{D'Hoker:2008qm}. The isometry
in both cases is given by $SO(2,2) \oplus SO(4) \oplus SO(4)$ and hence
$M_{1}=AdS_{3},M_{2}=M_3=S^{3}$. The base manifold
$\Sigma$ is a 2-dimensional Riemann surface with boundary and the solution
is completely specified by one harmonic function on $\Sigma$,
and a certain integral transform of two further harmonic functions
\cite{DHoker:2008wc}. For the M5 brane these supergravity solutions are the
dual description of   the self-dual string solution of the M5 brane worldvolume
theory found in  \cite{Howe:1997ue}.

\sm

For the 1+1-dimensional planar defect/interface
supergravity solutions, the asymptotic behavior is locally $AdS_4 \times S^7$,
and the  solutions are invariant under the
$OSp(4|2,\bR) \oplus OSp(4|2,\bR)$ subalgebra of
$OSp(8|4,\bR)$, which in turn leaves $AdS_4 \times S^7$ invariant.
For the 2-dimensional surface operator supergravity solutions, the
asymptotic behavior is locally $AdS_7 \times S^4$,
and the solutions are invariant under the
$OSp(4^*|2) \oplus OSp(4^*|2)$ subalgebra of
$OSp(8^*|4)$, which in turn leaves $AdS_7 \times S^4$ invariant.

\sm

$\bullet$ ~
Remarkably, the above families of half BPS-solutions on $AdS_3 \times S^3 \times S^3$
warped over a Riemann surface $\Sigma$ correspond to special points $c=1,-2,-1/2$
in a  family of solutions parametrized by one real parameter $c$ \cite{DHoker:2008wc}.
The discussion of the asymptotics and superalgebra
aspects of these general values of $c$ will be postponed until Section 7.2 below.

\newpage

\section{Lie superalgebras and their subalgebras}
\label{liesub}
\setcounter{equation}{0}

A Lie superalgebra $\cG$ is a  $\bZ_{2}$-graded associative algebra which admits
a unique decomposition $ \cG = \cGb \oplus \cGf$ under the
$\bZ_2$-grading into an even (or bosonic) subspace $\cGb$,
and an odd (or fermionic) subspace $\cGf$, and is equipped with a (graded) commutator,
\bea
[A,B]=A B- (-1)^{\a \b } B A
\eea
which satisfies a generalized  Jacobi-identity
\bea
(-1)^{\a \g }[A,[B,C]]+ (-1)^{\a \b  }[B,[C,A]]+(-1)^{\b \g } [C,[A,B]]=0
\eea
Here, $\a, \b, \g$ are the $\bZ_2$-gradings of $A,B,C \in \cG$ respectively.
A complete classification of simple Lie superalgebras was
obtained in  \cite{Kac:1977qb,Kac:1977em,Scheunert:1976uf,Scheunert:1976ug}.
A dictionary of  Lie superalgebras may be found in \cite{sorba}, which
also contains an extensive bibliography on superalgebras.

\sm

Under the commutator operation, the bosonic subspace $\cGb$ forms an ordinary
Lie algebra, referred to as the {\sl maximal bosonic subalgebra}.
The fermionic subspace $\cGf$ transforms under a representation of $\cGb$,
induced by the commutator. If $\cG$ is simple, and the representation
of $\cGf$ under $\cGb$ is completely reducible, then $\cG$ is referred to as
a {\sl classical Lie superalgebra}. If $\cG$ is classical and also has a non-degenerate
invariant bilinear form, then $\cG$ is referred to as a {\sl basic Lie superalgebra}.
In this paper, we shall restrict attention to the {\sl basic Lie superalgebras},
since they are closest in properties to ordinary Lie algebras,  and probably of
greatest interest in physics. Their classification is also known \cite{Kac:1977qb,Kac:1977em,Scheunert:1976uf,Scheunert:1976ug}.

\sm

In this section, we will focus on the basic Lie superalgebras  $PSU(2,2|4)$,
$OSp(8^*|4)$, and $OSp(8|4,\bR )$, of which we shall classify all basic Lie
subalgebras (or direct sums thereof) which have 16 fermionic generators.
In  Section 6, we will turn our attention to conformal superalgebras,
which are basic Lie superalgebras whose maximal bosonic subalgebra includes
a conformal algebra in various dimensions. The above constitute the Lie superalgebras
of greatest interest to AdS/CFT dualities. Since these facts will be needed throughout
the paper, we shall begin by reviewing, in the next subsections, the classification of
basic Lie superalgebras, their real forms, and their matrix representations.

\subsection{Basic  Lie superalgebras over $\bC$}

We begin by listing, in Table \ref{table6} below, all basic superalgebras over $\bC$,
their maximal bosonic subalgebras, rank, and fermion representation content.

\begin{table}[htdp]
\begin{center}
\begin{tabular}{|c||c|c|c|c|} \hline
Superalgebra $\cG$  & bosonic subalgebra $\cGb$
& rank & fermion rep $\cGf$  &  $\#(\cGf)$
\\ \hline \hline
$SL(m|n)$, {\small $m \not= n$} & $SL(m) \oplus SL(n) \oplus \bC$ & $m+n-1$
& $(m,n^*) \oplus (m^*,n)$ & $ 2mn$
\\  \hline
$PSL(m|m)$ & $SL(m)\oplus SL(m)$ & $2m-2$ & $(m,m^*) \oplus (m^*,m)$  & $ 2m^2$
\\  \hline
$OSp(m|2n)$ & $SO(m)\oplus Sp(2n)$ & $[m/2]+n$ & $(m,2n)$ & $ 2mn$
\\  \hline
$G(3)$ & $SL(2)\oplus G_2$ & 3 & $(2,7)$ & $ 14$
\\  \hline
$F(4)$ & $SL(2)\oplus SO(7)$ & 4 & $ (2, 8_s)$ & $ 16$
\\  \hline
$D(2,1; c) $ & $SL(2)\oplus SL(2) \oplus SL(2)$ & 3 & $(2,2,2)$ & $ 8 $
\\  \hline
 \end{tabular}
\end{center}
\caption{Basic  Lie superalgebras over $\bC$.}
\label{table6}
\end{table}

\noindent
We have the following isomorphisms
of basic  Lie superalgebras over $\bC$,
\bea
SL(m|n) & = &  SL(n|m)
\no \\
OSp(2|2) & = &SL(1|2)
\no \\
OSp(4|2) & = & D(2,1;1)
\no \\
D(2,1;c') & = & D(2,1;c) \hskip 1in c' \in A(c)
\eea
where the set $A(c)$ is defined by,
\bea
A(c) \equiv \left \{ c, 1/c, -(1+c)^{\pm 1}, -(1+1/c)^{\pm 1} \right \}
\eea

\subsection{Real forms}

In physics, and in particular in supergravity, superalgebras act on various fields,
some of which are invariably real. This is the case of the space-time metric,
and the antisymmetric 4-form field in Type IIB. In M-theory, the 3-form field
is real, and the gravitino is subject to a Majorana condition. Thus, we
consider not just Lie superalgebras over $\bC$, as in Table \ref{table6},
but we need also their various real forms, such as $PSU(2,2|4)$, $OSp(8^*|4)$,
and $OSp(8|4,\bR )$.

\sm

A real form of a Lie superalgebra $\cG$ over $\bC$ may be defined,
and constructed, as follows (see for example \cite{Parker:1980af,sorba}).
Let $\phi$ be an involution from $\cG$ to $\cG$,
which preserves the $\bZ_2$-grading of $\cG$, the graded commutator,
and which is a semi-linear map,
\bea
\label{real1}
\phi (\cG_\alpha) & = & \cG_\alpha \hskip 1.9in \a \in \{ \bar 0, \bar 1\}
\no \\
\phi ( [X,Y]) & = & [\phi (X), \phi (Y) ] \hskip 1.15in  X,Y \in \cG
\no \\
\phi (aX+bY) & = & \bar a \phi (X) + \bar b \phi (Y)  \hskip 1in a,b \in \bC
\eea
Here, $\bar a, \bar b$ stand for the standard complex conjugates of the complex
numbers $a,b$. The last condition gives to $\phi$ the meaning of a generalized
complex conjugation operation. Given such a $\phi$, the {\sl real Lie subalgebra}
$\cG^\phi$ may be defined by,
\bea
\cG^\phi \equiv \{ X + \phi (X), ~ X\in \cG \}
\eea
It is readily checked that $\cG^\phi$ is a Lie subalgebra of $\cG$; 
it is real in the sense that
$\phi (\cG^\phi )=\cG^\phi $. One refers to $\cG^\phi$ as the 
{\sl real form of $\cG$ associated with $\phi$.}

\sm

The complete list of possible real forms of the basic Lie superalgebras is reviewed in
Table~\ref{table7} below (see \cite{Kac:1977qb,Parker:1980af,sorba}). 
For the case of the Lie superalgebras $SL(m|m)$,
and their associated real forms, the $\bR$- or $U(1)$- term is to be removed in
the last column. For the  definitions of the real forms $SU(2m^*)$ and $SO(2m^*)$,
and isomorphisms between ordinary Lie algebras of low rank, see also Appendix A.
Of special importance to us will be the real forms $PSU(2,2|4)$, $OSp(8^*|4)$,
and $OSp(8|4,\bR )$, whose maximal bosonic subalgebras are respectively
$SU(2,2) \oplus SU(4)$, $SO(2,6) \oplus Sp(4)$, and $SO(8) \oplus Sp(4,\bR)$.

\begin{table}[htdp]
\begin{center}
\begin{tabular}{|c||c|c|} \hline
Superalgebra $\cG$ &  real form $\cG^\phi$ &  maximal bosonic subalgebra $\cGb^\phi$
\\ \hline \hline
$SL(m|n)$ & $SL(m|n,\bR)$ & $SL(m,\bR) \oplus SL(n,\bR) \oplus \bR$
\\
$SL(m+m'|n+n')$ 	& $SU(m,m'|n,n')$ & $SU(m,m') \oplus SU(n,n') \oplus U(1)$
\\
$SL(2m|2n)$	& $SL(2m|2n, \bH)$ & $SU(2m^*) \oplus SU(2n^*) \oplus \bR$
\\ \hline
$OSp(m+m'|2n)$ & $ OSp(m,m'|2n,\bR)$ & $SO(m,m') \oplus Sp(2n,\bR)$
\\
$OSp(2m|2n+2n')$ & $OSp(2m^*|2n,2n')$ & $SO(2m^*) \oplus Sp(2n,2n')$
\\ \hline
$F(4)$ &  $F(4;p)$, $p=0,1,2,3$ & $SL(2,\bR) \oplus SO(p,7-p)$
\\  \hline
$G(3)$  & $G(3;p)$,  $p=0,1$  & $SL(2,\bR) \oplus G_{2,2p}$
\\  \hline
$D(2,1;c) $ &  $D(2,1;c;p)$,   $ p=0,1,2$ & $SL(2, \bR)\oplus SO(p,4-p)$
\\  \hline
 \end{tabular}
\end{center}
\caption{Real forms of the basic  Lie superalgebras.}
\label{table7}
\end{table}

\noindent
We have the following isomorphisms of real forms of low rank,
\bea
\label{iso1}
OSp(2^*|2) & = & SU(2|1)
\no \\
OSp(2|2,\bR) & = & SU(1,1|1)
\no \\
D(2,1;1;0) & = & OSp(4|2,\bR)
\no \\
D(2,1;c;0) & = & OSp(4^*|2) \hskip 0.8in \hbox{for} ~ c=-2,-1/2
\no \\
D(2,1;1;1) & = & OSp(1,3|2,\bR)
\no \\
D(2,1;1;2)& = & OSp(2,2|2,\bR)
\no \\
D(2,1;c;0) & = & D(2,1;1/c;0)
\no \\
D(2,1;c;2) & = & D(2,1;c';2) \hskip 0.65in \hbox{for} ~ c' \in A(c)
\eea
as well as  $SU(n|m)=SU(m|n)$, and $SL(m|n,\bR)=SL(n|m,\bR)$ for all $m,n$.

\subsection{Matrix representation of the $SL$, $SU$ and $OSp$ series}

The Lie superalgebras in the $SL$, $SU$, and $OSp$ series may be defined
in terms of matrix representations, which are especially useful in
defining the real forms of these algebras.  Let $\cV = \cV_{\bar 0} \oplus \cV_{\bar 1}$
be a $\bZ_2$-graded vector space, where $\dim (\cV_{\bar 0}) = m$ and
$\dim (\cV_{\bar 1}) = n$. A supermatrix $M$ is a linear map of $\cV$ into $\cV$
which  may be decomposed as follows,
\bea
M \left ( \matrix{ \cV_{\bar 0} \cr \cV_{\bar 1} \cr} \right )
\to \left ( \matrix{ \cV_{\bar 0} \cr \cV_{\bar 1} \cr} \right )
\hskip 0.6in
M = \left ( \matrix{ A & B \cr C & D \cr} \right )
\eea
where $A,D$ are even grading and of dimension $m\times m$ and $n\times n$
respectively, and where $B,C$ are odd grading and of dimension $m\times n$
and $n \times m$ respectively. The general linear Lie superalgebra $GL(m|n)$
is defined to be the set of all such matrices over $\bC$. The supertrace
is defined by $\str (M) \equiv \tr (A) - \tr (D)$, and the supertranspose is defined by
\bea
M^{st} \equiv \left ( \matrix{ A^t & C^t \cr -B^t & D^t \cr} \right )
\hskip 1in
\left ( M^{st} \right )^{st} = \left ( \matrix{ A & -B \cr -C & D \cr} \right )
\eea
Note that the double supertranspose takes the value $(-1)^\alpha$ on
$\cV_\alpha$. The matrix representation of  $SL(m|n)$ and $OSp(m|2n)$
over $\bC$ are given by,
\bea
SL(m|n) & = & \Big \{ M \in GL(m|n), ~ \str(M)=0 \Big \}
\no \\
OSp(m|2n) & = & \Big \{ M \in GL(m|2n), ~ M^{st} K_{m|2n} + K_{m|2n} M=0 \Big \}
\eea
Here, the matrix $K_{m|2n}$ is defined as follows,
\bea
\label{KJ}
K_{m|2n} \equiv \left ( \matrix{ I_m & 0 \cr 0 & J_{2n} \cr} \right )
\hskip 1in
J_{2n} \equiv \left ( \matrix{ 0 & I_n \cr -I_n & 0 \cr} \right )
\eea
where $I_m$ is the identity matrix in dimension $m$. Note that
the expression for $K_{m|2n}$ in the definition of $OSp(m|2n)$
is consistent with the fact that the double supertranspose
reverses the sign of the odd parts $B,C$. It is readily shown that
the matrix realization given here is equivalent to the definition
given in \cite{Kac:1977qb}.

\sm

The matrix representations of the  real forms may be obtained
in a similar way. To do so, we define the following matrices,
\bea
L_{m,m'|n,n'} & \equiv & \left ( \matrix{ I_{m,m'} & 0 \cr 0 & -i I_{n,n'} \cr} \right )
\hskip 0.9in
I_{m,m'} \equiv \left ( \matrix{ I_m & 0 \cr 0 & - I_{m'} \cr} \right )
\no \\
K_{m,m'|2n} & \equiv & \left ( \matrix{ I_{m,m'} & 0 \cr 0 & J_{2n} \cr} \right )
\hskip 1in
\tilde K _{2m|2n,2n'} \equiv \left ( \matrix{ J_{2m} & 0 \cr 0 & I_{2n, 2n'} \cr} \right )
\eea
The matrix representation of real forms is given in terms of the
semi-linear involutions $\phi$ by,\footnote{To exhibit the involution $\phi$
more clearly on the last line, we may use the following alternative definition,
$OSp(m,m'|2n, \bR) =  \{
M \in OSp(m+m'|2n ), ~  M= \phi(M); ~ \phi(M) \equiv  I_{m,m'|2n}  M^* I_{m,m'|2n}  \}$.
The two definitions are related by a (complex) change of basis,
given by the square root of $I_{m,m'|2n}$.}
\bea
\label{matrixrep}
SL(m|n;\bR) & = & \Big \{ M \in SL(m|n), ~ M=\phi(M); ~ \phi (M) \equiv M^* \Big \}
\no \\
SU(m,m'|n,n') & = & \Big \{ M \in SL(m+m'|n+n'), ~
M=\phi (M); ~ \phi(M)  \equiv - L^{-1} (M^*)^{st} L    \Big \}
\no \\
OSp(2m^*|2n,2n') & = & \Big \{ M \in OSp(2m|2n+2n'), ~
M=\phi(M) ; ~ \phi (M) \equiv - \tilde K ^{-1} (M^*)^{st} \tilde K \Big \}
\no \\
OSp(m,m'|2n, \bR) & = & \Big \{
M \in SL(m+m'|2n;\bR ), ~  M= - K^{-1} M^{st} K  \Big \}
\eea
where we have used the abbreviations $L= L_{m,m'|n,n'}$, 
 $\tilde K = \tilde K_{2m|2n,2n'}$, and $K=K_{m,m'|2n}$.
The matrix representations of the real forms $SU$ and $OSp$ given here are
equivalent to the matrix formulations of \cite{Kac:1977qb}.

\subsection{Subalgebras of the complex Lie superalgebras $SL$ and $OSp$}

A systematic classification of all Lie subalgebras, for a given basic Lie superalgebra,
does not appear to be available. Therefore, we shall have to obtain the classification
of all Lie subalgebras of $PSU(2,2|4)$, $OSp(8^*|4)$, and $OSp(8|4,\bR )$
with 16 fermionic generators by drawing together the known results and
deriving here the remaining cases ourselves. From the outset, we shall
restrict to considering only Lie subalgebras which are direct sums of
basic Lie superalgebras and possibly ordinary simple Lie algebras.
These appear to be the only cases that will be needed for AdS/CFT applications.

\begin{table}[htdp]
\begin{center}
\begin{tabular}{|c||c|} \hline
Superalgebra  & subalgebra 
\\ \hline \hline
$SL(m+m'|n+n')$ & $SL(m|n) \oplus SL(m'|n') \oplus \bC$ 
\\ \hline
\quad $OSp(m+m'|2n+2n')$ \quad & \quad $OSp(m|2n) \oplus OSp(m'|2n')$ \quad
\\  \hline
$SL(m|2n) $ & $OSp(m|2n)$ 
\\  \hline
$OSp(2m|2n)$ & $SL(m|n) \oplus \bC$ 
\\  \hline
 \end{tabular}
\end{center}
\caption{Maximal subalgebras of the $SL$ and $OSp$ series over $\bC$.}
\label{table8}
\end{table}

The known cases consist of the maximal regular subalgebras
(for which the rank of the subalgebra coincides with the rank of the
original algebra), and those singular
subalgebras that can be obtained by {\sl folding} the root diagram,
both cases being considered as Lie superalgebras over $\bC$.
(The distinction between regular and singular subalgebras has to do with the root
structure of the subalgebra \cite{sorba}, and will not be needed here.) The results
for subalgebras of the $SL$ and $OSp$ series (subalgebras of
the exceptional cases will not be needed here) are listed in Table \ref{table8} above.
This list includes the cases where $m'n'=0$ by setting $SL(m|0)=SL(m)$,
$OSp(m|0)=SO(m)$, and $OSp(0|2n)=Sp(2n)$. Using the matrix representations
of the Lie superalgebras given in Section 4.2, it is straightforward to check that
the embeddings of Table \ref{table8} indeed hold.

\subsection{Subalgebras of the real form Lie superalgebras $SU$ and $OSp$}

From the results on Lie subalgebras over $\bC$, as well as from the matrix representations
for the real forms of the superalgebras, given in Section 4.2, certain subalgebras
of the real forms in the $SU$ and $OSp$ series may be deduced. The derivation 
of some of these results will be given in Appendix B. To illustrate here the key
issue in the construction of real form subalgebras, we shall begin by deriving 
the following general result.

\sm

Let $\cH$ be a subalgebra of the Lie superalgebra $\cG$ defined as the invariance 
subalgebra of a (bosonic) element $T \in \cG$,
\bea
\cH = \left \{ M \in \cG, ~ [T, M]=0 \right \}
\eea
Let $\phi$ be a semi-linear involution of $\cG$ which preserves its $\bZ_2$ grading,
and graded commutator. The complete definition of $\phi$  was given in (\ref{real1}). 
If there exists a $\lambda \in \bC$, such that
\bea
\label{phiT1}
\phi (T) = \lambda T
\eea
then the semi-linearity and involutive properties of $\phi$ imply that $|\lambda |=1$.
Any  $\phi$ satisfying (\ref{phiT1}) produces a unique real form $\cH^\phi$ of $\cH$
which is a subalgebra of the real form $\cG^\phi$, and
\bea
\cH^\phi = \cG ^\phi \cap \cH
\eea
The proof is immediate upon use of
$\phi ([T,M]) = [\phi(T), \phi(M)]= \lambda [T, \phi(M)]$, for any $M \in \cG$.

\sm

It is important to realize that a given real form $\cG^\phi$ of a complex Lie 
superalgebra $\cG$ may induce several different real forms of a given 
complex subalgebra $\cH \subset \cG$, provided  $\cH$ has several 
different embeddings into $\cG$. When two different embeddings 
$\cH_{(1)}$, and $\cH_{(2)}$ of $\cH$ into $\cG$ are equivalent to one another, 
namely $\cH_{(2)} = S \cH_{(1)} S^{-1}$ for some $S$ in the maximal bosonic 
subgroup of $\cG$, the real forms $\cH^\phi _{(1)}$ and $\cH^\phi _{(2)}$
will be isomorphic provided $\phi (S) = \lambda S$. If this relation does not hold, however,
the two real forms $\cH^\phi _{(1)}$ and $\cH^\phi _{(2)}$ will generally be inequivalent, 
even though $\cH _{(1)}$ and $\cH _{(2)}$ are equivalent  over $\bC$.

\sm

Henceforth, we shall restrict to the series of real forms needed
in this paper, namely $SU(m,m'|n)$, $OSp(m|2n,\bR)$,
and $OSp(2m^*|2n)$. They are given in Table \ref{table9} below.
The results on the first three lines of Table \ref{table9} are immediate
consequences of the matrix representations for these superalgebras
given in (\ref{matrixrep}), and are available from \cite{sorba}.
The results on the fourth and fifth lines have been known to hold, when $pq=0$,
for some time in the context of the oscillator realization of superalgebras
\cite{Bars:1982ep, Gunaydin:1985tc, Gunaydin:1990ag}. As we are
not aware of complete proofs elsewhere, these results will be proven generally in
Appendix B, together with those of the last four lines of Table \ref{table9},
 for the sake of completeness.

\begin{table}[htdp]
\begin{center}
\begin{tabular}{|c||c|} \hline
Real form superalgebra  & real form subalgebra 
\\ \hline \hline
$SU(p+p', q+q'|n+n')$ & $SU(p,q|n) \oplus SU(p',q'|n') \oplus U(1)$
\\ \hline
$OSp(m+m'|2n+2n', \bR)$ & $OSp(m|2n, \bR) \oplus OSp(m'|2n', \bR)$ 
\\ \hline
\quad $OSp(2m+2m'{}^*|2n+2n')$ \quad &
\quad $OSp(2m^*|2n) \oplus OSp(2m'{}^*|2n')$ 
\\  \hline
$OSp(2m^*|2n)$ & ~ $SU(p,q|n) \oplus U(1)$, ~$p+q=m$; ~ $p,q\geq 0$ ~
\\ \hline
$ OSp(2m|2n,\bR)$ & $SU(m|p,q) \oplus U(1)$, ~$p+q=n$; ~ $p,q\geq 0$ 
\\ \hline
$SU(m|n,n)$ & $OSp(m|2n,\bR )$ 
\\ \hline
$SU(m,m|2n)$ & $OSp(2m^*|2n)$ 
\\ \hline
 \end{tabular}
\end{center}
\caption{Maximal  subalgebras of relevant real forms of the $SU$ and $OSp$ series.}
\label{table9}
\end{table}

In addition, it will be proven in Appendix C that none of the real forms
of the  exceptional basic Lie superalgberas $F(4)$, $G(3)$, and
$D(2,1;c)$ for $c \not= 1, -2,-1/2$, is a subalgebra of $PSU(2,2|4)$,
$OSp(8^*|4)$, and $OSp(8|4,\bR )$.

\subsection{Basic subalgebras with 16 fermionic generators}
\label{subalone}

All basic superalgebras over $\bC$ (which are by definition simple)
with 16 fermionic generators are readily identified
from Table \ref{table6}; they are  $SL(4|2)$,  $SL(8|1)$, $OSp(8|2)$, $OSp(4|4)$,
$OSp(2|8)$, and $F(4)$. The algebras $SL(8|1)$ and  $OSp(2|8)$ are manifestly
{\sl not  subalgebras} of either $PSL(4|4)$, or $OSp(8|4)$, since their bosonic
subalgebras do not fit. Also, the $SO(8)$ part of the maximal bosonic  subalgebra
of $OSp(8|2)$ cannot fit into $PSL(4|4)$, so that $OSp(8|2) \not \subset PSL(4|4)$.
In Appendix C it is shown that $F(4)$
is not a subalgebra of either $PSL(4|4)$, or $OSp(8|4)$.
All remaining basic subalgebras over $\bC$ are given by,
\bea
\label{subs3}
PSL(4|4) & \supset &  SL(4|2), ~ OSp(4|4)
\no \\
OSp(8|4) & \supset & OSp(8|2), ~ OSp(4|4) , ~ SL(4|2)
\eea
It follows from Table \ref{table8} that these embedding relations over $\bC$
hold in all cases of (\ref{subs3}). Note that some of these embeddings are not
maximal and admit purely bosonic Lie algebra enhancements in the cases
$PSL(4|4)  \supset   SL(4|2) \oplus SL(2)$, and $OSp(8|4) \supset OSp(8|2)
\oplus Sp(2)$, and $OSp(8|4) \supset OSp(4|4)\oplus SO(4)$.

\sm

Next, we shall list the associated real forms. Several of these simple subalgebras 
are not in fact maximal subalgebras. In those cases, there exists a complementary 
purely bosonic subalgebra, which we shall list below in its maximal 
form.\footnote{In the first version of this paper, the inclusions of $SU(1,3|2)$, $SU(2,2|2)$,
and $SU(1,1|4)$ were listed incorrectly; their correct inclusion will lead to cases IIIb, IIIc, 
and VIc in section 5.5. We are grateful to Jaume Gomis for questions that led us to find 
and correct these omissions.}
\bea
\label{incl1}
PSU(2,2|4) 
	& \supset &  SU(2,2|2) \oplus SU(2), ~  
	SU(2|4) \oplus SU(2), 
\no \\ &&
	SU(1,1|4) \oplus SU(1,1),
\no \\
	&  & OSp(4^*|4), 
OSp(4|4,\bR)
\no \\
OSp(8^*|4) 
	 & \supset & OSp(8^*|2) \oplus Sp(2),
\no \\
	&& OSp(4^*|4) \oplus SO(4^*),
\no \\
	&&  SU(4|2) \oplus U(1), 
\no \\
	&& SU(1,3|2) \oplus U(1),
\no \\
	&&  SU(2,2|2) \oplus U(1)
\no \\
OSp(8|4,\bR) 
	& \supset &  OSp(8|2, \bR) \oplus Sp(2,\bR)
\no \\
	&&OSp(4|4,\bR) \oplus SO(4), 
\no \\
	&& SU(4|2) \oplus U(1), 
\no \\
	&& SU(4|1,1) \oplus U(1)
\eea
The existence of these embeddings  follows immediately from the results
of Table \ref{table9}. The fact that no other real forms of the complex  algebras are
subalgebras may be shown by systematically considering all such possible 
real forms.
The following cases may be easily excluded, by checking that the maximal bosonic
subalgebras do not fit,
\bea
PSU(2,2|4) & \not \supset &  SU(1,3|2), ~ SU(1,3|1,1), ~
 SU(2,2|1,1), 
\no \\ &&
OSp(4^*|2,2), ~OSp(p,4-p|4,\bR),  p=1,2
\no \\
OSp(8^*|4) & \not \supset & OSp(8|2,\bR) , ~ OSp(p,4-p|4,\bR), \, p=0,1,2,
\no \\ &&
SU(2,2|1,1), ~ SU(1,3|1,1)
\no \\
OSp(8|4,\bR) & \not \supset &  OSp(8^*|2) , ~ OSp(4^*|4) , ~
OSp(p,8-p|2,\bR), \, p=1,\cdots, 4,
\no \\ &&
SU(2,2|2), ~ SU(1,3|2),~ SU(2,2|1,1), ~ SU(1,3|1,1)
\eea

\subsection{Direct sum subalgebras with 16  fermionic generators over $\bC$}
\label{subaltwo}

Subalgebras with two simple Lie superalgebra components or more offer further possibilities.
We begin by presenting these subalgebras over $\bC$.
We list all simple basic superalgebras with strictly fewer than 16 fermionic generators
(indicated in the left column below),
and whose maximal bosonic subalgebra fits into either $PSL(4|4)$
or $OSp(8|4)$. They are.\footnote{The Lie superalgebras $OSp(1|6)$, $OSp(1|8)$,
$OSp(1|10)$,  $OSp(1|12)$, $OSp(1|14)$, $OSp(2|6)$,  $SL(5|1)$, $SL(6|1)$,
and $SL(7|1)$ also have fewer than 16 fermionic generators,
but they manifestly do not fit into either $PSL(4|4)$  or $OSp(8|4)$.
Thus, they are excluded from the list.}
\bea
2 & \hskip 0.7in & OSp(1|2)
\no \\
4 & & SL(2|1)= OSp(2|2), ~ OSp(1|4)
\no \\
6 & & SL(3|1),  ~ OSp(3|2)
\no \\
8 & & PSL(2|2), OSp(4|2), ~ OSp(2|4), ~ SL(4|1),  ~  D(2,1;c)
\no \\
10 &&  OSp(5|2)
\no \\
12 &&  SL(3|2),  ~ OSp(3|4), ~ OSp(6|2)
\no \\
14 &&  OSp(7|2), ~ G(3)
\eea
In Appendix C, it will be proven that the exceptional superalgebras
$G(3)$, and $D(2,1;c)$ with $c \not= 1, -2, -1/2$,
are not subalgebras of $PSL(4|4)$ or $OSp(8|4)$.

\sm

To classify the non-exceptional cases, we shall make use of the subalgebra
embeddings given in Table \ref{table8}.
From the table, we deduce that no direct sum involving  $SL(4|1)$ can fit either
in $PSL(4|4)$ or $OSp(8|4)$, because the $SL(4)$ component
of $SL(4|1)$ is already maximal both in $PSL(4|4)$ and in $OSp(8|4)$.
The same arguments lead us to eliminate the possibility of
having $OSp(1|4)$, $OSp(2|4)$ and $OSp(3|4)$ as subalgebras.
The above considerations reduce the number of possible
subalgebra components  to the following list,
\bea
2 && OSp(1|2)
\no \\
4 & \hskip 1in & SL(2|1)=  OSp(2|2)
\no \\
6 & & SL(3|1), ~ OSp(3|2)
\no \\
8 & & PSL(2|2), ~  OSp(4|2)
\no \\
10 && OSp(5|2)
\no \\
12 &&  SL(3|2), ~ OSp(6|2)
\no \\
14 && OSp(7|2)
\eea
The possible subalgebra components may be enumerated by ensuring that
the total number of fermionic generators equals 16, and that the maximal bosonic
subalgebras fit. As was the case with the basic subalgebras, an extra bosonic 
subalgebra may be allowed. In the table below, we shall include the maximal
version of such a bosonic subalgebra, whenever it is allowed.  
Lie superalgebras with two or more components can only have $\bC$, $U(1)$, or 
$\bR$  as bosonic complements.

\sm

The resulting subalgebra embeddings are as follows,
\bea
\label{fact}
PSL(4|4)  & \supset &  PSL(2|2)\oplus PSL(2|2) \oplus \bC^2,
\no \\ && SU(2|3) \oplus SU(2|1)
\no \\
OSp(8|4) & \supset & OSp(4|2)  \oplus OSp(4|2),
\no \\ && OSp(5|2) \oplus OSp(3|2),
\no \\ && OSp(6|2) \oplus OSp(2|2)
\no \\ && OSp(7|2) \oplus OSp(1|2) ,
\eea
The non-trivial cases are eliminated as follows.
\begin{enumerate}
\item $SL(3|2) \oplus OSp(2|2)$ is not a subalgebra of $OSp(8|4)$. Note that
both $OSp(8|4)$ and $SL(3|2) \oplus OSp(2|2)$ are of rank $6$. If the embedding
existed, it would have to be a maximal regular subalgebra, but all such sub-algebras
of $OSp(8|4)$ are known, and $SL(3|2) \oplus OSp(2|2)$ is not amongst them.
(For example, $OSp(6|2) \oplus OSp(2|2)$ is a regular maximal subalgebra of
$OSp(8|4)$, but $SL(3|2)$ does not fit into $OSp(6|2)$.)
Finally, since the ranks of $SL(3|2) \oplus OSp(2|2)$ and $OSp(8|4)$ are equal,
the embedding cannot be that of a non-regular subalgebra either.
\item $PSL(2|2) \oplus OSp(4|2)$ is not a subalgebra of $OSp(8|4)$.
Note that the rank of $PSL(2|2)$ is only 2, so the total rank of
$PSL(2|2) \oplus OSp(4|2)$ is 5. This case is ruled out as follows.
While the maximal bosonic subalgebra $SO(4)\oplus Sp(2)
\oplus SL(2) \oplus SL(2)$ fits, the fermionic generators in $PSL(2|2)$ transform
in the spin 1/2 representation of the $SL(2)$ components. But the first
$SL(2)$ component is embedded into $SO(8)$ by the spin 1 representation of
that same algebra, so that the fermionic generators do not fit.
\end{enumerate}
Finally, subalgebras with three simple components or more are all ruled out by
analogous arguments. Thus, the cases of (\ref{fact})  exhaust all possible
subalgebras.

\subsection{Direct sum subalgebras with 16 real fermionic generators}

The associated real forms are as follows,
\bea
\label{incl2}
PSU(2,2|4)
& \supset &
	PSU(2|2) \oplus PSU(2|2) \oplus \bR^2 ,
\no \\ && PSU(1,1|2) \oplus PSU(1,1|2) \oplus U(1)^2 ,
\no \\ && SU(2|3) \oplus SU(2|1),
\no \\ && SU(1,1|3) \oplus SU(1,1|1),
\no \\ && SU(1,2|2) \oplus SU(1|2)
\no \\
OSp(8^{*}|4)  & \supset & OSp(4^*|2)  \oplus OSp(4^*|2),
\no \\
&& OSp(2^*|2)  \oplus OSp(6^*|2)\no \\
OSp(8|4, \bR) & \supset &
	OSp(7|2, \bR) \oplus OSp(1|2, \bR) ,
\no \\ &&  OSp(6|2, \bR ) \oplus OSp(2|2, \bR),
\no \\ &&  OSp(5|2, \bR ) \oplus OSp(3|2, \bR),	
\no \\ &&  OSp(4|2, \bR ) \oplus OSp(4|2, \bR)
\eea
Note that $OSp(2^*|2) \oplus OSp(6^*|2)$ is a subalgebra of $OSp(8^*|4)$.
The orthogonal parts of the bosonic subalgebras are respectively given by
$SO(2^*) \oplus SO(6^*) = SO(2) \oplus SU(1,3)$ and $SO(2,6)$. The
subalgebra $SU(1,3)$ is maximal in $SO(2,6)$, and the generator commuting
with $SU(1,3)$ inside $SO(2,6)$ is indeed a compact $SO(2)$, whence the result.

\subsection{Enhanced supersymmetries}

Finally, we note that in various cases, there exist subalgebras $\cH^*$ of
the superalgebras $PSU(2,2|4)$, $OSp(8^*|4)$, or $OSp(8|4,\bR)$ whose number of
fermionic generators actually exceeds 16, but remains less than 32
(or even less than 28). If $\cH^*$  contains one of the subalgebras $\cH$
with precisely 16 fermionic generators that were classified
in this section, then it may play the role of a symmetry algebra for half-BPS
solutions with an enhanced supersymmetry. The systematic classification
of such superalgebras $\cH^*$ will be carried out in Section 7.1,
where their significance to the existence of half-BPS solutions with enhanced
supersymmetry will be discussed as well.

\newpage

\section{Half-BPS supergravity solutions and superalgebras}
\setcounter{equation}{0}

\label{secfive}

In the present section, we shall spell out and prove generally the correspondence
between superalgebras $\cH$ and the existence of half-BPS solutions. The
classification of superalgebras $\cH$ with 16 fermionic generators which are
subalgebras of $PSU(2,2|4)$,  $OSp(8^*|4)$, and $OSp(8|4,\bR)$, obtained in
Section 4, allows us then to derive a classification of all possible families of half-BPS
solutions which are locally asymptotic to respectively $AdS_5 \times S^5$,
$AdS_7 \times S^4$, and $AdS_4 \times S^7$ for each associated supergravity.

\sm

All known families of half-BPS solutions may be identified in this classification.
The remaining cases, where solutions are not known at this time, will be discussed,
and arguments for either their existence or their absence will be developed
using input from the AdS/CFT correspondence and from intersecting brane and
probe brane dynamics.

\subsection{The superalgebra - half-BPS solution correspondence}

The precise statement of the correspondence is as follows.

\sm

{\sl If a half-BPS solution of Type IIB  is locally asymptotic to $AdS_5 \times S^5$,
or a half-BPS solution of M-theory is locally asymptotic to 
$AdS_7 \times S^4$ or $AdS_4 \times S^7$,
and invariant under a superalgebra $\cH$ (with 16 fermionic generators since
the solution is assumed to be half-BPS), then $\cH$ must be a subalgebra of
$PSU(2,2|4)$,  $OSp(8^*|4)$ or $OSp(8|4,\bR)$ respectively.}

\sm

By {\sl locally} asymptotic, we mean here that the boundary can consist of several
components, each of which is asymptotic to $AdS_5 \times S^5$,
$AdS_7 \times S^4$ or $AdS_4 \times S^7$.

\sm

The compact factors of the above space-times cause no difficulty in the definition
of this asymptotic behavior.  The precise definitions of asymptotically $AdS_3$ and
$AdS_4$ were given respectively in \cite{Brown:1986nw} and \cite{HennTeit},
and of asymptotically $AdS_{d+1}$, for general $d$, may be found, for example in 
\cite{Hollands:2005wt,Skenderis:2002wp}.
We shall reproduce it here for the sake of completeness and clarity. In local adapted
coordinates $x^\mu = (t,r,\theta ^i)$, for $\mu = 0,r,i$ and $i=1,\cdots, d-1$,
a locally asymptotically $AdS_{d+1}$ metric $ds^2$ on a manifold $M$ takes the form,
\bea
ds^2 & = & d\hat s ^2_R + h _{\mu \nu} dx^\mu dx^\nu
\no \\
d\hat s ^2_R & = & - \left ( 1 + { r^2 \over R^2} \right ) dt^2
+  \left ( 1 + { r^2 \over R^2} \right ) ^{-1} \! dr^2
+ r^2 d\hat s ^2 _{S^{d-1}}
\eea
Here, $d\hat s ^2 _R$ is the $SO(2,d)$-invariant metric on $AdS_{d+1}$
with radius $R$, expressed in terms of the $SO(d)$-invariant metric
$d\hat s ^2 _{S^{d-1}}$ on $S^{d-1}$ with radius 1.
The boundary of $AdS_{d+1}$, and thus of $M$, is located at $r= \infty$.

\sm

The metric $ds^2$ is locally asymptotic to the metric $d\hat s^2 _R$ if the deformations
away from $d\hat s^2 _R$ (which are parametrized by $h _{\mu \nu}$)
obey the following asymptotics as $r\to \infty$,
\bea
\label{asymph}
h_{tt} & = & r^{2-d} \, \g_{tt} (t,\theta ^i) + \cO \left ( r^{1-d} \right )
\no \\
h _{tr} & = & r^{-1-d} \, \g_{tr} (t,\theta ^i) + \cO \left ( r^{-2-d} \right )
\no \\
h _{ti}  & = & r^{2-d} \, \g_{ti} (t,\theta ^i) + \cO \left ( r^{1-d} \right )
\no \\
h_{rr}  & = & r^{-2-d} \, \g_{rr} (t,\theta ^i) + \cO \left ( r^{-3-d} \right )
\no \\
h _{ri} & = & r^{-1-d} \, \g_{ri} (t,\theta ^i) + \cO \left ( r^{-2-d} \right )
\no \\
h_{ij} & = & r^{2-d} \, \g_{ij} (t,\theta ^i) + \cO \left ( r^{1-d} \right )
\eea
The reduced metric functions $\g_{\mu \nu}$ are allowed to be arbitrary
functions of $t$ and $\theta ^i$.

\sm

The notion we need next is that of {\sl asymptotic symmetry} of $M$, a notion that
found its origins in gauge theory, and was extended long ago to general relativity
\cite{Abbott:1981ff,Brown:1986nw,HennTeit}. Simply put for the case at hand,
the asymptotic symmetry of the asymptotically $AdS_{d+1}$ manifold $M$ is $SO(2,d)$.
This may be understood as follows. Under the isometry $SO(2,d)$ of 
$AdS_{d+1}$, interior points of $AdS_{d+1}$ are mapped to interior points, 
while points on the boundary of $AdS_{d+1}$ are mapped to boundary 
points. The action of $SO(2,d)$ induces conformal transformations on 
the boundary of $AdS_{d+1}$. It therefore makes sense to associate
an asymptotic symmetry with any asymptotically $AdS_{d+1}$ space-time,
and this symmetry is precisely $SO(2,d)$. The group $SO(2,d)$ then
transforms the form of the boundary conditions (\ref{asymph}) into itself.

\sm

More mathematically,  the asymptotic symmetry may be viewed as the quotient
of the group of all diffeomorphisms {\sl Diff(M)} of $M$ by the group {\sl Diff}$_\infty (M)$
of all diffeomorphisms that leave the asymptotic conditions (\ref{asymph}) invariant.
Thus one has {\sl Diff}$(M)$/{\sl Diff}$_\infty(M)= SO(2,d)$. 

\sm

When the boundary of a connected locally asymptotically $AdS_{d+1}$ manifold
counts several disconnected  components, the asymptotic symmetry algebra
remains a single copy of $SO(2,d)$. This follows from the fact that the imposition
of asymptotic boundary conditions on the Killing vectors on one boundary
component fixes the Killing vectors completely on all other boundary components.

\sm

Extending the concept of asymptotic symmetry to include supersymmetry \cite{Brown:1986nw}
is achieved by collecting  Killing vectors and Killing spinors, and considering
the superalgebra of their asymptotic actions on the various fields on the manifold,
such as the metric. In this way, the asymptotic symmetry
of an asymptotically $AdS_5 \times S^5$ manifold in Type IIB theory, and of
asymptotically $AdS_7 \times S^4$ and  $AdS_4 \times S^7$ manifolds in
M-theory are $PSU(2,2|4)$, $OSp(8^*|4)$, and $OSp(8|4,\bR)$ respectively.

\sm

Global charges, which give a representation of the algebra of global symmetries
on the fields, are defined in terms of the asymptotic behavior of the fields
and are insensitive to the sub-leading behavior in (\ref{asymph}).
As a result, the global charges must span a subalgebra of the asymptotic
symmetry algebra. Thus, any symmetry of an asymptotically
$AdS_5 \times S^5$ solution in Type IIB theory, and of asymptotically
$AdS_7 \times S^4$ and  $AdS_4 \times S^7$ solutions in  M-theory
must be a subalgebra of  $PSU(2,2|4)$, $OSp(8^*|4)$, or $OSp(8|4,\bR)$ respectively.

\sm

Note that the converse of the correspondence does not hold. Namely, it
does not follow from the above arguments that every solution whose
symmetry superalgebra $\cH$ is a subalgebra of  $PSU(2,2|4)$, $OSp(8^*|4)$, 
or $OSp(8|4,\bR)$, is necessarily respectively asymptotically $AdS_5 \times S^5$,
asymptotically $AdS_7 \times S^4$ or asymptotically $AdS_4 \times S^7$,
and indeed, this converse property does not appear to hold.

\subsection{Construction of the space-time Ansatz}

Given the maximal bosonic subalgebra $\cH_{bos}$ of one of the subalgebras
$\cH$, the corresponding space-time Ansatz is generally not unique. It is always
possible to restrict, without loss of generality, the space-time Ansatz so that each
component space of the Ansatz consists of a single orbit of each simple or
$U(1)$-component of  $\cH_{bos}$. Families of different orbits may then be
assembled using the remaining free parameters of the Ansatz. 
For example, the algebra $SO(3)$ may be realized on either $S^2$, or on $S^3$,
since the latter is itself a group manifold, namely the double cover of $SO(3)$.
The realization on $S^3$, however, is a special case of the realization on $S^2$,
as $S^3$ may be viewed as the Hopf bundle over $S^2$. As a result, $SO(3)$
needs only be realized on $S^2$. Even after this
simplification, however, the space-time component in the Ansatz may not be
unique. Below we shall discuss precisely how this happens.

\sm

The simple and $U(1)$ components of $\cH_{bos}$ that will be needed in our
analysis may be read off from the classification of subalgebras $\cH$  in (\ref{incl1})
and  (\ref{incl2}), and their maximal bosonic subalgebras $\cH_{bos}$ in Table \ref{table9}.
They are $SO(n+1)$ for $n=0,\cdots, 7$, $SO(2,n)$ for $n=1,2,3,4,6$,
$SU(n+1)$ for $n=1,2,3$ and $SU(1,n)$ for $n=1,2,3$. (Here, we have
used the coincidences at low rank, listed in Appendix A, to replace the required
$Sp(2n), Sp(2n,\bR)$ and $SO(2n^*)$  by equivalent algebras in the $SO$ and
$SU$ series.) The standard  realizations of these symmetries on corresponding
homogeneous spaces are as follows,
\bea
 SO(n+1)  \hskip 1.1in  S^n & = & SO(n+1)/SO(n)
\no \\
SO(2,n)   \hskip 0.8in AdS_{n+1} & = & SO(2,n)/SO(1,n)
\no \\
SU(n+1)  \hskip 0.95in CP_n & = & SU(n+1)/S[U(1)\times U(n)]
\no \\
SU(1,n)  \hskip 1in CH_n & = & SU(1,n)/S[U(1)\times U(n)]
\eea
Here, $CP_n$ is (compact) complex projective space, while $CH_n$ is (non-compact)
complex hyperbolic space-time (see for example \cite{helgason}, \cite{Besse},
and especially \cite{KN}, page 282). Recall that
$CP_n$ and $CH_n$ for $n\geq 2$ are symmetric spaces but, unlike the manifolds
$S^n$ and  $AdS_{n+1}$, they are not maximally symmetric. As a result, $CP_n$
and $CH_n$ will support only half of the maximal number of Killing spinors.
This forces the dilaton to be constant, and the axion and 3-form field strengths
to vanish for the corresponding supergravity solutions.

\sm

Several of these low-rank algebras admit, however, several equivalent and
inequivalent realizations, which we shall now discuss.
The algebra $SO(4) = SO(3) \oplus SO(3)$  may be realized on either
$S^3$, or on $S^2 \times S^2$, and these cases need to be considered
separately, as they are inequivalent. Analogously, 
$SO(2,2) = SO(1,2) \oplus SO(1,2)$ could in principle be realized
in two inequivalent ways as well, namely on $AdS_3$ or on $AdS_2 \times AdS_2$.
Unlike its compact counterpart, $AdS_2 \times AdS_2$ cannot correspond to a
physically acceptable space-time manifold, and will therefore be excluded
from consideration. 

\sm

The realization of $SO(5)=Sp(4)$ on $S^4$ is unique, since we 
have $S^4= Sp(4)/[Sp(2) \times Sp(2)]$.
The realization of $SO(6)=SU(4)$ may be either on $S^5$ or on $CP_3$.  
However, in the case of Type IIB supergravity, $CP_3$ produces a space that 
is not asymptotically $AdS_5 \times S^5$, and is therefore to be excluded.  
In the case of M-theory, $CP_3$ is allowed for solutions asymptotic to 
$AdS_4 \times S^7$, but not to $AdS_7 \times S^4$.  Similarly, its hyperbolic 
counterpart may be used for solutions asymptotic to $AdS_7 \times S^4$,
but not to $AdS_4 \times S^7$. 

\sm

Similarly, $SU(3)$ can in principle be realized 
on the space $SU(3)/SO(3)$ instead of  $CP_2$. The space $SU(3)/SO(3)$
is a 5-dimensional compact manifold which is topologically and metrically different 
from $S^5$,  and can not lead to a solution which is asymptotic to 
$AdS_5 \times S^5$. The manifold  $SU(1,2)/SO(1,2)$ is  5-dimensional 
with signature $(++---)$ 
and is thus not a viable candidate either. The space  $SU(4)/SO(4)$, having dimension 9, 
is excluded on dimensional grounds. Finally, space-times which contain the products 
$CP_2 \times S^1$ and $CH_2 \times S^1$ may be warped into $S^5$ and $AdS_5$ respectively, so that the corresponding full space-time can be  asymptotic to 
$AdS_5 \times S^5$.

\subsection{Half-BPS solutions in Type IIB supergravity}
\label{sec61}

For half-BPS solutions asymptotic to $AdS_{5}\times S^{5}$ in Type IIB,
the relevant superalgebra is $PSU(2,2|4)$. In Table \ref{table10} we  list all simple
and direct sum subalgebras $\cH$ with 16 fermionic generators of $PSU(2,2|4)$, their
maximal bosonic subalgebra $\cH_{bos}$,  the corresponding Ansatz for the space-time 
manifold of the solution, as derived in subsection 5.2, and in the last column entitled  ``brane"
we list the corresponding intersecting and probe brane cases of Tables \ref{table1} 
and \ref{table2}.  In each case, there is a base
space of parameters which has either Euclidean or Minkowski
signature and over which the products of the symmetric spaces are fibered.
The Euclidean base spaces of dimension $d$ are denoted by $\Sigma_d$,
while the Minkowski ones are denoted by $M_d$.

\begin{table}[t]
\begin{center}
\begin{tabular}{|c|c|c|c|c|} \hline
case& $\cH$ & $\cH_{bos}$ &  space-time Ansatz &  {\small brane} 
\\ \hline \hline
{\small I}
& {\small $SU(2|4)$} 
& {\small $SO(3) \oplus  SO(6) \oplus  SO(2)$}
& {\small $M_2 \tim S^2 \tim S^5 \tim S^1$}
&
\\ \hline
{\small I$^*$}
& {\small $SU(2|4) \oplus SU(2) $} 
& {\small $SO(4) \oplus  SO(6) \oplus  SO(2)$}
& {\small $M_1 \tim S^3 \tim S^5 \tim S^1$}
& 
\\ \hline
{\small II}
& {\small $SU(1,1|4)$ }
& {\small  $SO(2,1) \oplus  SO(6) \oplus  SO(2)$}
& {\small $AdS_2  \tim S^5 \tim S^1 \tim \Sigma_2$}
& 
\\ \hline
{\small II$^*$}
& {\small $SU(1,1|4) \oplus SU(1,1) $} 
& {\small $SO(2,2) \oplus  SO(6) \oplus  SO(2)$}
& {\small $AdS_3  \tim S^5 \tim S^1 \tim \Sigma_1$}
& {\small 4}
\\ \hline
{\small III}
& {\small $SU(2,2|2)$} 
& {\small $SO(2,4) \oplus  SO(3) \oplus  SO(2) $}
& {\small $AdS_5  \tim S^2 \tim S^1 \tim \Sigma _2$}
& 
\\ \hline 
{\small III$^*$}
& {\small $SU(2,2|2) \oplus SU(2) $} 
& {\small $SO(2,4) \oplus  SO(4) \oplus  SO(2) $}
& {\small $AdS_5  \tim S^3 \tim S^1 \tim \Sigma _1$}
& {\small 6}
\\ \hline
{\small IV}
& {\small $OSp(4^*|4)$} 
& {\small $SO(2,1) \oplus  SO(3) \oplus  SO(5) $}
& {\small $AdS_2 \tim S^2 \tim S^4 \tim \Sigma_2 $}
& {\small 1,2}
\\ \hline
{\small Va}
& {\small $OSp(4|4,\bR)$} 
& {\small $SO(2,3) \oplus  SO(3)^2$}
& {\small $AdS_4 \tim S^2 \tim \! S^2 \tim \Sigma_2 $}
& {\small 5}
\\ \hline
{\small Vb}
& {\small $OSp(4|4, \bR)$} 
& {\small $SO(2,3)\oplus  SO(4) $}
& {\small $AdS_4 \tim S^3 \tim \Sigma _3$}
& 
\\ \hline
{\small VIa}
& {\small $PSU(2|2)^2 \oplus \bR $}  
& {\small $SO(4) \oplus  SO(4) \oplus \bR$}
& {\small $M_3 \tim \bR \tim S^3 \tim S^3 $}
& 
\\ \hline
{\small VIb}
& {\small $PSU(2|2)^2 \oplus \bR$}  
& {\small $SO(4) \oplus  SO(3)^2 \oplus \bR$}
& {\small $M_2 \tim \bR \tim S^3 \tim S^2 \tim S^2$}
& 
\\ \hline
{\small VIIa}
& {\small $PSU(1,1|2)^2 \oplus U(1)$} 
& {\small $SO(2,2)\oplus  SO(4) \oplus SO(2)$}
& {\small $AdS_3 \tim S^3 \tim S^1 \tim \Sigma_3$}
& {\small 3}
\\ \hline
{\small VIIb}
& {\small $PSU(1,1|2)^2 \oplus U(1) $} 
& {\small $SO(2,2) \! \oplus \! SO(3)^2 \! \oplus \! SO(2)$}
& {\small $AdS_3 \tim \! S^2 \! \tim S^2 \! \tim S^1 \! \tim \Sigma_2$}
& 
\\ \hline
{\small VIII}
& {\small $SU(2|3) \oplus  SU(2|1)$} 
& {\small $SU(3) \oplus  SO(3)^2 \oplus  SO(2)^2$}
& {\small $M_1 \tim CP_2\tim S^3 \tim S^1 \tim S^1$}
&
\\ \hline
{\small IX}
& {\small $\! SU(1,1|3) \! \oplus \! SU(1,1|1) \! $} 
& {\small $SO(2,2)\! \oplus \! SU(3) \! \oplus \! SO(2)^2$}
& {\small $AdS_3  \tim  CP_2 \tim \! S^1 \! \tim \! S^1 \! \tim \Sigma _1 $}
&
\\ \hline
{\small X}
& {\small $SU(1,2|2) \oplus  SU(1|2) $} 
& {\small $SU(1,2) \! \oplus \! SU(2)^2 \! \oplus \! SO(2)^2   $}
& {\small $  M_1 \tim CH_2 \tim S^3 \tim S^1 \tim S^1 $}
&
\\ \hline \hline
\end{tabular}
\end{center}
\caption{Subalgebras of $PSU(2,2|4)$ with 16 fermionic generators}
\label{table10}
\end{table}

The superalgebra $\cH$ is a {\sl maximal subalgebra} of $PSU(2,2|4)$ for cases 
I$^*$, II$^*$, III$^*$, IV, Va, Vb, VIII, IX,  and X. Cases I, II, and III may be obtained 
from cases I$^*$, II$^*$, III$^*$ by removing their respective purely bosonic invariant 
subalgebras $SU(2)$, $SU(1,1)$, and $SU(2)$. The space of solutions for cases 
I$^*$, II$^*$, III$^*$ forms a sub-family (with enhanced purely bosonic symmetries)
of the space of solutions for cases I, II, and III, respectively. Note that it is possible
as well to have solutions with partially enhanced bosonic symmetry, associated
with a non-trivial subalgebra of the corresponding purely bosonic algebra.
For $SU(2)$, we can have a $U(1)$ subalgebra, while for $SU(1,1)$, the subalgebra
can be $U(1)$, $SO(1,1)$ or even the 2-dimensional algebra consisting of a 
boost and a translation.

\sm

Cases VIa, VIb, and cases VIIa, VIIb have Abelian purely bosonic invariant subalgebras, 
and are listed in Table \ref{table10} with one factor of $\bR$ or $U(1)$ respectively. 
This factor is not required by the mandate of 16 supersymmetries, and could be dropped,
resulting in less bosonic symmetry for the space-time Ansatz and corresponding solutions.
On the other hand, solutions could exist with an extra factor of $\bR$ or $U(1)$,
since the maximal subalgebras of $PSU(2,2|4)$ contain two factors of $\bR$ or $U(1)$
respectively.

\sm

Several of the cases in Table \ref{table10} correspond to existing families of 
known fully back-reacted  half-BPS solutions in Type IIB supergravity. For other cases, 
where no non-trivial solutions are known yet,  the knowledge of a suitable 
superalgebra $\cH$ will lead us to suggest the existence of new families of 
half-BPS solutions. We shall discuss each case in turn below.

\begin{itemize}

\item Cases I and I$^*$: The exact supergravity solutions for case I are discussed in
\cite{Lin:2004nb,Lin:2005nh,vanAnders:2007ky,Shieh:2007xn}. They are dual to local 
gauge invariant half-BPS operators in $\cN=4$ super-Yang-Mills. Case I$^*$ would
correspond to a sub-family of solutions with enhanced bosonic symmetry; it is not
known whether any such non-trivial  solutions exist.

\item Cases II and II$^*$:  No non-trivial solutions are known to exist for either case.
The bosonic symmetries and space-time Ansatz of case II$^*$ precisely match 
those of the D7/D3 intersection case 4 in Table \ref{table1}, and corresponding
D7 probe case 4 of Table \ref{table2}. A more detailed discussion is postponed 
until subsection 5.4.

\item Cases III and III$^*$:  No non-trivial solutions are known to exist for either case,
but these cases are presently under investigation in \cite{yu}.
The bosonic symmetries and space-time Ansatz of case III$^*$ precisely match 
those of the D7/D3 intersection case 6 in Table \ref{table1}, and corresponding
D7 probe case 6 of Table \ref{table2}. A more detailed discussion is postponed 
until subsection 5.4.

\item Case IV:  This case corresponds to holographic duals of supersymmetric
Wilson loops \cite{Gomis:2006im,Gomis:2006sb}, and is associated with the  
intersecting brane and probe brane cases 1, and 2 in Tables \ref{table1} and 
\ref{table2}. The fully back-reacted supergravity solutions were obtained exactly in
\cite{DHoker:2007fq}, following earlier work in \cite{Yamaguchi:2006te,Lunin:2006xr}.
Applications to Wilson loops at strong coupling in the CFT is discussed in
\cite{Gomis:2008qa}.

\item Case V: Case Va corresponds to the holographic duals of 2+1 dimensional
defects, and is associated with the intersecting brane and probe brane case 5
of  Tables \ref{table1} and \ref{table2}. The fully back-reacted supergravity
solutions were obtained  exactly in  \cite{DHoker:2007xy,DHoker:2007xz},
following earlier work in \cite{Gomis:2006cu}. No non-trivial  solutions are known 
for Case Vb; since there is an $AdS_{4}$ factor, such solutions should 
be dual to an interface or defect.

\item Case VI: Case VIa is the ``bubbling AdS" solution of \cite{Lin:2004nb},
which is dual to local half-BPS operators.  No non-trivial half-BPS supergravity
solutions are known to exist to Case VIb. (Note that the supergroup can in principle be
realized on the space $M_2 \tim S^2 \tim S^2 \tim S^2 \tim S^2$, but
this space cannot be  asymptotic to $AdS_5 \times S^5$.)

\item Case VII: Case VIIa corresponds to the holographic dual of half-BPS surface
operators, and is associated with the intersecting brane and probe brane case 3 of 
Tables \ref{table1} and \ref{table2}.  The solutions have been
discussed  recently in \cite{Drukker:2008wr,Gomis:2007fi}.
No non-trivial half-BPS supergravity solutions are known to exist to Case VIIb.
(Type IIB solutions with an $AdS_3$, but fewer than 16 supersymmetries, 
have been studied in \cite{Kim:2005ez} and \cite{Gauntlett:2006ns}.)

\item Cases VIII, IX, and X: No supergravity solutions for these cases are presently 
known. If solutions exist,  they will be qualitatively different from those of the other 
cases  of Table \ref{table10}. Since the spaces $CP_{2}$ or $CH_2$, which appear 
in their space-time Ans\"atze, support only half the maximal number of Killing spinors, the 
solutions are restricted to have constant dilaton and axion, and vanishing 3-form 
field strength. Since the parameter spaces $\Sigma_1$ of $M_1$ are 
only 1-dimensional, the reduced BPS equations may be too restrictive
for solutions to exist. We leave  this problem for future work.
\end{itemize}

\sm

A general argument \cite{SSJ}, using the AdS/CFT correspondence with local 
operators, shows\footnote{We are grateful to Shahin Sheikh-Jabbari for 
sharing this argument \cite{SSJ}, and agreeing to its use here.} 
that the invariance superalgebra $\cH$ for half-BPS solutions
which are genuinely asymptotic to $AdS_5 \times S^5$ (and not just {\sl locally}
asymptotic to $AdS_5 \times S^5$) must have rank strictly less than the rank 6
of $PSU(2,2|4)$. The use of this argument leads us to conclude 
that cases I$^*$, II$^*$, III$^*$, VIII, IX, and X cannot have solutions that
are genuinely asymptotic to $AdS_5 \times S^5$, though there appears
to be no obstruction to the existence of {\sl locally} asymptotic solutions.
The argument also clarifies the natural occurrence in cases VIa, VIb of one 
factor of $\bR$ (which  gives rank($\cH$)=5), but not of two factors of $\bR$,
since such solutions would correspond to rank 6. A similar clarification emerges for 
the $U(1)$ factors of cases VIIa, and VIIb.

\subsection{On the existence of fully back-reacted D7 brane solutions}
\label{secfivefour}

Cases related to D7 branes in Type IIB appear to be especially intricate, and we shall
devote the present subsection to an analysis of their existence.
This concerns the  intersections of D3 and D7 branes denoted as cases 4 and 6 in Table  
\ref{table1}, and the related probe D7 brane cases 4 and 6 in Table  \ref{table2}.  
Case 6 is especially interesting since probe D7 branes may be used to introduce flavor 
into the $AdS/CFT$ correspondence \cite{Karch:2002sh}.

\sm

On the one hand, the nature of near-horizon limits of 1+1- or 3+1-dimensional 
intersections of D7 branes with D3 branes, or of D7 branes viewed as probes in
the $AdS_5 \times S^5$ background provides evidence for the existence of 
corresponding  half-BPS solutions. Case 4 in Table \ref{table1} 
corresponds to a 1+1-dimensional intersection of D7 branes and D3 branes. 
The enhancement of the symmetries in the near-horizon limit, as well as the 
symmetry of the probe D7 brane of case 4 in Table \ref{table2}, suggest a 
solution with  bosonic symmetry  $SO(2,2) \oplus SO(6) \oplus SO(2)$.
Case 6 in Table \ref{table2} corresponds to a 3+1-dimensional intersection of 
D7 branes and D3 branes, and similarly suggests the existence of a solution with  
bosonic symmetry  $SO(2,4) \oplus SO(4) \oplus SO(2)$. On the other hand, D7 branes 
produce flavor multiplets in the 
fundamental representation of the gauge group in the dual CFT, which produce
a non-vanishing renormalization group $\beta$-function, break scaling and conformal 
invariance, and thus vitiate solutions with $AdS_3$ or $AdS_5$ factors on the gravity side.
Arguments have indeed been presented in 
\cite{Kirsch:2005uy,Buchbinder:2007ar,Harvey:2008zz} (following earlier work 
on D7 branes in \cite{Aharony:1998xz,Grana:2001xn})  that no such fully back-reacted 
solutions corresponding to D7 branes should exist. 

\sm

Our superalgebra analysis  has revealed the existence of case II$^*$ and case III$^*$ 
in Table \ref{table10}, whose global bosonic symmetries, supersymmetries, and 
space-time structure all exactly match those suggested by D7 probe or D7/D3 intersecting 
brane analysis. The superalgebra, bosonic subalgebra, and space-time Ansatz for these 
cases are given as follows, 
\bea
{\rm II}^* \, & \quad  SU(1,1|4) \oplus SU(1,1) & \quad  SO(2,2) \oplus SO(6) \oplus SO(2)
\qquad AdS_3 \tim S^5 \tim S^1 \tim \Sigma _1
\no \\
{\rm III}^* & \quad  SU(2,2|2) \oplus SU(2) & \quad  SO(2,4) \oplus SO(4) \oplus SO(2)
\qquad AdS_5 \tim S^3 \tim S^1 \tim \Sigma _1
\no
\eea
Both superalgebras above are subalgebras of $PSU(2,2|4)$, suggesting that 
fully back-reacted half-BPS solutions for the near-horizon limit of D7/D3 branes can exist. 

\sm

Closer analysis of the superalgebra story reveals a further layer of subtlety. 
Both subalgebras above are not ``minimal", but possess  purely bosonic invariant 
subalgebras (respectively $SU(1,1)$ and $SO(3)$), which are not 
required by the mandate of 16 supersymmetries. Their removal 
leads to corresponding ``general  cases", namely II and III of Table \ref{table10},
\bea
{\rm II} \, & \quad  SU(1,1|4) & \quad  SO(2,1) \oplus SO(6) \oplus SO(2)
\qquad AdS_2 \tim S^5 \tim S^1 \tim \Sigma _2
\no \\
{\rm III} & \quad  SU(2,2|2) & \quad  SO(2,4) \oplus SO(3) \oplus SO(2)
\qquad AdS_5 \tim S^2 \tim S^1 \tim \Sigma _2
\no
\eea
for which, based on superalgebra arguments alone, solutions should exist.
Fully back-reacted half-BPS solutions for the near-horizon limit of D7/D3 
branes would then have to emerge as a sub-family of solutions with enhanced 
bosonic symmetry. It is unclear whether such bosonic enhancement is fully consistent.

\sm

A further subtlety for case II$^*$,
is that the bosonic invariant Lie subalgebra $SU(1,1)$ is part of the isometry algebra
of $AdS_3$, and thus part of the conformal algebra of the CFT dual. Removing
this factor will destroy part of the conformal symmetry. Also, the superconformal
algebra of case II$^*$ is realized in a left-right asymmetric (or heterotic) way,
with right-movers carrying all of the 16 supersymmetries. In the full string theory
sigma model, this asymmetry will produce anomalies, and they in turn may
destroy the solution at the string theory level.  

\sm

Perhaps the most convincing evidence that cases II$^*$ and III$^*$ cannot 
support half-BPS solutions with genuine asymptotic $AdS_5 \times S^5$ behavior
result from applying the arguments of \cite{SSJ} (see the last paragraph of section 5.3). 
The Lie superalgebras $\cH$ corresponding
to cases II$^*$ and III$^*$ both have rank 6, which is not allowed for 
genuine $AdS_5 \times S^5$ asymptotics by the arguments of \cite{SSJ}.
The only way to lower the rank is to remove completely the corresponding 
purely bosonic invariant subalgebra, thus lowering the symmetries to those
of cases II and III respectively. But in doing so, the removal of the bosonic
invariant subalgebra vitiates the existence of the corresponding fully
back-reacted near-horizon D7 brane solutions. For case III, the search for
solutions is presently being carried out in \cite{yu}.

\subsection{Half-BPS solutions in M-theory}
\label{sec43}

The space-time Ans\"atze in M-theory are constructed in a similar manner
to those of Type IIB supergravity discussed in section 5.2.
The symmetries of the $AdS_7\times S^4$ space-time are encoded in the
\sa\ of $OSp(8^{*}|4)$. The simple and direct sum subalgebras with 16
fermionic generators were derived in Section \ref{liesub} and are listed in
Table \ref{table11}.  The notation is the same as in the previous
section dealing with the Type IIB case.

\begin{table}[htdp]
\begin{center}
\begin{tabular}{|c|c|c|c|c|} \hline
case& $\cH$ & $\cH_{bos}$  & space-time Ansatz & {\small brane}
\\ \hline \hline
{\small I}  &  {\small $OSp(8^*|2)$} & {\small $SO(2,6) \oplus SO(3) $}
& {\small $AdS_7 \tim S^2 \tim \Sigma _2$} & 
\\ \hline
{\small I$^*$} & {\small$OSp(8^*|2) \oplus Sp(2)$} & {\small $SO(2,6) \oplus SO(4) $}
& {\small $AdS_7 \tim S^3 \tim \Sigma _1$} & 
\\ \hline
{\small II} & {\small $OSp(4^*|4)$}  & {\small $SO(2,1) \oplus SO(3) \oplus SO(5) $}
& {\small $AdS_2 \tim S^2 \tim S^4 \tim \Sigma _3$} & 
\\ \hline
{\small II$^*$} & {\small$OSp(4^*|4) \oplus SO(4^*)$} 
&  {\small $SO(2,2) \oplus SO(4) \oplus SO(5) $}
& {\small $AdS_3 \tim S^3 \tim S^4 \tim \Sigma _1$} & 
\\ \hline
{\small IIIa} & {\small $SU(4|2)$}  & {\small $SO(6) \oplus SO(3) \oplus SO(2) $}
& {\small $  M_3 \tim S^5 \tim S^2 \tim S^1 $} & 
\\ \hline
{\small IIIb} & {\small $SU(1,3|2)$}  & {\small $SU(1,3) \oplus SO(3) \oplus SO(2) $}
& {\small $  M_2 \tim CH_3 \tim S^2 \tim S^1 $} & 
\\ \hline
{\small IIIc} & {\small $SU(2,2|2)$}  & {\small $SO(2,4) \oplus SO(3) \oplus SO(2) $}
& {\small $  AdS_5 \tim S^2 \tim S^1 \tim \Sigma _3 $} & {\small 3}
\\ \hline
{\small IV} & {\small $OSp(4^*|2) \oplus OSp(4^*|2)$} 
& {\small $SO(2,2) \oplus SO(4) \oplus SO(4)  $}
& {\small $AdS_3 \tim S^3 \tim S^3 \tim \Sigma_2$} & {\small 1,2,5}
\\ \hline
{\small V} & {\small $OSp(2^*|2) \oplus OSp(6^*|2)$}  
& {\small $SU(1,3) \oplus SO(4) \oplus SO(2) $}
& {\small $ M_1 \tim CH_3 \tim S^3 \tim S^1$} &
\\ \hline \hline
\end{tabular}
\caption{Subalgebras of $OSp(8^{*}|4)$ with 16 fermionic generators.}
\label{table11}
\end{center}
\end{table}

The  symmetries of $AdS_4\times S^7$ space-time are encoded in the \sa\ of
$OSp(8|4, \bR)$. Its simple and direct sum subalgebras with 16 fermionic generators
are listed in Table \ref{table12}. As in the case of Type IIB, there are several cases
where corresponding half-BPS solutions are known. There are also cases, however,
where the existence of solutions suggested by probe-brane arguments is not
backed up by the superalgebra structure. The remaining cases provide evidence for the
existence of new half-BPS solutions in M-theory. We shall now discuss these cases in turn.

\begin{table}[htdp]
\begin{center}
\begin{tabular}{|c|c|c|c|c|} \hline
case& $\cH$ & $\cH_{bos}$ &  space-time Ansatz & {\small brane}
\\ \hline \hline
{\small VIa} 
& {\small $SU(4|2) $} 
& {\small $SO(6) \! \oplus \! SO(3) \!  \oplus \! SO(2)  $}
& {\small $M_3 \tim S^5 \tim S^2 \tim S^1 $}
& 
\\ \hline
{\small VIb} 
& {\small $SU(4|2) $} 
& {\small $SO(6) \! \oplus \! SO(3) \! \oplus \! SO(2)  $}
& {\small $M_2 \tim CP_3 \tim S^2 \tim S^1 $}
& 
\\ \hline
{\small VIc} 
& {\small $SU(4|1,1) $} 
& {\small $SO(6) \! \oplus \! SO(1,2) \! \oplus \! SO(2)  $}
& {\small $AdS_2 \tim S^5 \tim S^1 \tim \Sigma _3 $}
& {\small 4}
\\ \hline
{\small VII} 
& {\small $OSp(4|2, \bR) \! \oplus \! OSp(4|2, \bR)$} 
& {\small $ SO(2,2) \! \oplus \! SO(4) \! \oplus \! SO(4) $}
& {\small $AdS_3 \tim S^3 \tim S^3 \tim \Sigma_2$}
& {\small 1,2,5}
\\ \hline
{\small VIII} 
& {\small $OSp(4|4, \bR)$} 
& {\small $ SO(4) \oplus SO(2,3)$}
& {\small $AdS_4 \tim S^2 \tim \! S^2 \tim \Sigma_3 $}
& 
\\ \hline
{\small VIII$^*$} 
& {\small $OSp(4|4, \bR) \oplus SO(4)$} 
& {\small $ SO(4)^2 \oplus SO(2,3)$}
& {\small $AdS_4 \tim S^3 \tim \! S^3 \tim \Sigma_1 $}
&  
\\ \hline
{\small IX} 
& {\small $OSp(8|2, \bR )$} 
& {\small $SO(8) \oplus SO(2,1) $}
& {\small $AdS_2 \tim S^7 \tim \Sigma_2 $}
& 
\\ \hline
{\small IX$^*$} 
& {\small $OSp(8|2, \bR ) \oplus Sp(2,\bR) $} 
& {\small $SO(8) \oplus SO(2,2) $}
& {\small $AdS_3 \tim S^7 \tim \Sigma_1 $}
&  
\\ \hline
{\small X} 
& {\small $OSp(5|2, \bR) \! \oplus \! OSp(3|2, \bR)$} 
& {\small $ SO(2,2) \! \oplus \! SO(5) \! \oplus \! SO(3)$}
& {\small $AdS_3 \tim S^4 \tim   S^2 \tim \Sigma_2$}
& 
\\ \hline
{\small XIa} 
& {\small $OSp(6|2, \bR) \! \oplus \! OSp(2|2, \bR)$} 
& {\small $ SO(2,2) \! \oplus \! SO(6) \! \oplus \! SO(2)$}
& {\small $AdS_3 \tim S^5 \tim S^1 \tim \Sigma_2$}
& 
\\ \hline
{\small XIb} 
& {\small $OSp(6|2, \bR) \! \oplus \! OSp(2|2, \bR)$} 
& {\small $ SO(2,2) \! \oplus \! SO(6) \! \oplus \! SO(2)$}
& {\small $AdS_3 \tim CP_3 \tim \! S^1\! \tim \Sigma_1$}
&  
\\ \hline
{\small XII} 
& {\small $OSp(7|2, \bR) \! \oplus \! OSp(1|2, \bR)$} 
& {\small $ SO(2,2) \oplus SO(7) $}
& {\small $AdS_3 \tim S^6 \tim \Sigma_2$}
& 
\\ \hline \hline
\end{tabular}
\caption{Subalgebras of  $OSp(8|4, \bR)$ with 16 fermionic generators  }
\label{table12}
\end{center}
\end{table}

The superalgebra $\cH$ is a {\sl maximal subalgebra} of $OSp(8^*|4)$ for cases 
I$^*$, II$^*$,  IV, and V, and  of $OSp(8|4,\bR)$ for 
cases VII, VIII$^*$, IX$^*$, X, XIa, XIb, and XII. Cases I, II, VIII, and IX may be obtained 
from cases I$^*$, II$^*$, VIII$^*$, and IX$^*$ by removing their respective purely 
bosonic invariant subalgebras $Sp(2)$, $SO(4^*)$, $SO(4)$, and $Sp(2,\bR)$. 
The space of solutions for cases I$^*$, II$^*$, VIII$^*$, and IX$^*$ thus form sub-families 
with enhanced purely bosonic symmetries. Note solutions may enjoy partial enhancement, associated with a non-trivial subalgebra, for example of $SO(4^*)$ or $SO(4)$.
For cases IIIa, IIIb, IIIc, VIa, VIb, VIc, the subalgebra $\cH$ is not maximal,
and allows for enhancement by an Abelian $U(1)$ or $\bR$ algebra.

\begin{itemize}
\itemsep 0.1in
\item Cases I and I$^*$:  No non-trivial half-BPS supergravity solutions are known.
These cases are analogous to Case III of Type IIB in Table \ref{table10}.

\item Cases II and II$^*$: No non-trivial half-BPS supergravity solutions are known.
Solutions in M-theory with an $AdS_2$-factor have been obtained in \cite{Gauntlett:2006ns}, 
but their number of supersymmetries is fewer than 16.

\item Case III:  Case IIIa  corresponds to the ``bubbling AdS" solution of
\cite{Lin:2004nb}, which are dual to local half-BPS operators in the 6-dimensional CFT. 
Whether the supergravity solution corresponding to case IIIb exists is not known 
at this point. Case IIIc corresponds to a four dimensional BPS defect in the 
6-dimensional CFT. These solutions should be related to the ones of case IIIa,  
but have not been explicitly  constructed. 

\item Case IV corresponds to the half-BPS defect solutions found
in \cite{DHoker:2008wc} which are associated with two dimensional defects
in the six dimensional CFT and from the perspective of the (0,2) theory are
the analog of the Wilson loop operators of Type~IIB. This case is associated
with the intersecting brane and probe branes cases 1, 2 of Tables \ref{table3}
and \ref{table4}.

\item Case V:  No non-trivial half-BPS supergravity solutions are known to exist.
This case is analogous to case X of Type IIB in Table \ref{table10},
and both the fact that the  $CH_3 $ restricts the supersymmetry and the base
space is only one dimensional indicate that the solution if it exists,
will be highly constrained.

\item Case VI: Case VIa corresponds to the ``bubbling AdS" solutions of
\cite{Lin:2004nb}, which are dual to local half-BPS operators in the
three dimensional CFT associated with the decoupling limit of the M2 brane
worldvolume theory. 
Whether the supergravity solution corresponding to case VIb exists is presently not known.
The supergravity solution corresponding to case VIc was found in \cite{Lunin:2007ab}.

\item Case VII: This case  also corresponds to the half-BPS defect solutions
found in \cite{DHoker:2008wc} which are associated with two dimensional
defects in the three dimensional CFT, they are the analog of the half-BPS
interface solutions of Type IIB. This case is associated with the
intersecting brane case 2 in Table \ref{table3} and the probe brane case 5 in
Table \ref{table5}.

\item Cases VIII and VIII$^*$: No non-trivial half-BPS supergravity solutions are known to exist.
This case is analogous  to Case III of Type IIB in Table \ref{table10}. Its dual
interpretation is a deformation of the M2 brane CFT which preserves conformal
symmetry but breaks half the supersymmetry.

\item Cases IX and IX$^*$: No non-trivial half-BPS supergravity solutions are known to exist.
Progress in this direction has been made, however, in \cite{FigueroaO'Farrill:2007gj}.
Solutions in M-theory with an $AdS_2$-factor have been obtained in 
\cite{Gauntlett:2006ns,Kim:2006qu,MacConamhna:2006nb}, but with fewer than 
16 supersymmetries.

\item Case X-XII: No non-trivial half-BPS supergravity solutions are known to exist.
(Solutions in M-theory with an $AdS_3$ factor, but with fewer than 16 supersymmetries,
have been obtained in \cite{Gauntlett:2006qw,Figueras:2007cn}.)
Cases X-XII are similar to Case VII as they all have an $AdS_{3}$ factor
and a two dimensional base base. They differ by the choice of sphere factors.
If these solutions exist they would be dual to supersymmetric defects in the
M2-brane CFT preserving different R-symmetries. 

\end{itemize}

The general argument of \cite{SSJ}, which we have already used for Type IIB
solutions in the last paragraphs of both sections 5.3, and 5.4, applies to
M-theory as well. The ranks of both superalgebras $OSp(8^*|4)$ and $OSp(8|4,\bR)$
is 6. The argument of \cite{SSJ} implies that solutions corresponding to 
superalgebras $\cH$ whose rank is equal to 6 cannot have genuinely 
asymptotic $AdS_7 \times S^4$ or $AdS_4 \times S^7$ behavior, and rules out 
such solutions for the cases I$^*$, II$^*$, IV, V, VII, VIII$^*$, IX$^*$, XIa, and XIb.
Of course, solutions which are {\sl locally asymptotic} to $AdS_7 \times S^4$ or 
$AdS_4 \times S^7$ are allowed to exist. In fact, they are clearly known to exist, 
since they were derived, in explicit and exact local form in \cite{DHoker:2008wc} 
for cases IV and VII, and  in global form in \cite{D'Hoker:2008qm} for case IV . 

\newpage

\section{Conformal superalgebras with 16 supersymmetries}
\setcounter{equation}{0}

The AdS/CFT correspondence maps the global symmetries of an AdS space-time
solution onto the conformal supersymmetries of the dual CFT, and vice-versa.
In particular, for the maximally supersymmetric case, the superalgebras $PSU(2,2|4)$,
$OSp(8^*|4)$ and $OSp(8|4,\bR)$ are the symmetry algebras of respectively the
$AdS_5 \times S^5$, $AdS_7 \times S^4$, and $AdS_4 \times S^7$  solutions,
but are also the conformal superalgebras of their respective dual CFTs,
namely $\cN=4$ SYM, CFT$_6$, and CFT$_3$. The AdS/CFT map between
global symmetries extends to any half-BPS supergravity solution 
(as classified in Sections 4 and 5), and any such solution
should have a field theoretic dual with matching symmetry. The dual 
field theory can be a genuine CFT, or it can exhibit 
renormalization group flow behavior. In the first case, the symmetry 
of the half-BPS solution will translate to a superconformal symmetry
on the CFT side.  All possible superconformal symmetries in arbitrary space-time
dimensions have been classified in \cite{VanProeyen:1986me} in terms of
{\sl conformal superalgebras $\cC$}. A conformal superalgebra $\cC$ in space-dimension
$d$ is such that its maximal bosonic subalgebra has a factor of the conformal
algebra $SO(2,d)$.

\sm

In this section, we shall review the classification of conformal superalgebras in
\cite{VanProeyen:1986me}, and derive its implications for  the half-BPS case.
In particular, we shall show that all  of the solutions whose existence
was advocated in Section 5 should have genuine CFT duals, although some 
solutions will be dual to the somewhat special case of 
``0-dimensional conformal field theory".
It will be convenient to separate the cases with $d\geq 3$, $d=2$ and $d=1$.

\subsection{CFT space-time dimension $d \geq 3$}

For CFT space-time dimension $d$ larger than 2, Table \ref{table13} below lists the
possible conformal superalgebras $\cC$ with 16 fermionic generators of
\cite{VanProeyen:1986me}, together with their maximal bosonic subalgebra
$\cC_{bos}$, as a function of  $d$.

\begin{table}[htdp]
\begin{center}
\begin{tabular}{|c||c|c||c||c|} \hline
Dimension & $\cC$  & $\cC_{bos}$ & Type IIB & M-theory
\\ \hline \hline
$d=3$ & $OSp(4|4, \bR)$ & $SO(2,3) \oplus  SO(4) $ &  Va,Vb &  VIII, VIII$^*$
\\ \hline
$d=4$ & $SU(2,2|2)$ & $SO(2,4) \oplus SO(3) \oplus SO(2)$ & III, III$^*$ & IIIc
\\ \hline
$d=5$ & $F(4;2)$ & $SO(2,5) \oplus SL(2, \bR)$ & none & none
\\ \hline
$d=6$ & $OSp(8^{*}|2)$ & $SO(2,6) \oplus SO(3)$ &  & I, I$^*$
\\  \hline \hline
 \end{tabular}
\end{center}
\caption{Conformal superalgebras in $d \geq 3$ and corresponding cases of Section 5.}
\label{table13}
\end{table}

In the rightmost two columns of Table \ref{table13} we list the cases of Type IIB
supergravity and M-theory solutions of Section 5 on which each conformal
superalgebra $\cC$ may be realized as invariance algebra (namely the
superalgebra $\cH=\cC$ of the classification in Section 5).

\sm

For $d=5$, the requirement that the superalgebra be conformal forces the unique
real form $F(4;2)$ of the exceptional superalgebra $F(4)$.
The presence, for this real form, of the extra non-compact algebra $SL(2,\bR)$
implies that $F(4;2)$ cannot be realized on $AdS_6 \times M$ with $M$ compact.
(One compact option is excluded as follows. The algebra $SL(2, \bR)$ is, of course,
the isometry of the non-compact Poincar\'e upper half plane $H_2$. The quotient of
$H_2$ by a hyperbolic Kleinian group  produces a compact Riemann surface,
but the quotienting destroys the supersymmetry.)  Since no supergravity solution
of the form $AdS_6 \times M$ with
$M$ compact exists, we write ``none" in Table \ref{table13} for this case.

\sm

When no corresponding case of Section 5 is available, the relevant entry
in either one of the last two columns in Table \ref{table13} is left blank.
Table \ref{table13} reveals that in $d=6$ for the Type IIB, and in $d=4$ for
M-theory, no cases of the classification of Section 5 are available, so that
no corresponding half-BPS solutions exist which are asymptotic respectively to 
$AdS_5 \times S^5$, $AdS_4 \times S^7$, or $AdS_7 \times S^4$, and any 
supergravity solutions corresponding to these cases must exhibit different 
asymptotics. On the CFT side, this implies result
implies that the dual CFT cannot be obtained as a deformation of the vacuum
via a ``sufficiently local operator" (which is an operator defined to
obey cluster decomposition with any set of local operators).

\subsection{CFT space-time dimension $d =2$}

For CFT space-time dimension $d=2$, the conformal algebra is not simple, but
instead given by $SO(2,2) = SO(1,2) \oplus SO(1,2)$.  As a result, the corresponding
conformal superalgebras $\cC$ with 16 fermionic generators are direct sums
of a left superalgebra $\cC^L$, and a right superalgebra $\cC^R$.
As is well-known, these finite-dimensional conformal and superconformal
algebras for $d=2$ are enhanced to  their full infinite-dimensional counterparts
\cite{VanProeyen:1986me}, with corresponding infinite-dimensional asymptotic
symmetry of the associated supergravity solutions \cite{Brown:1986nw}.

\begin{table}[htdp]
\begin{center}
\begin{tabular}{|c||c|c|c|} \hline
$\cC^{L,R}$   & $\cC^{L,R}_{bos}$ & \# fermions
\\ \hline \hline
$OSp(m|2, \bR)$ & $SO(2,1) \oplus  SO(m) $ &  $2m$
\\ \hline
$OSp(4^*|2m) $ & $SO(2,1) \oplus SO(3) \oplus Sp(2n)$ & $8m$
\\ \hline
$SU(1,1|m)$ & $SO(2,1) \oplus SU(m) \oplus SO(2)$ & $4m$
\\ \hline
$G(3;0)$ & $SO(2,1) \oplus G_2 $ &  $14$
\\ \hline
$F(4;0)$ & $SO(2,1) \oplus SO(7) $ &  $16$
\\ \hline
$D(2,1;c;0)$ & $SO(2,1) \oplus SO(4) $ &  $8$
\\  \hline \hline
 \end{tabular}
\end{center}
\caption{Conformal superalgebra factors for $d =2$}
\label{table14a}
\end{table}

We begin by recalling, in Table \ref{table14a}, the classification of the conformal
superalgebras of \cite{VanProeyen:1986me} for $d=2$, with arbitrary number of
fermionic generators. The most general $d=2$ finite-dimensional conformal
superalgebra $\cC$ is given by a direct sum $\cC= \cC^L \oplus \cC^R$, with
both $\cC^L$ and $\cC^R$ selected from Table \ref{table14a} so that
the total number of fermionic generators is 16. 

Note that $\cC^L$ and $\cC^R$ may differ from one another in this construction,
as they indeed do for the first seven cases in Table \ref{table14}. One might refer 
to these cases as ``heterotic", since $\cC^L$ and  $\cC^R$
act on left-moving and right-moving degrees of freedom on the boundary CFT. 
Note that we find cases where both $\cC^L$ and  $\cC^R$ are superalgebras,
such as case II$^*$ in M-theory, as well as other cases, like II$^{*}$ in Type IIB, 
where $\cC^R$  is the a purely bosonic algebra.

\sm

In Table \ref{table14}, we lists  conformal superalgebras $\cC$, together with
their maximal bosonic subalgebras $\cC_{bos}$, for which there exists at least
one corresponding case of half-BPS solutions in Section 5, either for Type IIB or
for M-theory. The corresponding cases are indicated in the rightmost two
columns of Table \ref{table14}. Blank entries signify again that any supergravity
half-BPS solutions must have asymptotics different from the respective maximally supersymmetric space-times, and that no such solutions are known at present.

\begin{table}[htdp]
\begin{center}
\begin{tabular}{|c||c|c||c||c|} \hline
$\cC = \cC_L \oplus \cC_R$  & $\cC_{bos}$ & Type IIB & M-theory
\\ \hline \hline
$SU(1,1|4) \oplus SU(1,1)$ & $SO(2,2) \oplus SO(6) \oplus SO(2)$ & II$^*$ & 
\\ \hline
$SU(1,1|3) \oplus SU(1,1|1)$ & $SO(2,2) \oplus SU(3) \oplus SO(2)^2$
	& IX &
\\ \hline 
\quad $OSp(4^*|4) \oplus SO(4^*)$ \quad &
\quad $SO(2,2) \oplus  SO(4) \oplus Sp(4) $ \quad
	&  & II$^*$
\\ \hline
\quad $OSp(5|2, \bR) \oplus OSp(3|2, \bR)$ \quad & $SO(2,2) 
\oplus  SO(5) \oplus SO(3) $
	&  & X
\\ \hline
\quad $OSp(6|2, \bR) \oplus OSp(2|2, \bR)$ \quad & $SO(2,2) 
\oplus  SO(6) \oplus SO(2) $
	&  & XIa,XIb
\\ \hline
\quad $OSp(7|2, \bR) \oplus OSp(1|2, \bR)$ \quad & $SO(2,2) \oplus  SO(7)  $
	&  & XII
\\ \hline
\quad $OSp(8|2, \bR) \oplus Sp(2, \bR)$ \quad & $SO(2,2) \oplus  SO(8)  $
	&  & IX$^*$
\\ \hline \hline
$PSU(1,1|2) \oplus PSU(1,1|2)$ & $SO(2,2) \oplus SO(3) \oplus SO(3)$
	& VIIa,VIIb &
\\ \hline
\quad $OSp(4|2, \bR) \oplus OSp(4|2, \bR)$ \quad &
\quad $SO(2,2) \oplus  SO(4) \oplus SO(4) $ \quad
	&  & VII
\\ \hline 
\quad $OSp(4^*|2) \oplus OSp(4^*|2)$ \quad &
\quad $SO(2,2) \oplus  SO(4) \oplus SO(4) $ \quad
	&  & IV
\\ \hline
$ D(2,1;c) \oplus D(2,1;c)$ & $SO(2,2) \oplus SO(4) \oplus SO(4)$ &  & Sect. 7.2
\\  \hline \hline
 \end{tabular}
\end{center}
\caption{Conformal superalgebras in $d =2$ with corresponding cases of Section 5.}
\label{table14}
\end{table}

On the last line of Table \ref{table14}, $c$ is an arbitrary real non-vanishing parameter.
The case $c = -2,-1/2$ coincides with the first line of Table \ref{table14}, since
$D(2,1;c;0)=OSp(4^*|2)$ for these values of $c$, while the case $c=1$ coincides
with the second line in Table \ref{table14}, since $D(2,1;1;0) =  OSp(4|2,\bR)$.
The cases $c\not= 1,-2,-1/2$ in M-theory correspond to known solutions
\cite{Boonstra:1998yu,DHoker:2008wc}, which will be discussed in detail
in Section 7.2.

\sm

The remaining conformal superalgebras $\cC= \cC^L \oplus \cC^R$
with 16 fermionic generators  are,
{\small
\bea
PSU(1,1|2) \oplus D(2,1;c;0) & \qquad
	OSp(4|2,\bR) \oplus PSU(1,1|2) \qquad &
	OSp(4|2,\bR) \oplus OSp(4^*|2)
\no \\
OSp(4|2,\bR) \oplus D(2,1;c;0) & \qquad
	OSp(2|2,\bR) \oplus SU(1,1|3) \qquad &
	  OSp(1|2,\bR) \oplus G(3;0) 
\no \\
 OSp(4^*|2) \oplus D(2,1;c;0)  & \qquad
	 OSp(4^*|2) \oplus PSU(1,1|2)  \qquad &
	SO(2,1) \oplus F(4;0) 
\eea}
There are no known corresponding cases of Section 5, so that any CFTs
with these symmetries cannot be obtained from the maximally supersymmetric
ones by the insertion of a ``sufficiently local operator".

\subsection{CFT space-time dimension $d=1$}

Table \ref{table15} lists the possible conformal superalgebras for $d=1$ of
\cite{VanProeyen:1986me}, together with their maximal bosonic subalgebras,
and the AdS-dual cases, if any, they correspond to.

\begin{table}[htdp]
\begin{center}
\begin{tabular}{|c||c|c||c||c|} \hline
$\cC$   & $\cC_{bos}$ & IIB case & M case
\\ \hline \hline
$OSp(8|2, \bR)$ & $SO(2,1) \oplus  SO(8) $
	&  &  IX
\\ \hline
$OSp(4^*|4) $ & $SO(2,1) \oplus SO(3) \oplus SO(5)$ & IV & II
\\ \hline
$SU(1,1|4)$ & $SO(2,1) \oplus SO(6) \oplus SO(2)$
	& II & VIc
\\ \hline
$F(4)$ & $SO(2,1) \oplus SO(7) $ &  &
\\  \hline \hline
 \end{tabular}
\end{center}
\caption{Conformal superalgebras in $d =1$ with corresponding cases
of Section 5.}
\label{table15}
\end{table}

\subsection{CFT in dimension 0}

Cases I, I$^*$, VIa, VIb, and VIII and X of Type IIB in Table \ref{table10}, and cases 
IIIa, IIIb, V, VIa, and VIb of M-theory in Tables \ref{table11} and \ref{table12}, may 
be interpreted as corresponding to 0-dimensional conformal field theory. 
The factor $M_d$ then contains the Minkowski signature time-direction. The 
solutions may be time-independent such as in \cite{Lin:2004nb}, or become cosmological
type solutions in view of time-dependent warping.

\subsection{Examples from $\cN=4$ SYM/Type IIB duality}
\label{secknowncft}

As discussed in Section \ref{sectwo}, the superconformal symmetry of a CFT
can be reduced by the presence of operators of varying worldvolume dimension,
including local operators, Wilson loops, and operators associated with defects,
interfaces, and domain walls. An operator of worldvolume dimension $d$
is expected to preserve the conformal algebra $SO(2,d)$, associated with
the conformal transformations which leave the operator invariant. For half-BPS
operators, the symmetry algebra has to contain 16 fermionic generators  and
form a superalgebra. Perturbing the CFT by any such operator will cause
a deformation of the dual supergravity solution, with an associated reduction
of the invariance superalgebra to a subalgebra $\cH$.
A CFT with conformal superalgebra $\cH$ which is a subalgebra of $PSU(2,2|4)$
should be a deformation of $\cN = 4$ SYM by the insertion of some 
``sufficiently local operator". (Similarly a CFT with conformal superalgebra 
$\cH$ which is a subalgebra of $OSp(8^*|4)$, $OSp(8|4, R)$ should be a 
deformation of CFT$_6$, and CFT$_3$ respectively.)

\sm

For a conformal superalgebra $\cH$ which is not a subalgebra of $PSU(2,2|4)$,
the AdS/CFT correspondence predicts that no CFT deformation of $\cN = 4$
SYM exists which can break $PSU(2,2|4)$  down to $\cH$.  This is because the
dual AdS solution would then have to be locally asymptotic to $AdS_5 \times S^5$,
which in turn would contradict the results of Sections \ref{liesub} and \ref{secfive}. While supergravity
solutions with these symmetries may exist, they cannot be locally asymptotically
$AdS_5 \times S^5$.  One immediate result from the below analysis is that there are
no deformations of $\cN = 4$ SYM which yield a CFT of dimension greater than 4.

\sm

The known  half-BPS operators in $\cN=4$ SYM provide the following examples,

\begin{itemize}
\item $PSU(2|2) \oplus PSU(2|2) \oplus \bR$
corresponds to deformations of $\cN = 4$ SYM by local half-BPS operators.
In \cite{Berenstein:2004kk} it was argued that the
$PSU(2|2) \oplus PSU(2|2) \oplus \bR$ half-BPS sector of
$\cN = 4$ SYM can be described by the quantum mechanics of a system of
fermions, which can be though of as a zero dimensional conformal field theory.
In \cite{Lin:2004nb}, it was shown that the dual geometries can be described in
the same manner.
\item $OSp(4^*|4)$ corresponds to deformations of $\cN =4$ SYM
by half-BPS Wilson loop operators.  In \cite{Gomis:2006sb}, it was argued that
such operators are described by coupling a 1-dimensional defect action to
$\cN = 4$ SYM.  The authors essentially then reduced the problem to the
1-dimensional field theory of a fermion.
\item $OSp(4|4, \bR)$ corresponds to the interface/defect CFTs discussed in
\cite{DHoker:2006uv,DeWolfe:2001pq,Gaiotto:2008sa,Gaiotto:2008sd}.
The 3-dimensional behavior is the most explicit in \cite{Gaiotto:2008sa}
where the CFT is constructed from three dimensional superfields and so
the set of such field theories is governed by the space of three dimensional
conformal SYM theories.  The fourth direction is introduced as a gauge theory
direction.  The M-theory CFT has not been studied yet, although a three
dimensional CFT is expected to exist.
\item $SU(2,2|2)$ corresponds to coupling $\cN = 4$ SYM to
additional hyper-multiplets \cite{DeWolfe:2001pq,Karch:2002sh}.
This corresponds to introducing flavor into the AdS/CFT correspondence.
The resulting CFT is 4-dimensional.
\end{itemize}

\newpage

\section{Discussion}
\setcounter{equation}{0}

In this section, we shall discuss two further issues which are directly
related to the correspondence between Lie superalgebras and half-BPS 
supergravity solutions.
The first is an analysis of when the 16 supersymmetries of a half-BPS
solution may be enhanced. The second is an analysis of the family
of M-theory solutions of \cite{DHoker:2008wc} in the context of
the Lie superalgebra -- half-BPS supergravity solution correspondence.

\subsection{Enhanced supersymmetries}

On the one hand, all Type IIB and M-theory supergravity solutions with the maximal
number of 32 supersymmetries are known to be given by the
solutions $AdS_5 \times S^5$, $AdS_4 \times S^7$, and $AdS_7 \times S^4$,
with their associated invariance superalgebras respectively $PSU(2,2|4)$,
$OSp(8|4,\bR)$ and  $OSp(8^*|4)$. On the other hand, we have classified in this
paper all possible basic Lie superalgebras which are subalgebras of $PSU(2,2|4)$,
$OSp(8|4,\bR)$ and $OSp(8^*|4)$ with 16 fermionic generators, and argued that
these provide a classification for half-BPS supergravity solutions asymptotic
to $AdS_5 \times S^5$, $AdS_4 \times S^7$, and $AdS_7 \times S^4$ respectively.
Finally, we know that any solution which has 28 supersymmetries or more
will automatically have the maximal number of 32 supersymmetries \cite{Gran:2007eu}.

\sm

Taken together, the above results raise the question as to whether any
of the half-BPS solutions advocated in this paper may actually exhibit enhanced
supersymmetry, and possess a number of supersymmetry generators
larger than 16, but smaller than 28.

\sm

The existence of such solutions
within a family of half-BPS solutions with invariance superalgebra $\cH$
will require the existence of a superalgebra $\cH^*$ which is contained
in the corresponding superalgebra with maximal symmetry, and obeys
$\cH \subset  \cH^*$. This condition is necessary for the existence
of regular supergravity solutions, but  we are not guaranteed
that any non-trivial such solutions will exist.

\sm

It is straightforward to classify all such possible basic Lie superalgebras $\cH^*$;
remarkably, they are all simple superalgebras (i.e. they involve no
non-trivial direct sums). The result is given in Table \ref{table16}.

\sm

Note that there may exist subalgebras of $PSU(2,2|4)$, $OSp(8|4,\bR)$, or
$OSp(8^*|4)$, with more than 16 fermionic generators, but which do not
contain a basic Lie superalgebra (or direct sum thereof) with 16 fermionic
generators. For example, we have the embedding
$SU(1,2|3) \subset PSU(2,2|4)$, but $SU(1,2|3) $
has no direct sum of basic Lie subalgebra with 16 fermionic generators.
As a  result, we do not expect $SU(1,2|3)$ to show up as an enhanced symmetry
of a half-BPS solution, and the corresponding case is omitted from Table \ref{table16}.

\begin{table}[htdp]
\begin{center}
\begin{tabular}{|c|c|c|c|c|} \hline
$ \cH  $  &  $ \cH ^* $ & $\cG$    & $ \cH^* _{bos}$ & space-time
\\ \hline \hline
$ SU(2,2|2)$   &$ SU(2,2|3)$ & $PSU(2,2|4)$ & $SO(2,4) \oplus U(3) $
& $AdS_5 \times CP_2 \times S^1$
\\ \hline
$ SU(2|4)$ & $ SU(1,2|4)$ & $PSU(2,2|4)$ & $SU(1,2) \oplus U(4) $
& $ CH_2 \times S^5 \times S^1$
\\ \hline
$ OSp(4^*|4)$ & $ OSp(6^*|4)$ & $OSp(8^*|4)$ & $SU(1,3) \oplus SO(5) $
& $M_1 \times CH_3 \times S^4$
\\ \hline
$ OSp(4|4,\bR)$ & $ OSp(5|4,\bR)$ & $OSp(8|4,\bR)$ & $SO(2,3) \oplus SO(5) $
& $AdS_4 \times S^4 \times \Sigma _3$
\\ \hline
$ OSp(4|4,\bR)$ & $ OSp(6|4,\bR)$ & $OSp(8|4,\bR)$ & $SO(2,3) \oplus SO(6) $
& $AdS_4 \times S^5 \times \Sigma _2$
\\ \hline
$ OSp(4|4,\bR)$ & $ OSp(6|4,\bR)$ & $OSp(8|4,\bR)$ & $SO(2,3) \oplus SO(6) $
&  $AdS_4 \times CP_3 \times \Sigma _1$
\\  \hline
 \end{tabular}
\end{center}
\caption{Lie superalgebras and associated space-times for enhanced half-BPS solutions}
\label{table16}
\end{table}

The case on the fourth line in Table \ref{table16} with $\cH^*=OSp(5|4,\bR)$
has 20 fermionic generators, while all other cases have 24; they will correspond to
supergravity solutions with respectively 20 and 24 supersymmetries
(the latter being so-called three-quarter-BPS). Note that also $OSp(7|4,\bR)$ is
a subalgebra of $OSp(8|4,\bR)$, but is has 28 supersymmetries, and will always
lead to solutions invariant under the full $OSp(8|4,\bR)$ by \cite{Gran:2007eu}.

\sm

The space-time forms on the first two lines involve genuine direct products
of the components, as the total dimension 10 leaves no room for any extra
free parameters. To recover the Minkowski metric, the $S^1$ component 
of the second line must be the time direction, and the solutions will have 
closed time-like curves.
The space-times on the last four lines in Table \ref{table16} are generally
warped over the parameter spaces $M_1, \Sigma _1$, $\Sigma _2$ and $\Sigma_3$.
The case of the last line of Table \ref{table16} is related to the well-known
$AdS_4 \times CP_3$ solution to Type IIA supergravity \cite{Nilsson:1984bj},
which has risen to prominence recently in view of its importance to the
$AdS_4/CFT_3$ correspondence with maximal supersymmetry (see for
example \cite{Aharony:2008ug,Klebanov:2008vq}). In the $AdS_4 \times CP_3$
solution to Type IIA supergravity, the product is direct, and no warping occurs.
The relation with M-theory thus involves
choosing $\Sigma _1$ to be simply $S^1$, and carrying out standard
Kaluza-Klein reduction of M-theory on $S^1$ to Type IIA supergravity.
It is an interesting question as to whether other such solutions exist as well,
including those with actual warped products.

\subsection{Symmetries of the exact M-theory solutions}

The  half BPS-solutions on $AdS_3 \times S^3 \times S^3$, either as a direct product
with a flat Euclidean surface $E_2$ of \cite{Boonstra:1998yu}
(see also \cite{Gauntlett:1998kc,deBoer:1999rh}), or warped
over a Riemann  surface $\Sigma$  of  \cite{DHoker:2008wc}, actually consist of families
of solutions indexed by a real parameter $c$. These families provide a beautiful
illustration of the interplay between boundary asymptotic behavior of half-solutions
and their superalgebra invariance, which we shall bring to the fore in this subsection.

\sm

To show how the parameter $c$ emerges, we exhibit the metric of the solutions
of \cite{DHoker:2008wc},
\bea
ds^2 = {\tilde f_1 ^2 \over c_1^2} \, ds^2 _{AdS_3} +
 {\tilde f_2 ^2 \over c_2^2} \, ds^2 _{S^3} +
 {\tilde f_3 ^2 \over c_3^2} \, ds^2 _{S^3} + ds^2 _\Sigma
 \eea
Here, $ds^2 _{AdS_3}$ and $ds^2 _{S^3}$ stand for the invariant metrics
on the corresponding spaces, normalized to unit radius. The metric factors
$\tilde f_i$, as well as all the other supergravity fields of the solution, may
be expressed in terms of the supersymmetry parameters  $\ep$ (or Killing
spinors) of the solution  (see Appendix D for the detailed relations).

\sm

The parameters  $c_i$, for $i=1,2,3$, are real. As a result of the BPS equations
they must be constant, and subject to the relation, $c_1 + c_2 + c_3=0$.
Since a common multiplicative factor in the $c_i$ corresponds to an overall
scale of the solution, there remains only a single intrinsic combination of the
$c_i$, which we may take to be $c=c_2/c_3$. All the supergravity fields
have a non-trivial dependence on $c$.

\sm

In Appendix D, it will be shown that the invariance superalgebra of any solution of
\cite{DHoker:2008wc} is $D(2,1;c;0) \oplus D(2,1;c;0)$ for any arbitrary value of 
$c\not= 0$. Here, $D(2;1;c;0)$ is
the real form of the exceptional superalgebra $D(2,1;c)$ whose maximal
bosonic subalgebra is $SO(1,2) \oplus SO(4)$ (see Section 4.2).
For general values of $c$, the supergravity solutions involve a factor of $AdS_3$,
and are thus dual to a 1+1-dimensional CFT. For each value of $c$,
the superalgebra
$D(2,1;c;0) \oplus D(2,1;c;0)$ indeed corresponds to one of the superconformal
algebras allowed in 1+1 dimensions in the classification of \cite{VanProeyen:1986me}.

\sm

\begin{figure}[tbph]
\begin{center}
\epsfxsize=3.5in
\epsfysize=3in
\epsffile{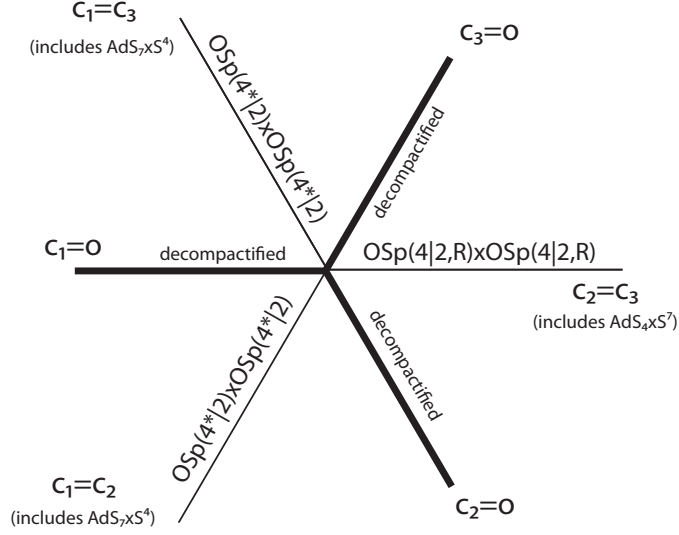}
\label{figure1}
\caption{The space of parameters $c_1,c_2,c_3$ and corresponding limiting superalgebras.}
\end{center}
\end{figure}

The dual 1+1-dimensional CFT may be enlarged, however, to a defect/interface
CFT in 2+1 dimensions for $c=1$, and to a 2-dimensional surface operator
CFT in 5+1 dimensions for $c=-2,-1/2$. The corresponding supergravity solutions
are precisely those discussed in the second bullet point of Section 3.6; they are
summarized in Table \ref{table17a}, and associated parameter 
diagram in Figure \ref{figure1}. {\small [Erratum: some of the superalgebras 
of the corresponding figure in the JHEP version of \cite{DHoker:2008wc} 
were labeled incorrectly, and have been corrected here.]}

\begin{table}[htdp]
\begin{center}
\begin{tabular}{|c||c|c|c|c |} \hline
case  & $c$  & includes & superalgebra of solution & maximal superalgebra
\\ \hline \hline
$c_2=c_3$ & $1$ & $AdS_4 \times S^7$ & $OSp(4|2,\bR) \times OSp(4|2,\bR) $
&  $OSp(8|4,\bR)$
\\ \hline
$c_3=c_1$ & $-2$ & $AdS_7 \times S^4$ &  $OSp(4^*|2) \times OSp(4^*|2) $
& $ OSp(8^*|4)$
\\ \hline
$c_1=c_2$ & $-1/2$ & $AdS_7 \times S^4$ &  $OSp(4^*|2) \times OSp(4^*|2) $
& $ OSp(8^*|4)$
\\ \hline
\end{tabular}
\end{center}
\label{table17a}
\caption{M-theory solutions asymptotic to$AdS_4 \times S^7$ or $AdS_7 \times S^4$.}
\end{table}

This may be seen as follows.
Since the $c_i$ are constants, they are determined by the asymptotic behavior
of the solution. We have the following special cases.

\begin{itemize}
\item For $c=1$, the solutions are asymptotic to $AdS_4 \times S^7$, and their
invariance superalgebra  is $OSp(4|2,\bR) \oplus OSp(4|2,\bR)$,
which coincides with $D(2,1;c;0) \oplus D(2,1;c;0)$ for $c=1$ (see Section 4.2),
and is indeed contained in $OSp(8|4,\bR)$.
\item For $c=-2,-1/2$, the solutions are asymptotic to $AdS_7 \times S^4$, and their
invariance superalgebra  is $OSp(4^*|2) \oplus OSp(4^*|2)$,
which coincides with $D(2,1;c;0) \oplus D(2,1;c;0)$ for $c=-2,-1/2$ (see Section 4.2),
and is indeed contained in $OSp(8^*|4)$.
\item For $c \not= 1, -2, -1/2 $,  the  solutions are asymptotic to
neither  $AdS_4 \times S^7$ nor  $AdS_7 \times S^4$.
It will be shown in Appendix C that, correspondingly,  the superalgebra of the solutions
$D(2,1;c;0) \oplus D(2,1;c;0)$ is then  a subalgebra
of neither $OSp(8|4,\bR)$ nor $OSp(8^*|4)$.
\end{itemize}

For each value of $c$, there exists a special solution for which 
$\tilde f_{1}=\tilde f_{2}=\tilde f_{3}=1$, and the metric $ds^2 _\Sigma$ is flat
Euclidean  \cite{Boonstra:1998yu}. For these solutions, the
product $AdS_3 \times S^3 \times S^3 \times \Sigma$ is not warped  over
$\Sigma$, so that the space-time geometry is genuinely a direct product
of these factors. The asymptotic behavior of the solution is  
that of $AdS_3$,
and is dual to a 1+1-dimensional CFT. It would appear that this $AdS_3$
solution cannot be extended to be dual to an interface or surface operator,
unlike for the case of $c=1,-2,-1/2$.  It was shown in
\cite{Gauntlett:1998kc,deBoer:1999rh} that these special solutions
are invariant under $D(2,1;c;0) \oplus D(2,1;c;0)$ for all values of $c$.

\newpage

\noindent{\Large \bf Acknowledgments }

\medskip

\noindent 
We gratefully acknowledge helpful conversations and correspondence with
Murat  G\"unaydin, Per Kraus,  David Mateos, Kostas Skenderis,
and Joris Van der Jeugt. We have benefited from various important questions
and comments on the initial version of this paper by Jose Figueroa-O'Farrill, 
Jerome Gauntlett, Andy Royston, and especially Jaume Gomis and 
Shahin Sheikh Jabbari, and we wish to extend our warm thanks them.

\sm

E.D. and P.S. would like to thank the Laboratoire de Physique Th\'eorique et
Hautes Energies du CNRS at Jussieu, Paris, and especially Boris Pioline and
Olivier Babelon for their warm hospitality while  part of this work was carried out.
E.D. wishes to thank the Aspen Center for Physics, where some of this work
was done. M.G. gratefully acknowledges the hospitality of the International
Center for Theoretical Science at the Tata Institute, Mumbai and the
Department of Physics and Astronomy, Johns Hopkins University
during the course of this work.

\newpage

\appendix

\section{Lie algebras, real forms, coincidences at low rank}
\setcounter{equation}{0}

In this section, we shall briefly review the definitions of the classical complex Lie
algebras, their real forms, various useful intersection and inclusion relations,
and we shall give a summary of their coincidences at low rank (see for example
\cite{helgason}). The starting point for the definitions of the ordinary Lie algebras
is $SL(m,\bC)$, which consists of $m \times m$ traceless matrices with complex entries.
We have the following complex Lie algebras,
\bea
SO(m,\bC) & \equiv & \{ A \in SL(m,\bC), ~ A^t  +  A =0 \}
\no \\
Sp(2m,\bC) & \equiv & \{ A \in SL(m,\bC), ~ A^t J_{2m} +  J_{2m} A =0 \}
\eea
and their real forms are as follows,
\bea
SU(m,n) & \equiv & \{ A \in SL(m+n,\bC), ~ A^\dagger I_{m,n} + I_{m,n} A =0 \}
\no \\
SO(m,n) & \equiv & \{ A \in SL(m+n,\bR), ~ A^t I_{m,n} + I_{m,n} A =0 \}
\no \\
Sp(2m,2n) & \equiv & Sp(2m+2n, \bC) \cap SU(2m,2n)
\no \\
SO(2m^*) & \equiv & \{ A \in SO(2m,\bC), ~ A ^\dagger J_{2m} +  J_{2m} A=0  \}
\no \\
SU(2m^*) & \equiv & \{ A \in SL(2m,\bC), ~ A J_{2m} - J_{2m} A^* =0 \}
\eea
The matrices $I_{m,n}$ and $J_{2m}$ are defined by,
\bea
I_{m,n}= \left ( \matrix{ I_m & 0 \cr 0 & -I_n \cr} \right )
\hskip 1in
J_{2m}= \left ( \matrix{ 0 & I_m \cr -I_m & 0 \cr} \right )
\eea
and $I_m$ is the identity matrix in dimension $m$. The Lie algebras
$SU(m) = SU(m,0)$, $SO(m)=SO(m,0)$, $SO(2^*)=SO(2)$,
and $Sp(2m)=Sp(2m,0)$ correspond to compact Lie groups;
all other cases correspond to non-compact Lie groups.

\subsection{Intersections and inclusions}

It will be useful to have the following intersection and inclusion relations
at our disposal,
\bea
SO(2n) \cap Sp(2n) & = & U(n)
\no \\
SO(2n^*) \cap U(2n) & = & U(n)
\no \\
Sp(2n, \bR) \cap U(2n) & = & U(n)
\eea

\subsection{Coincidences at low rank}

Coincidences at low rank of the complex Lie algebras are as follows,
\bea
SO(3,\bC) & = & SL(2,\bC) = Sp(2,\bC)
\no \\
SO(4,\bC) & = & SL(2,\bC) \oplus SL(2, \bC)
\no \\
SO(5,\bC) & = & Sp(4,\bC)
\no \\
SO(6,\bC) & = & SL(4,\bC)
\eea
Coincidences at low rank of compact real forms are
\bea
SO(3) & = &  SU(2) = Sp(2)
\no \\
SO(4) & = & SO(3) \oplus SO(3)
\no \\
SO(5) & = &  Sp(4)
\no \\
SO(6) & = & SU(4)
\eea
Coincidences at low rank of non-compact real forms are,
\bea
SO(1,2) & = & SU(1,1)=Sp(2,\bR) = SL(2,\bR)
\no \\
SO(4^*) & = & SO(1,2) \oplus SO(3)
\no \\
SO(1,3) & = & SL(2,\bC)
\no \\
SO(2,2) & = & SO(1,2) \oplus SO(1,2)
\no \\
SO(1,4) & = & Sp(2,2)
\no \\
SO(2,3) & = &  Sp(4,\bR)
\no \\
SO(6^*) & = & SU(1,3)
\no \\
SO(1,5) & = & SU(4^*)
\no \\
SO(2,4) & = & SU(2,2)
\no \\
SO(3,3) & = & SL(4,\bR)
\no \\
SO(8^*) & = & SO(2,6)
\eea
Proofs of these equalities may be found in \cite{helgason}.

\newpage

\section{Non-exceptional embeddings}
\setcounter{equation}{0}

A number of cases involve embeddings of real forms of the $SU$ type into real
forms of the $OSp$ type, and vice versa, and are not simply settled by inspection
of known subalgebra relations, or by simple arguments involving their maximal
bosonic subalgebras. Some progress on the remaining cases may be made
using Dynkin diagram techniques  \cite{sorba2}. Inspection of the Dynkin
diagrams allows one to recognize the regular subalgebras by deleting
appropriate roots such as, for example, in the embedding
$SL(m|n) \subset OSp(2 m|2n)$. Dynkin diagrams also allow one to obtain
some of the singular subalgebras by using the ``folding trick" such as, for
example, in the embedding $OSp(2m|2n) \subset SL(2m|2n)$. These
methods are, however, generally not sufficient  for complete proofs when
dealing with the real forms of the Lie superalgebras.
In this section, we shall provide careful proofs of the embeddings of
real forms of the $SU$ type into real  forms of the $OSp$ type, and vice versa,
using the matrix representations of these superalgebras of (\ref{matrixrep}).

\subsection{Maximal $SU$ subalgebras of $OSp(2m^*|2n)$}

We shall show the following maximal subalgebra relations,
\bea
\label{rel1}
SU(p,q|n) \oplus U(1) \subset OSp(2m^*|2n) \hskip 1in p+q=m, ~~ p,q \geq 0
\eea
We begin by recalling the definition of $M \in OSp(2m^*|2n)$: 
\bea
\label{B1}
M^{st} K + KM=0 & \hskip 1in &
K = \left ( \matrix{I_{2m} & 0 \cr 0 & J_{2n} \cr } \right )
\no \\
(M^*)^{st} \tilde K + \tilde K M=0 & \hskip 1in &
\tilde K = \left ( \matrix{J_{2m} & 0 \cr 0 & I_{2n} \cr } \right )
\eea
The presence of a $U(1)$ factor guarantees that the 
$SU(p,q|n)$ Lie superalgebra can be realized as the invariance algebra
of a bosonic element $T$ in $OSp(2m^*|2n)$. The embedding of the $SU(n)$ subalgebra
of $SU(p,q|n)$ into the $Sp(2n)$ subalgebra of $OSp(2m^*|2n)$ is {\sl unique},
and is produced as the invariance subalgebra of the generator 
$J_{2n}$. The $SU(p,q)$ subalgebra of $SO(2m^*)$ will be obtained
as the invariance subalgebra of an as yet undetermined generator $T_1 \in SO(2m^*)$.
Thus, the $SU(p,q|n)$  is specified by,
\bea
\label{B2}
[T,M]=0 \hskip 1in T = \left ( \matrix{T_1 & 0 \cr 0 & J_{2n} \cr } \right )
\eea
We shall now search for a matrix representation of this $SU(p,q|n)$ 
by investigating the simultaneous solution to the conditions of (\ref{B1}) and 
(\ref{B2}) on $M$. To do so, we decompose $M$ into blocks, 
\bea
M= \left ( \matrix{A & B \cr C & D \cr } \right )
\eea
where, as usual, $A, B, C, D$ are matrices respectively of dimension
$2m \times 2m$, $2m \times 2n$, $2n \times 2m$, and $2n \times 2n$.
In terms of the blocks $A,B,C,D$, the first condition of (\ref{B1}) becomes, 
\bea
 A^t + A  = 0 & \hskip 1in & B =  - C^t J_{2n}
\no \\
D^t J_{2n} + J_{2n} D  = 0 &&
\eea
It will be convenient to recast the second condition of (\ref{B1}) by using the first
condition of (\ref{B1}) to eliminate $M^{st}$ in favor of $M$. The combined 
condition then takes the form $M^* = K^{-1} \tilde K M \tilde K^{-1} K$, and decomposes 
as follows into the block components of $M$,
\bea
A^* =  - J_{2m} A J_{2m} & \hskip 1in & B^* =  + J_{2m} B J_{2n}
\no \\
D^* =  -J_{2n} D J_{2n} && C^* =  + J_{2n} C J_{2m}
\eea
Finally, the condition (\ref{B2}) becomes, 
\bea
\label{B3}
[T_1, A]  =  0 & \hskip 1in &  BJ_{2n} - T_1 B = 0
\no \\
{} [J_{2n}, D] =  0 && J_{2n} C - C T_1 = 0
\eea
The combined conditions on the matrix $D$ imply that it is indeed an 
element of $SU(n) \oplus U(1)$. 

\subsubsection{Solving for the fermionic generators}

Next, we analyze the conditions on the fermionic generators, represented 
by the matrices $B$ and $C$. Eliminating $B$ in terms of $C$, using equation
$B= - C^t J_{2n}$ of (\ref{B1}), gives  $C^* = J_{2n} C J_{2m}$
(which is identical to the third equation in (\ref{B2})), and $J_{2n} C = - C T_1 ^t$. 
The remaining independent equations  for $C$ are thus, 
\bea
\label{B4}
C^* & = &  J_{2n} C J_{2m}
\no \\
J_{2n} C & = &  + C T_1
\no \\
J_{2n} C & = &  - C T_1 ^t
\eea
In any embedding of $SU(p,q|n)$ into $OSp(2m^*|2n)$ with $p+q=m$, 
the number of fermionic generators of $SU(p,q|n)$  is fixed to be $2mn$
real generators, out of the $4mn$ real generators of $OSp(2m^*|2n)$,
thus preserving half thereof. As a generator of the algebra $OSp(2m^*|2n)$,
the matrix $C$ has $4mn$ real independent generators, so that the second and third equations
of (\ref{B4}) must be responsible for precisely halving that number down to $2mn$.
This condition forces the matrix $T_1$ to be of maximum rank $2m$, and the 
matrix $C$ to have no non-trivial null space. The relations of (\ref{B4}) then imply 
further restrictions on $T_1$, which may be obtained as follows.
Eliminating $J_{2n} C$ between the first and second equations, we get 
$C^* = C T_1 J_{2m}$. Iterating this equation, as well as the middle equation
above, eliminating $J_{2n}C$ between the last two equations,  and using the fact that
$C$ has no non-trivial null space, we derive the following requirements on $T_1$, 
\bea
\label{B5}
(T_1) ^2 & = &  - I_{2m}
\no \\
T_1 J_{2m} T_1^* J_{2m} & = & + I_{2m}
\no \\
T_1 ^t &  = & - T_1
\eea
Note that the second condition is equivalent to $J_{2m} T_1^* = T_1 J_{2m}$,
upon using the first condition.
Any such matrix $T_1$ will support $2mn$ real fermionic generators.

\subsubsection{Solving for the $SU(p,q)$ subalgebra of $SO(2m^*)$}

The remaining conditions on the matrix $A$, with  $T_1$ satisfying (\ref{B5}), 
are given by,
\bea
\label{B6}
A^t + A & = & 0
\no \\
J_{2m} A^* + A J_{2m} & = & 0 
	\hskip 1in 
	J_{2m} = \left ( \matrix{0 & I_m \cr -I_m & 0 \cr} \right )
\no \\
T_1 A - A T_1 & = & 0
\eea
It is standard to solve the first two conditions above combined (see for example
\cite{helgason}, p. 446), in terms of $m \times m$ matrices $Z_1$ and $Z_2$, 
\bea
\label{B7}
A = \left ( \matrix{ Z_1 & Z_2 \cr - Z_2 ^* & Z_1^* \cr} \right )
\hskip 1in 
\left \{ \matrix{ Z_1 ^t = - Z_1 \cr Z_2 ^t = Z_2^* \cr} \right.
\eea
It is actually more convenient to decompose $Z_1$ and $Z_2$
into real matrices $X_1, Y_1, X_2$, and $Y_2$,
\bea
Z_1 = X_1 + i Y_1 & \hskip 1in & X_1 ^t = - X_1 \hskip 0.4in Y_1 ^t = - Y_1
\no \\
Z_2 = X_2 + i Y_2 & \hskip 1in & X_2 ^t = + X_2 \hskip 0.4in Y_2 ^t = - Y_2
\eea
so that we have 
\bea 
A = \left ( \matrix{ X_1  & X_2  \cr - X_2  & X_1 \cr} \right )
+ \left ( \matrix{  i Y_1 &  i Y_2 \cr  i Y_2 & -i Y_1 \cr} \right )
\eea
The first term on the right hand side is anti-hermitean, while the 
second is hermitean, corresponding respectively to the compact and 
non-compact directions of Lie algebra. Setting $Y_1=Y_2=0$ gives the 
$U(m)$ subalgebra of $SO(2m^*)$, which may be realized by demanding
that $A$ commute with $T_1= J_{2m}$. This value for $T_1$ satisfies all three 
conditions (\ref{B5}), thereby providing a proof of (\ref{rel1}) for the special 
case where  $pq=0$, namely $SU(m|n) \subset OSp(2m^*|2n)$.
Note that, in view of the remarks on the (in)equivalence of real forms of section 
4.5, the basis in which $T_1$ takes the form $J_{2m}$ must coincide with  
the basis used for the  matrix $J_{2m}$ of (\ref{B6}).

\sm

To prove the subalgebra relations (\ref{rel1}) for  $pq \not= 0$, we concentrate 
first on the  compact (i.e. anti-hermitean) part of the maximal bosonic subalgebra. 
The embedding $SU(p,q|n) \oplus U(1) \subset OSp(2m^*|2n)$ with $p+q=m$, 
requires the compact subalgebra $SU(p) \oplus SU(q) \oplus U(1)$ to be a subalgebra 
of the compact subalgebra $SU(m)$ on $SO(2m^*)$, whose existence was established 
in the preceding paragraph, and whose matrix form corresponds to $Y_1=Y_2=0$ in (\ref{B7}).
The subalgebra $SU(p) \oplus SU(q) \oplus U(1)$ may be realized inside
$SU(m)$ as the invariance subalgebra of $T_1$ of the form
\bea
\label{T1}
T_1 = \left ( \matrix{ r I_{p,q} & s I_{p,q} \cr - s I_{p,q} & r I_{p,q} \cr} \right )
\hskip 1in 
I_{p,q}= \left ( \matrix{ I_p & 0 \cr 0 & -I_q \cr} \right )
\eea
for any pair of real numbers $r,s$, not both zero. To satisfy the first and
the last conditions (\ref{B5}), we must have $r=0$, and  $s= \pm 1$.
The second condition of (\ref{B5}) is then automatic. This proves that
the embedding (\ref{rel1}) exists for all $p,q \geq 0$.

\sm

It is straightforward to exhibit the fermionic generator content. Parametrizing the 
$2n \times 2m$ matrix $C$, which satisfies the condition $C^* = J_{2n} C J_{2m}$
of (\ref{B4}), in terms of the block matrices $\a, \b$ of dimension $n \times m$, gives
the following explicit representation for $B$ and $C$,
\bea
B = \left ( \matrix{ \b^\dagger & - \a^t \cr -\a^\dagger & - \b^t \cr} \right )
\hskip 1in 
C = \left ( \matrix{ \a & \b \cr \b^* & - \a^* \cr} \right )
\eea
The remaining conditions on $C$ of (\ref{B4}) restrict the blocks $\a,\b$ as follows,
\bea
\a ^* & = & - \a I_{p,q}
\no \\
\b ^* & = & - \b I_{p,q}
\eea
The number of linearly independent generators adds up to $2mn$, as required.

\sm

We may also exhibit the realization of the subalgebra $SU(p,q) \oplus U(1)$ inside 
$SU(m)$, by deriving an explicit parametrization of the solutions to (\ref{B6})
for the matrix $A$ in terms of $Z_1$ and $Z_2$ of (\ref{B7}). The requirement
$[T_1,A]=0$ with $T_1$ given by (\ref{T1}) is solve by, 
\bea
Z_1 = \left ( \matrix{ a_1 & i b_1 \cr -i b_1^t  & c_1 \cr} \right ) 
\hskip 1in 
Z_2 =  \left ( \matrix{ a_2 & ib_2 \cr -i b_2 ^t  & c_2 \cr} \right )
\eea
The dimensions of $a_i,b_i,c_i$, for $i=1,2$, are respectively $p \times p$, $p \times q$, 
and $q \times q$. The matrices $a_i,b_i, c_i$ are real; $a_1,d_1$ are anti-symmetric, 
while $a_2,d_2$ are symmetric.
Assembling $Z_1$ and $Z_2$ into the matrix $A$ using (\ref{B7}), interchanging 
the second and third columns of $A$, and then interchanging also the 
second and third rows, gives the following expression,
\bea
A \sim \left ( \matrix{
a_1		& 	a_2 	&	i b_1		& ib_2 \cr 
- a_2 	& 	a_1 	& 	i b_2 	& -ib_1 \cr
-i b_1^t 	&	-ib_2^t	& 	d_1	& 	d_2 \cr
-ib_2^t 	& 	i b_1^t	& -d_2	& d_1\cr} \right )
\sim 
\left ( \matrix{ a_1+i a_2 & b_1 - i b_2 \cr b_1 ^t + i b_2^t & c_1 + i c_2 \cr } \right )
\eea
It is immediate that the generators $a_1 + ia_2$ and $c_1+ic_2$ are 
anti-hermitean and correspond to the $SU(p) \oplus SU(q) \oplus U(1)$
subalgebra of $SU(p,q) \oplus U(1)$, while $b_1-ib_2$ and $b_1 ^t + i b_2^t$
correspond to its non-compact directions.

\subsection{Maximal $SU$ subalgebras of $OSp(2m|2n, \bR)$}

We shall show the following maximal embedding relations,
\bea
\label{rel2}
SU(m|p,q) \oplus U(1)  \subset OSp(2m|2n, \bR) \hskip 1in p+q=n, ~~ p,q \geq 0
\eea
We begin by recalling the definition of $M \in OSp(2m|2n,\bR)$; we have $M^*=M$ and 
\bea
\label{B11}
M^{st} K + KM=0 \hskip 1in 
K = \left ( \matrix{I_{2m} & 0 \cr 0 & J_{2n} \cr } \right )
\eea
The presence of a $U(1)$ factor guarantees that the 
$SU(m|p,q)$ Lie superalgebra can be realized as the invariance algebra of a bosonic 
element $T$ in $OSp(2m|2n,\bR)$. The embedding of the $SU(m)$ subalgebra
of $SU(m|p,q)$ into the $SO(2m)$ subalgebra of $OSp(2m|2n,\bR)$ is {\sl unique},
and is produced as the invariance subalgebra of the generator 
$J_{2m}$. The $SU(p,q)$ subalgebra of $Sp(2n)$ is  obtained
as the subalgebra of an as yet undetermined generator $T_2 \in Sp(2n,\bR)$.
Thus, the $SU(m|p,q)$ Lie superalgebra is specified by,
\bea
\label{B12}
[T,M]=0 \hskip 1in
T = \left ( \matrix{J_{2m} & 0 \cr 0 & T_2 \cr } \right )
\eea
We shall now seek a matrix representation by investigating the three
simultaneous conditions of $M^*=M$, (\ref{B11}) and (\ref{B12}).
To do so, we decompose $M$ into blocks, 
\bea
M= \left ( \matrix{A & B \cr C & D \cr } \right )
\eea
with $A,B,C,D$ real matrices of respective dimensions $2m \times 2m$,
$2m \times 2n$, $2n \times 2m$, and $2n \times 2n$. Condition (\ref{B11}) 
requires that 
\bea
\label{B13}
A^t + A  = 0 & \hskip 1in & B =  - C^t J_{2n}
\no \\
D^t J_{2n} + J_{2n} D = 0 &&
\eea
Condition (\ref{B12}) amounts to
\bea
\label{B14}
[J_{2m} , A ] =  0 & \hskip 1in & BT_2 - J_{2m} B  =  0
\no \\
{} [ T_2 , D ]  =  0 && CJ_{2m} - T_2 C  =  0
\eea
The combined conditions on the matrix $A$ imply that $A$ is indeed an 
element of $SU(m) \oplus U(1)$. 

\subsubsection{Solving for the fermionic generators}

Next, we analyze the conditions on the matrices $B$ and $C$. 
Eliminating $B$ in terms of $C$, using $B= - C^t J_{2n}$
of the last line of (\ref{B13}),  gives two equations for $C$,
\bea
\label{B15}
CJ_{2m} & = &  T_2 C
\no \\
C & = &  - J_{2n} T_2 ^t J_{2n} C J_{2m}  
\no \\
& = & - J_{2n} T_2 ^t J_{2n} T_2 C
\eea
To have the correct number $2mn$ of real fermionic generators for the superalgebra
$SU(m|p,q)$ with $p+q=n$, the matrix $T_2$ must have maximal rank. 
Iterating the first equation in (\ref{B15}), and using also the last one, we retain the 
following conditions on $T_2$, 
\bea
\label{B20}
(T_2)^2 & = & - I_{2n}
\no \\
T_2 ^t & = & J_{2n} T_2 J_{2n}
\eea
The last condition ensures that $T_2$ is of the form of the matrix $D$,
so that $T \in OSp(2m|2n,\bR)$.

\subsubsection{Solving for the $SU(p,q)$ subalgebra of $Sp(2n,\bR)$}

The remaining conditions on the real matrix $D$ are as follows, 
\bea
\label{B21}
D^t J_{2n} + J_{2n} D & = & 0 
\no \\
{}[T_2, D] & = & 0
\eea
It is standard to solve the first condition in terms of real $m \times m$ 
matrices $a,b,c$, so that
\bea
\label{B22}
D = \left ( \matrix{ a & b \cr c & - a^t \cr } \right )
\eea
with $b^t = b$ and $c^t=c$, and no symmetry restrictions on $a$.

\sm

Choosing $T_2 = J_{2n}$, and enforcing the second condition in (\ref{B21}),
gives the further restriction $a^t=-a$ and $c= -b$.  The matrix $D$ is now real 
and anti-symmetric and, using a standard change of basis, the combination $a+ib$ 
then parametrizes the compact subalgebra $SU(n) \oplus U(1)$. This value of $T_2$ 
satisfies the conditions of (\ref{B20}), thereby providing a proof of (\ref{rel2})
for the special case where $pq=0$, namely $SU(m|p,q) \subset OSp(2m|2n,\bR)$.

\sm

To prove the subalgebra relations (\ref{rel2}) for general $p,q$, we use the matrix,
\bea
\label{T2}
T_2 = \left ( \matrix{0 & I_{p,q} \cr -I_{p,q} & 0 \cr } \right )
\eea
$T_2$ clearly satisfies conditions (\ref{B20}), and thus will support $2mn$ real
fermionic generators needed for the $SU(m|p,p)$ subalgebra. It remains to check
that $T_2$ indeed gives the correct real form $SU(p,q)$. The condition $[T_2,D]=0$
of (\ref{B21}) imposes the following relations between $a,b,c$ of (\ref{B22}),
\bea
a I_{p,q} + a^t I_{p,q} & = & 0
\no \\
b I_{p,q} + I_{p,q} c & = & 0
\eea
whose general solution may be parametrized by the real block matrices
$a_1,a_2,a_3,b_1,b_2,b_3$,
\bea
a = \left ( \matrix{ a_1 & a_2 \cr a_2^t & a_3 \cr } \right )
 \hskip 0.8in  
b = \left ( \matrix{ b_1 & b_2 \cr b_2^t & b_3 \cr } \right )
\eea
whose dimensions are $p \times p$ for $a_1,b_1$, $p \times q$
for $a_2,b_2$, and $q \times q$ for $a_3,b_3$, 
with $a_1, a_3$ anti-symmetric, and $b_1,b_3$ symmetric.
Assembling these expressions into the full matrix $D$ of (\ref{B22}),
interchanging the second and third rows of this matrix $D$, and then its second and
third columns, we obtain the following representation, 
\bea
D \sim \left ( \matrix{
a_1 & b_1 & a_2 & b_2 \cr
-b_1 & a_1 &  b_2 & -a_2 \cr
a_2^t & b_2^t & a_3  & b_3 \cr
b_2^t & -a_2^t & -b_3 & a_3 \cr } \right )
\sim 
\left ( \matrix{ a_1+ib_1 & a_2 + i b_2 \cr a_2^t - i b_2 ^t & a_3 + i b_3 \cr} \right )
\eea
It is immediate that the anti-hermitean generators $a_1+ib_1$ and $a_3 + i b_3$ 
produce the compact subalgebra $SU(p)\oplus SU(q) \oplus U(1)$ of 
$SU(p,q) \oplus U(1)$, while the hermitean generators $a_2+i b_2$ and 
$a_2^t - i b_2^t$ produce its non-compact part. 

\subsection{The embedding $OSp(m|2n,\bR) \subset SU(m|n,n)$}

Embedding of the compact part of the maximal bosonic subalgebras,
$SO(m) \subset SU(m)$, is standard, while that of the non-compact part requires
$Sp(2n,\bR) \in SU(n,n)$. To realize the latter embedding explicitly, we use
the following change of basis,
\bea
\label{JS}
J_{2n} = \left ( \matrix{0 & I_n \cr -I_n & 0 \cr } \right )
= - i S I_{n,n} S^{-1}
\hskip 1in
I_{n,n} = \left ( \matrix{I_n & 0 \cr 0 & -I_n \cr } \right )
\eea
A convenient choice for $S$ is given by
\bea
S = {1 \over \sqrt{2}} \left ( \matrix{I_n & -iI_n \cr -iI_n & I_n \cr } \right )
\hskip 1in S^* = S^{-1}
\eea
By definition, any matrix $D \in Sp(2n,\bR)$ satisfies $D^*=D$, and
$D^t J_{2n} + J_{2n} D=0$, from which it follows that $\tr (D)=0$.
Using the above change of basis for $J_{2n}$, we have equivalently,
$D^t S I_{n,n} S^{-1} + S I_{n,n} S^{-1} D =0$.
Setting $D_s= S^{-1} D S$ then implies,
\bea
D_s ^\dagger I_{n,n} + I_{n,n} D_s=0 \hskip 1in \tr (D_s)=0
\eea
which is the defining relation for $D_s \in SU(n,n)$.

\sm

By definition, any matrix $M \in OSp(m|2n,\bR)$ satisfies $M^*=M$ and
$M^{st} K + K M=0$, from which it follows that $\str (M)=0$.
The matrix $K=K_{m|2n}$ was defined in (\ref{KJ})
and may be recast as follows, using the change of basis (\ref{JS}),
\bea
K = \left ( \matrix{I_m & 0 \cr 0 & J_{2n} \cr } \right ) =  S_1 L S_1^{-1}
\hskip 1in
S_1 = \left ( \matrix{I_m & 0 \cr 0 & S \cr } \right )
\eea
The resulting matrix $L$ is given by
\bea
L = \left ( \matrix{I_m & 0 \cr 0 & -i I_{n,n} \cr } \right )
\eea
and coincides with the matrix $L_{m|n,n}$, defined in (\ref{KJ}).
Using the above change of basis, the orthosymplectic
condition on $M$ becomes $M^{st} S_1LS_1^{-1} + S_1LS_1^{-1} M=0$.
Setting $M_s = S_1^{-1} M S_1$ and using the fact that $(M_s^*)^{st}
= S_1^{-1} M^{st} S_1$, implies that
\bea
(M_s^*)^{st} L + L M_s=0 \hskip 1in \str (M_s)=0
\eea
which is the defining relation for $M_s \in SU(m|n,n)$. This concludes the proof.

\subsection{The embedding $OSp(2m^*|2n) \subset SU(m,m|2n)$}

The embedding of the compact part of the maximal bosonic subalgebras,
$Sp(2n) \subset SU(2n)$, holds by definition of $Sp(2n)$, while that of the
non-compact part requires $SO(2m^*) \subset SU(m,m)$. The latter is
realized explicitly via a change of basis of (\ref{JS}). By definition,
any matrix $A \in SO(2m^*)$ satisfies $A^t + A=0$ and $A^\dagger J_{2m}
+ J_{2m} A=0$, from which it follows that $\tr (A)=0$.
Performing the change of basis  given by $J_{2m} = - i S I_{n,n} S ^{-1}$,
and defining $A_s = S^{-1} A S$ then implies,
\bea
A_s ^\dagger I_{n,n} + I_{n,n} A_s=0 \hskip 1in \tr (A_s)=0
\eea
which are the defining relations of $A_s \in SU(m,m)$.

\sm

By definition, a matrix $M \in OSp(2m^*|2n)$ satisfies
$M=- K^{-1} M^{st} K=- \tilde K^{-1} (M^*) ^{st} \tilde K$ from which it follows that
$\str (M)=0$.
The matrices $K=K_{2m|2n}$ and $\tilde K = \tilde K_{2m|2n}$
were defined in (\ref{KJ}), and may be recast as follows, using
the change of basis (\ref{JS}),
\bea
K = \left ( \matrix{I_m & 0 \cr 0 & J_{2n} \cr } \right ) =  S_2 K S_2^{-1}
& \hskip 1in &
S_2 = \left ( \matrix{S & 0 \cr 0 & I_{2n} \cr } \right )
\no \\
\tilde K = \left ( \matrix{J_{2m} & 0 \cr 0 & I_{2n} \cr } \right ) =   S_2 L S_2^{-1}
& \hskip 1in &
L = \left ( \matrix{-iI_{m,m} & 0 \cr 0 &  I_{2n} \cr } \right )
\eea
In the new basis, the relation involving $\tilde K$ becomes
\bea
S_2 L S_2 ^{-1} M + (M^*)^{st} S_2 L S_2 ^{-1}=0
\eea
In terms
of $M_s = S_2 ^{-1} M S_2$ this implies
\bea
(M_s^*)^{st} L + L M_s =0 \hskip 1in \str (M_s)=0
\eea
which is the defining relation of $M_s \in SU(m,m|2n)$.
This concludes the proof.

\newpage

\section{Classification of exceptional sub-superalgebras}
\setcounter{equation}{0}

In this appendix, we shall show that none of the exceptional simple basic  Lie
\sa s $F(4)$, $G(3)$, or $D(2,1;c)$ with $c\not= 1, -2, -1/2$, or any of their
real forms, is a subalgebra of either the complex superalgebras $OSp(8|4)$ 
and $PSL(4|4)$, or of their real forms $OSp(8^*|4)$, $OSp(8|4,\bR)$ or $PSU(2,2|4)$.

\sm

Dealing with the exceptional superalgebras, just as dealing with the exceptional
Lie algebras, is most easily done with the help of the roots in the Cartan formalism.
Therefore, we shall first give a brief review of the root systems of the
algebras $ SL(m|n)$, $OSp(2m|2n)$,
and the exceptional superalgebras $F(4), G(3)$ and $ D(2,1 ;c)$.

\subsection{Root systems of the basic classical superalgebras}

Given a superalgebra $\cG$, its maximal bosonic subalgebra will be
denoted $\cG_{\bar 0}$, and its fermionic subspace by $\cG_{\bar 1}$. The algebra
$\cG$ is {\sl classical} if $\cG$ is simple, and if the representation of $\cG_{\bar 0}$
on $\cG_{\bar 1}$ is completely reducible. The sets of all bosonic and fermionic
roots will be denoted by $\Delta _{\bar 0}$ and $\Delta _{\bar 1}$, respectively,
and the set of all roots is then $\Delta = \Delta _{\bar 0} \cup \Delta _{\bar 1}$.
For the superalgebras relevant here, they are given as follows,

\begin{table}[htdp]
\begin{center}
\begin{tabular}{|c||c|c|c|} \hline
Superalgebra  & $\Delta _{\bar 0}$ & $\Delta _{\bar 1}$ & normalizations
\\ \hline \hline
$ SL(m|n)$
	& $\ep_i - \ep_j$, ~~~ {\small $1 \leq i \not=j \leq  m$}
	& $ \pm (\ep_i - \delta _a)$
	&$\ep _i \cdot \ep_j = + \delta _{ij}$
\\ 	& $\delta _a - \delta _b$, ~~~ {\small $1 \leq a \not= b \leq  n$} &
	& $\delta _a \cdot \delta _b = - \delta _{ab}$
\\ \hline
$OSp(2m|2n) $
	& $\pm \ep_i \pm \ep_j$,~~~ {\small $1 \leq i \not=j \leq  m$}
	& $\pm \ep_i \pm \delta _a$
	& $\ep _i \cdot \ep_j = + \delta _{ij}$
\\ 	& $\pm \delta _a \pm \delta _b, \, \pm 2 \delta _a$,  
{\small $1 \leq a \not= b \leq  n$} &
	& $\delta _a \cdot \delta _b = - \delta _{ab}$
\\ \hline
{\small $OSp(2m\! + \! 1|2n) $}
	& $\pm \ep_i \pm \ep_j, \, \pm \ep _i,$ ~ {\small $1 \leq i \not=j \leq  m$}
	& $\pm \ep_i \pm \delta _a, ~ \pm \delta _a$
	& $\ep _i \cdot \ep_j = + \delta _{ij}$
\\ 	& $\pm \delta _a \pm \delta _b, \, \pm 2 \delta _a$,  {\small $1 \leq a \not= b \leq  n$} &
	& $\delta _a \cdot \delta _b = - \delta _{ab}$\\ \hline
$F(4)$
	& $\pm \ep_i \pm \ep_j, ~ \pm \ep_i$,~ {\small $1 \leq i \not=j \leq  3$}
	& $\half(\pm  \ep_1 \! \pm \! \ep_2 \! \pm \! \ep_3 \! \pm \! \delta)$
	& $\ep _i \cdot \ep_j = + \delta _{ij}$
\\ 	& $ \pm \delta$  && $\delta ^2 =-3$
\\ \hline
$G(3)$ & $ \ep_i - \ep_j, \, \pm \ep_i$,~~~ {\small $1 \leq i \not=j \leq  3$}
	& $\pm \ep_i \pm  \pm \delta, \, \pm \delta$
	& $\ep _i \cdot \ep_j = 1-3 \delta _{ij}$
\\ 	& $ \pm 2 \delta$ &
	& $\ep_1+\ep_2+\ep_3=0$
\\ 	&&& $\delta ^2 =2$
\\ \hline
$D(2,1;c)$
	& $ \pm 2 \ep_i$, ~~~ {\small $i=1,2,3$}
	& $\pm \ep_1 \pm \ep_2 \pm \ep_3$
	& $\ep_i \cdot \ep_j=0, ~  i \not= j $
\\	&&& $2\ep_1^2=-1-\alpha$
\\ 	&&& $ 2 \ep_2^2 =1, \, 2 \ep_3^2 = \alpha$
\\  \hline \hline
 \end{tabular}
\end{center}
\caption{Root systems of the basic classical  Lie superalgebras.}
\label{table17}
\end{table}

The vectors $\ep_i$ and $\delta_a$ form a basis for the space of roots.
For all cases, we have $\ep \cdot \delta =0$. Notice that, as a result, the
inner product on the vector spaces generated by $\ep, \delta$ is in each
case of indefinite signature. In particular, notice that for $D(2,1;c)$,
we have $\ep_1^2 + \ep_2^2 + \ep_3^2=0$.

\sm

Because of the indefinite signature inner product, the choice of a
simple root system is not unique, unlike in the case of ordinary Lie
algebras. As a result, a given basic classical Lie superalgebra can
be described by different Dynkin diagrams.

\subsection{Strategy}

To show that $\cH=F(4), \, G(3)$, and $D(2,1;c)$  for
$c \not= 1,-2, -1/2$, are  not subalgebras of $\cG = OSp(8|4), \, PSL(4|4)$
or of the real forms $OSp(8^*|4), \, OSp(8|4,\bR)$, or $PSU(2,2|4)$,
we shall use the strategy below. The maximal bosonic subalgebras will be
denoted $\cH_{\bar 0}$ and $\cG_{\bar 0}$, while their fermionic
counterparts will be denoted by $\cH_{\bar 1}$ and $\cG_{\bar 1}$

\begin{enumerate}
\itemsep -0.05in
\item Exhibit the finite number of possible embeddings of  $\cH_{\bar 0}$
into $\cG_{\bar 0} $. If we have $\cH_{\bar 0} \not \subset \cG_{\bar 0}$, then
it follows that $\cH$ is  not a subalgebra of $\cG$.
\item We now assume that $\cH_{\bar 0} \subset \cG_{\bar 0}$. Exhibit the
finite number of possible embeddings of $\cH_1$ into $\cG_{\bar 1}$, in the
representation of  $\cH_{\bar 0}$ on $\cH_{\bar 1}$ known on general grounds
by the Lie superalgebra classification theorem. If no such embedding of $\cH_1$ in
$\cG_{\bar 1}$ exists, it follows that $\cH$ is not a subalgebra of $\cG$.
\item In the remaining cases where  $\cH_{\bar 0} \subset \cG_{\bar 0}$, and
$\cH_{\bar 1}$ can be embedded in $\cG_{\bar 1}$ in a suitable
representation of $\cH_{\bar 0}$, one proceeds to compute the
anti-commutators of  $\cH_{\bar 1}$. \\
${}\qquad \bullet$  If $\{ \cH_{\bar 1}, \cH_{\bar 1} \} \subset \cH_{\bar 0}$,
then $\cH$ is a subalgebra of $\cG$; \\
${}\qquad \bullet$ If $\{ \cH_{\bar 1}, \cH_{\bar 1} \} \not \subset \cH_{\bar 0}$,
then $\cH$ is not a subalgebra of $\cG$;
\end{enumerate}

\subsection{The \sa\ $PSU(2,2|4)$}

In the paragraphs
below, we shall show that $PSU(2,2|4)$ does not have $F(4)$ and $G(3)$
as subalgebra, and that $D(2,1;c)$ is a subalgebra if and only if $c = 1,-2,-1/2$.

\subsubsection{$F(4)$ is not a subalgebra of $PSU(2,2|4)$}

None of the maximal bosonic subalgebras $SL(2,\bR) \oplus SO(p,7-p)$
of the reals forms $F(4;p)$ for $p=0,1,2,3$ of $F(4)$ is a subalgebra of the
maximal bosonic subalgebra $SU(2,2) \oplus SU(4)$ of $PSU(2,2|4)$,
because $SO(7)$ is not a subalgebra of $SU(4)=SO(6)$, and $SO(p,7-p)$
is not a subalgebra of $SU(2,2)=SO(2,4)$. Hence none of the real forms
of $F(4)$ is not a subalgebra of $PSU(2,2|4)$.

\subsubsection{$G(3)$ is not a subalgebra of $PSU(2,2|4)$}

Neither one of the maximal bosonic subalgebras $SL(2,\bR) \oplus G_{2,2p}$,
for $p=0,1$, of the reals forms $G(3;p)$ of $G(3)$ is a subalgebra of $PSU(2,2|4)$,
since neither $G_{2,2p}$ for $p=0,1$ is a subalgbera of $SU(2,2) \oplus SU(4)$.
As a result, $G(3)$ is not a subalgebra  of $PSU(2,2|4)$.

\subsubsection{$D(2,1;c)$ subalgebra of $PSU(2,2|4)$ implies $c = 1, -2, -1/2$}

The case of $D(2,1;c)$ is more intricate. To investigate this case systematically,
we shall follow the steps given in the  Strategy section.

\sm

1. The maximal bosonic subalgebra of the real form $D(2,1;c;p)$, for $p=0,1,2$,
is given by $SL(2,\bR) \oplus SO(p,4-p)$.
For $p=1,2$, this algebra is respectively $SO(1,2) \oplus SO(1,3)$, and
$SO(1,2) \oplus SO(2,2)$, neither of which is  a subalgebra of
$SO(2,4) \oplus SO(6)$.  This leaves only the case $p=0$, for which the
algebra $\cS=SU(1,1) \oplus SU(2) \oplus SU(2)$
is a subalgebra of  $SU(2,2) \oplus SU(4)$, with $SU(1,1) \subset SU(2,2)$, and
$SU(2) \oplus SU(2) \in SU(4)$. Several possible embedding exist for both.

2.  The fermionic part of $D(2,1;c;0)$ must transform under $\cS$
in the representation $(2,2,2)$. This requirement restricts the embedding of
$SU(1,1)$ in $ SU(2,2)$ to be such that the ${\bf 4}$ of $SU(2,2)$
decomposes under $SU(1,1)$ as ${\bf 4} = {\bf 2} + {\bf 1} + {\bf 1}$.
Similarly, the ${\bf 4} $ of $SU(4)$ must decompose under $SU(2) \oplus SU(2)=SO(4)$
as ${\bf 4} = ({\bf 2}, {\bf 2})$,  the vector representation
of $SO(4)$. Thus, the embedding of $\cS$ into $SU(2,2) \oplus SU(4)$,
is unique and may be realized as follows in terms of the root generators of $PSU(2,2|4)$,
\bea
\label{cS}
\cS \left \{ \matrix{
E_{\pm (\delta _2 - \delta _3)}  \cr
E_{\pm (\ep_1 - \ep_2)} + E_{\pm (\ep_3 - \ep_4)} \cr
E_{\pm (\ep_1 - \ep_3)} + E_{\pm (\ep_2 - \ep_4)}  \cr} \right  \}
\hskip 1in
\matrix{ SL(2) \cr SU(2) \cr SU(2) \cr}
\eea
The 32 fermionic generators of $PSU(2,2|4)$, namely $E_{\pm (\ep_i - \delta _a)}$
with $i,a=1,2,3,4$, transform under $\cS$ in a reducible representation. The 16
generators  $\{ E_{\pm (\delta_1-\ep_i)}, \, E_{\pm (\delta _4 - \ep_i)} \}$ for
$i=1,2,3,4$ are singlets under $SL(2)$, and cannot belong to $D(2,1;c;0)_{\bar 1}$.
The remaining 16 generators  $\{ E_{\pm (\delta_2-\ep_i)}, \, E_{\pm (\delta _3 - \ep_i)} \}$
for  $i=1,2,3,4$ transform non-trivially under $\cS$, in a reducible representation.
This is established by noticing that the involution $\rho$, defined by,
\bea
\rho \left \{ \matrix{
\delta _1 & \leftrightarrow & - \delta _4 \cr
\delta _2 & \leftrightarrow & - \delta _3 \cr} \right .
\hskip 1in
\rho \left \{ \matrix{
\ep _1    & \leftrightarrow & - \ep _4 \cr
\ep _2    & \leftrightarrow & - \ep _3 \cr} \right .
\eea
commutes with the generators of $\cS$ in (\ref{cS}). Thus, the 16-dimensional
representation of generators $\{ E_{\pm (\delta_2-\ep_i)}, \, E_{\pm (\delta _3 - \ep_i)} \}$
for  $i=1,2,3,4$ may be decomposed according to the eigenspaces of $\rho$
associated with the eigenvalues $\sigma = \pm 1$,
\bea
R_\sigma  \left \{ \matrix{
E_{\pm (\ep_1 - \delta _2)} + \sigma E_{\pm (\delta _3 - \ep _4)} \cr
E_{\pm (\ep_1 - \delta _3)} + \sigma E_{\pm (\delta _2 - \ep _4)} \cr
E_{\pm (\ep_2 - \delta _2)} + \sigma E_{\pm (\delta _3 - \ep _3)} \cr
E_{\pm (\ep_2 - \delta _3)} + \sigma E_{\pm (\delta _2 - \ep _3)} \cr } \right \}
\eea
Both  spaces $R_\pm$ are 8-dimensional. The representation
of $\cS$ under which $R_+$ and $R_-$ transform is $({\bf 2},{\bf 2},{\bf 2})$.
To form the superalgebra $D(2,1;c;0)$ from the bosonic
subalgebra $\cS$ we have two, and only two choices for the fermionic
generators, namely $R_+$ or $R_-$.

3. The structure relations of the fermionic generators are now easily
worked out and are found to reproduce either those of $OSp(4^*|2)=D(2,1;c;0)$,
for $c=-2,-1/2$, or those of $OSp(4|2,\bR) = D(2,1;1;0)$
We note that the embedding of $OSp(4^*|2)$ into $PSU(2,2|4)$ proceeds
through the following chain,
\bea
OSp(4^*|2) \subset OSp(4^*|4) \subset PSU(2,2|4)
\eea
where the last embedding corresponds to the real form of
$OSp(4|4) \subset PSL(4|4)$ of Table 8.

\subsection{The \sa s $OSp(8^*|4)$ and $OSp(8|4,\bR)$}

Next, we shall show that neither $F(4), G(3), D(2,1;c)$,  nor any one
of their real forms, are subalgebras of $OSp(8^*|4)$ or  $OSp(8|4,\bR)$.
We recall the respective maximal bosonic subalgebras of these real forms
of $OSp(8|4)$; they are given by
\bea
OSp(8^*|4) ~~ & \hskip 1in & \, SO(8^*) \oplus \, Sp(4) ~ = SO(2,6) \oplus SO(5)
\no \\
OSp(8|4, \bR) & \hskip 1in & SO(8) \oplus Sp(4, \bR) = SO(8) \oplus SO(2,3)
\eea
We note that  $SO(2,6)$ and $Sp(4,\bR)$ are precisely the non-compact
forms of respectively $SO(8)$ and $Sp(4)$ which appear naturally when describing
these real forms in the Cartan-Weyl basis. Using this basis will allow us to
study both non-compact real forms $OSp(8^*|4)$ and $OSp(8|4,\bR)$
simultaneously.

\sm

The system of all roots of  $OSp(8|4)$ is given by
\bea
( \pm \delta _1 \pm \delta _2) \qquad & ( \pm 2 \delta _a) \qquad &
Sp(4)
\no \\
(\pm \ep _i \pm \ep _j) \qquad & i \not= j \qquad & SO(8)
\no \\
(\pm \delta _a \pm \ep _i) \qquad & & {\rm fermionic}
\eea
The root vector space has a non-degenerate indefinite inner product for
which $\delta _a \cdot \ep_i=0$, $\delta _a \cdot \delta _b = - \delta _{ab}$,
and $\ep_i \cdot \ep_j=  \delta _{ij}$, and we have $a,b=1,2$, and $i,j=1,2,3,4$.

\subsubsection{$F(4)$ is not a subalgebra of $OSp(8^*|4)$}

The maximal bosonic subalgebra of the real form $F(4;p)$ of $F(4)$
is $SO(1,2) \oplus SO(p,7-p)$ for $p=0,1,2,3$. Clearly, we have
$SO(1,2) \oplus SO(p,7-p) \not \subset SO(2,6)$.
When $p=0$, the component $SO(7)$ fits neither in $SO(2,6)$, nor in $Sp(4)=SO(5)$.
When $p\not=0$, both components of $SO(1,2) \oplus SO(p,7-p)$ are non-compact;
since they do not simultaneoulsy  fit into $SO(2,6)$, it  follows that
$F(4;p)$ is not a subalgebra of $OSp(8^*|4)$ for $p=0,1,2,3$.

\subsubsection{$F(4)$ is not a subalgebra of $OSp(8|4,\bR)$}

When $p \not=0$, the maximal bosonic algebra of $F(4;p)$,
namely $SO(1,2) \oplus SO(p,7-p)$, has two non-compact components,
which do not simultaneously fit into $Sp(4,\bR)$, so that $F(4;p)$
is not a subalgebra of $OSp(8|4,\bR)$ for $p=1,2,3$. The
only case left  is the real form $F(4;0)$.

\sm

1. To realize the bosonic subalgebra  $Sp(2,\bR) \oplus SO(7)$ of $F(4;0)$,
the $SO(7)$ part must be a subalgebra of $SO(8)$, and the $Sp(2,\bR)$
part must be a subalgebra of  $Sp(4,\bR)$. There is a single embedding
of $SO(7)$ into $SO(8)$ (up to the triality of $SO(8)$), which may be
obtained by ``folding" any two of the orthogonal simple roots of $SO(8)$,
to be specific we fold the roots $\ep_3 \pm \ep_4$.
The resulting simple root $\beta$ of $SO(7)$ corresponds to the following
linear combination of generators of $SO(8)$,
\bea
E_\beta \sim E_{\ep_3- \ep_4} + E_{\ep_3 + \ep_4}
\eea
(Note that by this same ``folding", one obtains the subalgebra
$OSp(7|4)$ from $OSp(8|4)$.)
Using also the remaining two simple root generators of $SO(7)$,
namely $E_{\ep_1-\ep_2}$ and $E_{\ep_2-\ep_3}$, we deduce all 18
roots of $SO(7)$, expressed in the $SO(8)$ basis,
\bea
E_{\pm \ep _1 \pm \ep _2} &   \hskip 1in &
	E_{\ep _3 - \ep _4} + E_{\ep _3 + \ep_4}
\no \\
E_{\pm \ep _2 \pm \ep _3} &  \hskip 1in &
	E_{\ep _1 - \ep _4} + E_{\ep _1 + \ep_4}
\no \\
E_{\pm \ep _1 \pm \ep _3} & \hskip 1in &
	E_{\ep _2 - \ep _4} + E_{\ep _2 + \ep_4}
\eea
where the $\pm$ signs in the left column are all uncorrelated.
To realize the full maximal bosonic subalgebra $Sp(2,\bR) \oplus SO(7)$
of $F(4;0)$ as a subalgebra of $SO(8) \oplus Sp(4,\bR)$, we need to
realize also $Sp(2,\bR)$ as a subalgebra of $Sp(4,\bR)$.
There are 3 possible inequivalent embeddings of $Sp(2,\bR)=SO(1,2)$
into $Sp(4,\bR)=SO(2,3)$, corresponding to the following decompositions
of the ${\bf 5}$ of $SO(2,3)$, and $Sp(4,\bR) $ respectively:
\bea
\label{F4cases}
(a) & \hskip 1in & {\bf 5} = {\bf 2} \oplus {\bf 2} \oplus {\bf 1}
	 \hskip 1in  {\bf 4} = {\bf 2} \oplus {\bf 1} \oplus {\bf 1}
\no \\
(b) & \hskip 1in & {\bf 5} = {\bf 3} \oplus {\bf 1} \oplus {\bf 1}
	 \hskip 1in  {\bf 4} = {\bf 2} \oplus {\bf 2}
\no \\
(c) & \hskip 1in & {\bf 5} = {\bf 5}
	 \hskip 1.6in  {\bf 4} = {\bf 4}
\eea
The associated root generators of $Sp(2,\bR)$ in the $Sp(4,\bR)$ basis
for these three cases are given as follows, (up to equivalences),
\bea
\label{F4cases1}
(a) & \hskip 1in & E_{ 2 \delta _1} , \, E_{- 2 \delta _1}
\no \\
(b) & & E_{2  \delta _1} + E_{2 \delta _2} , \, E_{ -2  \delta _1} + E_{-2 \delta _2}
\no \\
(c) &&  E_{\delta _1 + \delta _2} , \, E_{-\delta _1 - \delta _2}
\eea
Thus, in total, the maximal bosonic subalgebra $SL(2) \times SO(7)$
of $F(4;0)$ can be embedded in three inequivalent ways in the
maximal bosonic subalgebra of $OSp(8|4,\bR)$, according to the three
cases listed above.

\sm

2. The fermionic generators of $F(4;0)$ must transform under the
representation $(2,8_s)$ of $Sp(2,\bR) \times SO(7)$, where the ${\bf 8}_s$
is the {\sl irreducible} spinor representation of $SO(7)$. Since we have
explicit expressions for the generators of $SO(7)$ in the $SO(8)$
Cartan basis, we may compute directly how the fermionic generators
$E_{\pm \ep _i \pm \delta _a}$ transform under this $SO(7)$.
For every $\delta \in \{ \pm \delta _1, \pm \delta _2\}$, we have
\bea
{\bf 7} ~ \Bigg  \{ \matrix{ E_{\pm \ep _i + \delta} & i=1,2,3 \cr
E_{\ep_4 + \delta } + E_{-\ep _4 + \delta } \cr}
\hskip 1in
{\bf 1} ~ \Bigg \{ \matrix{
E_{\ep_4 + \delta } - E_{-\ep _4 + \delta } \cr}
\eea
Hence, it is impossible to realize the 8-dimensional
irreducible spinor representation of $SO(7)$.
As a result, $F(4;0)$ cannot be  a subalgebra of $OSp(8|4,\bR)$,
irrespective of the realization of the $Sp(2,\bR)$ into $ Sp(4,\bR)$.

\subsubsection{$G(3)$ is not a subalgebra of $OSp(8^*|4)$ or $OSp(8|4,\bR)$}

The maximal bosonic subalgebra $SL(2,\bR) \oplus G_{2,2}$ of the real form
$G(3;1)$ cannot fit into either $SO(2,6) \oplus Sp(4)$ or $SO(8) \oplus Sp(4,\bR)$,
and hence $G(3;1)$ is not a subalgebra of either $OSp(8^*|4)$ or $OSp(8|4,\bR)$.
Also, the maximal bosonic subalgebra $SL(2,\bR) \oplus G_2$ of the real form
$G(3;0)$, with $G_2$ compact, cannot fit into $SO(2,6) \oplus Sp(4)$,
and hence $G(3;0)$ is not a  subalgebra of $OSp(8^*|4)$.
The only case left  is $G(3;0)$ as a subalgebra of $OSp(8|4,\bR)$.

\sm

1. To realize the maximal bosonic subalgebra $SL(2, \bR) \oplus G_2$ of the real
form $G(3;0)$ as a subalgebra of $SO(8) \oplus Sp(4,\bR)$, $G_2$ must be
in $SO(8)$ and $SL(2,\bR)$ must be in $Sp(4.\bR)$.
There is a single embedding of $G_2$ into $SO(8)$, (up to the triality of $SO(8)$),
realized as $G_2 \subset SO(7) \subset SO(8)$, which may
be obtained by ``folding" all three orthogonal simple roots of $SO(8)$,
$\ep_1-\ep_2$, $\ep_3-\ep_4$, and $\ep_3+\ep_4$.
The resulting simple root $\beta$ of $G_2$ corresponds
to the following linear combination of generators of $SO(8)$,
\bea
E_\beta  \sim  E_{\ep_1-\ep_2} + E_{\ep_3- \ep_4} + E_{\ep_3 + \ep_4}
\eea
Using the  remaining simple root $E_{\ep_2-\ep_3}$ of $G_2$,
we deduce the remaining  root system of generators
of $G_2$ in the $SO(8)$ basis of generators,
\bea
E_{\pm(\ep_2 - \ep_3)} & \hskip 1in &
	E_{\pm (\ep _1 - \ep_2)} + E_{\pm (\ep _3 - \ep_4)} + E_{\pm (\ep _3 + \ep_4)}
\no \\
E_{\pm(\ep_1 + \ep_2)} & \hskip 1in &
	E_{\pm (\ep _1 - \ep_3)} + E_{\pm (\ep _2 - \ep_4)} + E_{\pm (\ep _2 + \ep_4)}
\no \\
E_{\pm(\ep_1 + \ep_3)} & \hskip 1in &
	E_{\pm (\ep _2 + \ep_3)} + E_{\pm (\ep _1 - \ep_4)} + E_{\pm (\ep _1 + \ep_4)}
\eea
where the $\pm$ signs in the right column are correlated on each line.
To realize the full maximal bosonic subalgebra $SL(2,\bR) \oplus G_2$
of $G(3;0)$ as a subalgebra of $SO(8) \oplus Sp(4)$, we need to
realize also $SL(2,\bR)$ as a subalgebra of $Sp(4,\bR)$.
This problem was already encountered when dealing with the case of $F(4)$.
There are 3 possible inequivalent embeddings, corresponding to the
decompositions of the ${\bf 5}$ of $SO(2,3) $, given
in (\ref{F4cases}), with associated generators given in (\ref{F4cases1}).
In total, the maximal bosonic subalgebra $SL(2) \times G_2$
of $G(3;0)$ can be embedded in three inequivalent ways in the
maximal bosonic subalgebra of $OSp(8|4,\bR)$.

\sm

2. The fermionic generators of $G(3;0)$ transform under the representation
$(2,7)$ of $SL(2,\bR) \times G_2$. We shall now investigate whether fermionic
generators can be realized in terms of linear combinations of
the fermionic   generators of $OSp(8|4,\bR)$, for each of the three embeddings
of (\ref{F4cases}), and (\ref{F4cases1}).
\begin{description}
\item[(a)] Under $Sp(2,\bR) \oplus G_2$, the fermionic generators should transform
as a $({\bf 2}, {\bf 7})$, leaving the following unique realization,
\bea
\label{G2fermion}
&& E_{\pm \ep _i \pm \delta _1} \hskip 2in  i=1,2,3
\no \\
&& E_{\pm(\ep _4 + \delta _1) } + E_{\pm(- \ep _4 + \delta _1)}
\eea
The $\pm$ signs on the second line are correlated, while those on the
first line are not.
The generators $E_{\pm(\ep _4 + \delta _1) } - E_{\pm(- \ep _4 + \delta _1)}$
are  singlets under $G_2$, while $E_{\pm \ep_i \pm \delta _2}$
with $i=1,2,3,4$ are singlets under $Sp(2,\bR)$. The 14 generators of
(\ref{G2fermion}) have indeed the correct transformation properties
under $SL(2,\bR) \times G_2$ and constitute the only viable candidates
for the representation of the fermionic generators of $G(3;0)$ inside
$OSp(8|4,\bR)$.
\item[(b)] In this case, the representation of $Sp(2,\bR)$ is a sum of two doublets
under $Sp(2,\bR)$. This leaves  the following possible realizations
of the positive root fermionic generators,
\bea
\label{G2fermion1}
&& a E_{\pm \ep _i + \delta _1} + b E_{\pm \ep _i + \delta _2},  \hskip 1in  i=1,2,3
\no \\
&& a \left ( E_{\ep _4 + \delta _1 } + E_{- \ep _4 + \delta _1} \right )
+ b \left ( E_{\ep _4 + \delta _2 } + E_{- \ep _4 + \delta _2} \right )
\eea
Here,  $a$ and $b$ are two constants which we may choose to satisfy
$|a|^2 + |b|^2=2$. The fact that the same constants occur in the seven
linear combinations above is required by  $G_2$ covariance.
It is manifest that these generators also form 7 doublets under the action
of the $Sp(2,\bR)$ root generators $E_{\pm 2 \delta _1} + E_{\pm 2 \delta _2}$,
for any values of $a,b$.
\item[(c)] The representation of $Sp(2,\bR)$ is irreducible and
of dimension 4, and cannot be used to realize the doublet needed in
the representation $({\bf 2},{\bf 7})$.
Hence the fermionic generators of $G(3;0)$ cannot be realized in this case.
\end{description}

\sm

3. Having now determined the structure of the possible embeddings of
the maximal bosonic subalgebras and the fermionic generators, it remains
to analyze the structure of the anti-commutators of these generators,
and see whether they close onto the bosonic generators of $G(3;0)$;
we shall show that they do not.
\begin{description}
\item[(a)] The anticommutators $\{ E_{\pm \ep _i + \delta _1},
E_{\pm \ep _j - \delta _1} \} \sim E_{\pm \ep _i \pm \ep _j}$
for all $i\not= j=1,2,3$ produce the roots of the full orthogonal
algebra $SO(6)$, which is, however, not a subalgebra of $G_2$.
(The anti-commutators $\{ E_{\pm \ep _i + \delta _1},
E_{ \ep _4 - \delta _1} + E_{-\ep_4 - \delta _1} \} \sim E_{\pm \ep _i + \ep _4}
+ E_{\pm \ep _i - \ep _4}$ further enlarges this algebra.)
As a result, the fermion generators do not properly close
to form $G(3)$.
\item[(b)] The following anti-commutators
\bea
\{ aE_{\pm \ep _i + \delta _1} + bE_{\pm \ep _i + \delta _2},
aE_{\pm \ep _j - \delta _1} + bE_{\pm \ep _j - \delta _2} \}
\sim
E_{\pm \ep _i \pm \ep _j} \hskip 0.7in i\not= j
\eea
again produces all of $SO(6) \not \subset G_2$. Similarly, the anti-commutators
of the fermionic generators do also not properly close onto $Sp(2,\bR)$.
Thus, the fermion generators do not properly close to form $G(3;0)$.
\end{description}
This concludes the proof that $G(3;0)$ cannot be a subalgebra of $OSp(8|4,\bR)$.

\subsubsection{$D(2,1;c)$ a subalgebra of $OSp(8|4)$ implies $c =1,-2$, or $-1/2$}

The maximal bosonic subalgebra of the real form $D(2,1;c;p)$
of $D(2,1;c)$ for $p=0,1,2$ is given by $SL(2,\bR) \oplus SO(p,4-p)$.

\sm

1. We have the following results for the embeddings into the maximal
bosonic subalgebras of $OSp(8^*|4)$ and $OSp(8|4,\bR)$,
\bea
\label{subs1}
SL(2,\bR) \oplus SO(4) & \subset & SO(2,6) \oplus Sp(4)
\no \\
SL(2,\bR) \oplus SO(4) & \subset & SO(8) \oplus Sp(4,\bR)
\no \\
SL(2,\bR) \oplus SO(1,3) &  \subset & SO(2,6) \oplus Sp(4)
\no \\
SL(2,\bR) \oplus SO(1,3) & \not \subset & SO(8) \oplus Sp(4,\bR)
\no \\
SL(2,\bR) \oplus SO(2,2) & \not \subset & SO(2,6) \oplus Sp(4)
\no \\
SL(2,\bR) \oplus SO(2,2) & \not \subset & SO(8) \oplus Sp(4,\bR)
\eea

2. Next, we show that two of the remaining possible embeddings
\bea
SL(2,\bR) \oplus SO(4) & \subset & SO(2,6)
\no \\
SL(2,\bR) \oplus SO(1,3) & \subset & SO(2,6)
\eea
are ruled out for the following reason. In either case, the bosonic subalgebra $\cS$
must be realized in $SO(2,6)$ in such a way that the fermionic generators transform
under $\cS$ in the representation $({\bf 2}, {\bf 2}, {\bf 2})$. The representation
of the fermionic generators would have to involve an entire ${\bf 8}$ multiplet of
$SO(2,6)$, with generators $E_{\pm \ep _i + \delta}$ for $i=1,2,3,4$ for some $\delta$.
Since the desired subalgebra $D(2,1;c)$ is to be simple, the opposite roots
must be contained in the subalgebra as well, so we would also need to include the
fermion contents $E_{\pm \ep _i - \delta}$ for $i=1,2,3,4$, giving a total of 16
fermionic generators, which does not fit with  $D(2,1;c)$.

\sm

Next, we show that two more of the remaining possible embeddings
\bea
SO(3) \oplus SO(3) & \subset & Sp(4)
\no \\
SL(2,\bR) \oplus SO(3) & \subset & Sp(4,\bR)
\eea
are also ruled out. In both cases, the fermion contents must include a
full ${\bf 4}$ of $Sp(4)$ or $Sp(4,\bR)$, given by the generators
$E_{\pm \ep \pm \delta _a}$ with $a=1,2$. Under anti-commutation,
however, these fermionic generators will reproduce the entire $Sp(4)$,
or $Sp(4,\bR)$,  and not just the desired bosonic subalgebras.

\sm

Assembling the results from the above arguments, we see that the case
on the third line of (\ref{subs1}) is ruled out altogether, so that the only
real form that remains to be considered is $D(2,1;c;0)$, with the
following possible embeddings of the maximal bosonic subalgebras,
\bea
\label{twocases}
OSp(8|4,\bR) && \left \{ \matrix{
	SO(3) \oplus SO(3) \subset SO(8) \cr   SL(2,\bR) \subset Sp(4,\bR) \cr } \right .
\no \\
OSp(8^*|4) ~ && \left \{ \matrix{
	SL(2,\bR) \oplus SO(3) \subset SO(2,6) \cr SO(3) \subset Sp(4) \cr} \right .
\eea
The two cases of (\ref{twocases}) may be viewed as different real forms of the same
embedding structure over $\bC$. We shall discuss the first case in detail; the second
case is completely analogous.

\sm

For the first embedding case in (\ref{twocases}), there are two further different possible
embeddings of $SO(3) \oplus SO(3)$ into $SO(8)$,
which may be distinguished by the branching rules of the fundamental representation
${\bf 8}$ of $SO(8)$,
\bea
(1) & & {\bf 8} = {\bf 4} \oplus {\bf 1} \oplus {\bf 1} \oplus {\bf 1} \oplus {\bf 1}
\no \\
(2) && {\bf 8} = {\bf 4} \oplus {\bf 4}
\eea
Similarly, there are three possible embeddings of $SL(2,\bR)=Sp(2,\bR)$
into $Sp(4,\bR)$, which already listed for the case of $F(4)$ in (\ref{F4cases}).
It is immediate, however, that case (c) is ruled out, as we seek
a fermionic representation transforming under the  ${\bf 2}$ of this $Sp(2,\bR)$,
not an irreducible ${\bf 4}$. Combining all, we have 4 remaining possible cases,
which we can label (1a), (1b), (2a), (2b), according to the labels for embedding
into $SO(8)$ and $Sp(4,\bR)$.

\sm

3. We shall now work out these cases one by one.

\sm

$\bullet$ In case (a1), the generators produce the superalgebra
$OSp(4|2,\bR)=D(2,1;1;0)$.

\sm

$\bullet$ In case (1b), the bosonic generators are $E_{\pm \ep _1 \pm \ep _2}$
for the $SO(4)$ part, and $E_{\pm 2 \delta _1} + E_{\pm 2 \delta _2}$
for the $SL(2,\bR)$ part. The fermionic generators must then be,
\bea
a E_{\pm \ep _i + \delta _1} + b E_{\pm \ep _i + \delta _2} & & i=1,2
\no \\
a E_{\pm \ep _i - \delta _1} + b E_{\pm \ep _i - \delta _2} & & i=1,2
\eea
where $a,b$ are two constants. The anti-commutators of the
generators with opposite $\delta _a$ correctly reproduce
the bosonic generators of $SO(4)$. But the anti-commutators
\bea
\{ a E_{\ep  + \delta _1} + b E_{\ep  + \delta _2} , \,
a E_{-\ep  + \delta _1} + b E_{-\ep  + \delta _2} \}
= a^2 E_{2 \delta _1} + b^2 E_{2 \delta _2} + 2 ab E_{\delta _1 + \delta _2}
\eea
do not correctly reproduce the generators of $SL(2,\bR)$ for any choice
of $a,b$. The absence of the generators $E_{\delta _1 +\delta _2}$ would
require $ab=0$, while reproducing $E_{\pm 2 \delta _1} + E_{\pm 2 \delta _2}$
would require $a=b$,  implying $a=b=0$,
and thus no fermionic generators. This rules out case (1b).

\sm

$\bullet$ In case (2a), the argument is similar to the one given in case (1b), but the roles
of the bosonic subalgebras $SO(4)$ and $SL(2,\bR)$ are interchanged.
The bosonic generators are now $E_{\pm 2 \delta _1}$
for the $SL(2,\bR)$ part, and $E_{\pm \ep _1 \pm \ep_2}
+ E_{\pm \ep _3 \pm \ep _4}$ (with correlated $\pm$ signs
between the two terms). The fermionic generators must form a
doublet under $SL(2,\bR)$, and an irreducible ${\bf 4}$ under $SO(4)$,
which fixes them to be
\bea
a E_{\pm \ep _i + \delta _1} + b E_{\pm \ep _{2+i} + \delta _1} && i=1,2
\no \\
a E_{\pm \ep _i - \delta _1} + b E_{\pm \ep _{2+i} - \delta _1} && i=1,2
\eea
where $a,b$ are two constants. The anti-commutators for opposite
$\ep$ correctly reproduce the generators $E_{\pm 2 \delta _1}$, but
the commutators for opposite $\delta$ give, for example,
\bea
&&
\{ a E_{\ep _1 + \delta _1} + b E_{\ep _3 + \delta _1}, \,
a E_{\ep _2 -  \delta _1} + b E_{\ep _4 - \delta _1} \}
\no \\ && \qquad a^2 E_{\ep _1 + \ep_2} + b^2 E_{\ep_3 + \ep_4}
+ ab (E_{\ep_1 + \ep_4} - E_{\ep _2 + \ep_3}) \quad
\eea
Absence of the generators $E_{\ep_1 + \ep_4}$
and $E_{\ep _2 + \ep_3}$ requires $ab=0$, while recovering
the correct generators of the type $E_{\pm \ep _1 \pm \ep_2}
+ E_{\pm \ep _3 \pm \ep _4}$ requires $a^2=b^2$, which implies
$a=b=0$, and hence no fermionic generators. This rules out case (2a).

\sm

$\bullet$ Finally, in case (2b), the bosonic generators of the $SO(4)$ part
are $E_{\pm \ep _1 \pm \ep_2} + E_{\pm \ep _3 \pm \ep _4}$,
while those of the $SL(2,\bR)$ part are
$E_{\pm 2 \delta _1} + E_{\pm 2 \delta _2}$. The fermionic generators
must be,
\bea
X^+_{\pm \, i} \equiv a E_{\pm \ep _i + \delta _1} + b E_{\pm \ep _i + \delta _2}
+ c E_{\pm \ep'_i + \delta _1 } + d E_{\pm \ep '_i + \delta _2}
&\hskip 0.6in & i=1,2
\no \\
X^-_{\pm \, i} \equiv a E_{\pm \ep _i - \delta _1} + b E_{\pm \ep _i - \delta _2}
+ c E_{\pm \ep'_i - \delta _1 } + d E_{\pm \ep '_i - \delta _2} && i=1,2
\eea
where $\ep'_1 = \ep _3$ and $\ep_2' = \ep _4$.
Here,  $a,b,c,d$ are constants which are independent of $i$ and of $\pm$
by $SO(4)$ invariance and must be the same on the two lines
in view of $SL(2,\bR)$-invariance. The anti-commutators for
opposite $\delta$, and opposite $\ep$ give, for example,
\bea
\{ X^+_{+ 1}, X^- _{+2} \}
& = &
(a^2 +b^2) E_{\ep_1 + \ep_2} + (c^2 + d^2) E_{\ep_3 + \ep_4}
+ ac E_{\ep_1 + \ep_3} + bd E_{\ep_2 + \ep_4}
\no \\
\{ X^+ _{+ \, 1}, X^+ _{- \, 1} \}
& = &
(a^2 +c^2) E_{2 \delta _1} + (b^2 + d^2) E_{2 \delta _2}
+ ( ab + cd) E_{\delta _1 + \delta _2}
\eea
Absence of the undesirable generators $E_{\ep_1 + \ep_3}$, and
$ E_{\ep_2 + \ep_4}$ on the first line requires that $ac=bd=0$.
Absence of the undesirable generators $E_{\delta _1 + \delta _2}$
on the second line requires $ab+cd=0$. Fully recovering the
closure onto the bosonic subalgebra requires the  conditions,
\bea
\label{conds}
0 & = & a^2 + b^2 - c^2 - d^2 = a^2 - b^2 + c^2 - d^2
\no \\
0 & = & ab+cd = ac=bd
\eea
The first line is equivalent to $a^2=d^2, b^2=c^2$.
If $a \not=0$, we have $c=b=0$ by (\ref{conds}). Choosing $a=d=1$, without loss of
generality, the fermionic generators reduce to
\bea
X^+_{\pm \, i} \equiv E_{\pm \ep _i + \delta _1} +  E_{\pm \ep '_i + \delta _2}
&\hskip 0.6in & i=1,2
\no \\
X^-_{\pm \, i} \equiv  E_{\pm \ep _i - \delta _1}  +  E_{\pm \ep '_i - \delta _2}
&& i=1,2
\eea
This gives the structure relations of $OSp(4|2,\bR)$,
embedded into $OSp(8|4,\bR)$ as the diagonal
subalgebra of $OSp(4|2,\bR) \oplus OSp(4|2,\bR)$, where the first term
is based on the root generators $E_{\pm \ep _1 \pm \delta _1}
+ E_{\pm \ep_3 \pm \delta _2}$ and the second on the
root generators $E_{\pm \ep _2 \pm \delta _1} + E_{\pm \ep_4 \pm \delta _2}$.
If $b \not=0$, we obtain an equivalent realization to the case
$a^2=d^2\not=0$, again giving $OSp(4|2,\bR)$, embedded into $OSp(8|4,\bR)$
as the diagonal subalgebra of $OSp(4|2,\bR) \oplus OSp(4|2,\bR)$.

\sm

In conclusion, cases (1b) and (2a) are ruled out,  while for cases (1a), and (2b)
the subalgebra is $D(2,1;1;0)=OSp(4|2,\bR)$.  In case (1a), there is room for also
a second copy, and the corresponding embedding is actually via two commuting
superalgebras,
\bea
\label{case11}
OSp(4|2,\bR) \oplus OSp(4|2,\bR) \subset OSp(8|4,\bR)
\eea
The proof for the second embedding case in (\ref{twocases}) is completely
analogous, and offers an outcome analogous to that of (\ref{case11}).
The result is that $D(2,1;c;0)$ can be embedded into $OSp(8^*|4)$ only when
$c=-2,-1/2$, in which case we have $D(2,1;c;0) = OSp(4^*|2)$ for $c=-2,-1/2$.
The embedding of $OSp(4^*|2)$ can be via a single block, or via two commuting
superalgebras,
\bea
\label{case12}
OSp(4^*|2) \oplus OSp(4^*|2) \subset OSp(8^*|4)
\eea
The exceptional basic Lie superalgebra $D(2,1;c)$, or any real form $D(2,1;c;p)$ thereof,
is never a subalgebra of $OSp(8|4,\bR)$ when $c \not= 1$,
and is never a subalgebra of $OSp(8^*|4)$ when $c\not= -2,-1/2$.

\newpage

\section{Superalgebra structure of the M-theory solutions}
\setcounter{equation}{0}

The infinitesimal symmetry generators of a particular half-BPS solution form a \sa\
with bosonic and fermionic generators.
On a supergravity solution we identify the bosonic generators  with isometries
of the metric and the fermionic generators with unbroken supersymmetries.
The BPS equations for M-theory are given by,
\bea
\label{BPS1}
\nabla_M \ep +{1\over 288} \Big(\Gamma_M{}^{NPQR}
- 8 \delta_M{}^N \Gamma^{PQR} \Big) F_{NPQR} \, \ep=0
\eea
We generally follow the conventions and notations of \cite{DHoker:2008wc}.
Here, $\ep$ is an 11-dimensional Majorana spinor, and $\nabla _M$ is
the covariant derivative with respect to the Levi-Civita connection for $g_{MN}$.
Given any $\ep$ and $\ep'$ satisfying (\ref{BPS1}),
it follows from the BPS equations that the vector $v_M$,
defined below, satisfies the Killing equation,
\bea
\label{Kill1}
v_M & \equiv & e_M {}^A \left ( \bar \ep  \Gamma _A \ep ' \right )
\no \\
0 & = &   \nabla _M v_N + \nabla _N v_M
\eea
This result is established by  computing the following expression using the
BPS equations,
\bea
3 \nabla _M v_N =   ( \bar \ep  \Gamma ^{QR} \ep '  ) F_{MNQR}
\eea
and symmetrizing in $M,N$.
The result may be interpreted as a composition law for supersymmetry
transformations $\ep$ and $\ep'$, which closes onto Killing
vector $v_M$, and may be used to give a constructive derivation
of the superalgebra  \cite{Gauntlett:1998kc}.

\sm

We apply the above methods to prove that the
invariance \sa\ of the M-theory half-BPS solutions of \cite{DHoker:2008wc} is
$D(2,1;c) \times D(2,1;c)$. Here, $D(2,1;c)$ is the
1-parameter family of exceptional basic classical Lie superalgebras,
and the real number $c$ will be related to certain parameters of the solutions.
We refer to \cite{DHoker:2008wc}  for additional details on this solution.

\subsection{Calculation of the Killing vectors}

The Ansatz for $SO(2,2) \times SO(4)_2 \times SO(4)_3$ -invariant solutions to
M-theory on a space-time  $AdS_3 \times S^3_2 \times S^3_3$
warped over a two-dimensional Riemann surface $\Sigma$ with boundary
is given as follows. The 11-dimensional metric $ds^2$ associated with
the Minkowski metric $\eta _{AB} = (- + \cdots +)$ on the orthonormal frame
$e^A$, with $A,B=0,1,2, \cdots, 9, \natural = 10$, is given by
$ds^2 = \eta _{AB} \, e^A \otimes e^B$.

\sm

The components of the orthonormal frame $e^A = dx^M \, e_M {}^A$ for this
Ansatz take the form,
\bea
\label{frame1}
e^{a_1} & = & f_1 \hat e ^{a_1} \hskip 1in a_1 = 0,1,2 \hskip 1in AdS_3
\no \\
e^{a_2} & = & f_2 \hat e ^{a_2} \hskip 1in a_2 = 3,4,5 \hskip 1.1in S_2^3
\no \\
e^{a_3} & = & f_1 \hat e ^{a_3} \hskip 1in a_3 = 6,7,8 \hskip 1.1in S_3^3
\no \\
e^a ~ && \hskip 1.35in a=9, \natural = 10 \hskip 1in \Sigma
\eea
An explicit expression for the invariant 4-form field strength $F$ may
be found in \cite{DHoker:2008wc} but will not be needed here.
A choice of well-adapted Dirac matrices is given by
\bea
\G^{a_1} = \g^{a_1} \otimes I_2 \otimes I_2 \otimes \s^1 \otimes \s^3
& \hskip 0.5in &
\s^2 = -i \g^0 = \g^3 = \g ^6 = \g^\natural
\no \\
\G^{a_2} =  I_2 \otimes \g^{a_2} \otimes I_2 \otimes \s^2 \otimes \s^3
& \hskip 0.5in &
\s^1 = \g^1 = \g^4 = \g^7 = \g^9
\no \\
\G^{a_3} =  I_2 \otimes I_2 \otimes \g^{a_3} \otimes \s^3 \otimes \s^3
& \hskip 0.5in &
\s^3 = \g^2 = \g^5 = \g^8
\no \\
\G^a ~ = ~ I_2 \otimes I_2 \otimes I_2 \otimes I_2  \otimes \g^a \, & &
\eea
where $ a_1 = 0,1,2$, $ a_2 = 3,4,5$,  $ a_3 = 6,7,8$, and $a=9,\natural$.
The supersymmetry parameter $\ep$ may be  decomposed on a basis of
$SO(2,2)\times SO(4)_2 \times SO(4)_3$ -invariant  Killing spinors
$\chi ^{\eta_1, \eta _2, \eta _3}$, with
$\eta _1, \eta _2 , \eta _3 = \pm$,
\bea
\ep = \chi ^{\eta _1 , \eta _2 , \eta _3} \zeta _{\eta_1, \eta _2, \eta _3}
\eea
In total, there is a 16-dimensional space of solutions $\ep$, which may
be parametrized by picking definite values for $\eta _1, \eta _2 , \eta _3 = \pm$,
and using a two-fold degeneracy of $\zeta$. It suffices to solve the BPS equations
for the above Ansatz only in part \cite{DHoker:2008wc}
to obtain the  metric factors $f_1,f_2,f_3$ in
terms of $\zeta$. They are given by,
\bea
\label{radii}
2 c_1 f_1 & = & \zeta ^\dagger (I_2 \otimes I_2) \zeta
\no \\
2 c_2 f_2 & = & - \zeta^\dagger (\sigma ^3  \otimes I_2) \zeta
\no \\
2 c_3 f_3 & = & \zeta ^\dagger (\sigma ^2 \otimes I_2)  \zeta
\eea
Here, $c_1, c_2, c_3$ are real integration constants, which are
related by $c_1+c_2+c_3=0$. (In fact, we shall show that the
constant $c$ in the invariance super algebra
$D(2,1;c) \times D(2,1;c)$ is given by $c=c_2/c_1$, or equivalently,
by any other ratio of two $c_i$.)

\sm

The 8 linearly independent Killing spinors $\chi ^{\eta_1, \eta _2, \eta _3}$
may be normalized as follows,
\bea
\label{spinornorm}
\chi ^{\eta_1, \eta _2, \eta _3} & = &
\chi _1^{\eta_1} \otimes \chi _2 ^{ \eta _2} \otimes \chi _3 ^{ \eta _3}
\no \\
\bar \chi _1 ^{\eta_1}  \chi _1 ^{\eta_1'}
& = & i \delta _{\eta _1, \eta _1'}
\no \\
\left ( \chi _i ^{\eta_i} \right )^\dagger  \chi _i ^{\eta_i'}
& = & \delta _{\eta _i, \eta _i'} \hskip 1in i=2,3
\eea
where we define, as usual for Minkowski signature spinors, $\bar \chi _1 ^\eta =
(\chi _1 ^\eta)^\dagger \g^0$.
The Killing vectors of the M-theory solutions may now be evaluated
by considering the combinations $\bar \ep ' \Gamma ^A \ep$ (with
Lorentz frame indices). It will be convenient to decompose these
according to the product $AdS_3 \times S^3_2 \times S^3_3 \times \Sigma$.
One finds,
\bea
\bar \ep  \Gamma ^{a_1} \ep '
& = &
\left ( \bar \chi _1 ^{\eta_1}  \gamma ^{a_1} \chi _1^{\eta_1'} \right ) \,
\delta _{\eta _2, \eta _2'} \,
\delta _{\eta _3 , \eta _3'} \,
\left ( \zeta ^\dagger I_2 \otimes I_2 \zeta \right )
\no \\
\bar \ep   \Gamma ^{a_2} \ep '
& = &
\delta _{\eta _1, \eta _1'} \,
\left ( \left ( \chi _2 ^{\eta_2} \right ) ^\dagger \gamma ^{a_2} \chi _2 ^{\eta_2'} \right ) \,
 \delta _{\eta _3 , \eta _3'} \,
\left ( \zeta ^\dagger (i \, \sigma ^3) \otimes I_2 \zeta \right )
\no \\
\bar \ep   \Gamma ^{a_3} \ep '
& = &
\delta _{\eta _1, \eta _1'} \,
\delta _{\eta _2 , \eta _2'} \,
\left ( \left ( \chi _3 ^{\eta_3} \right ) ^\dagger \gamma ^{a_3} \chi _3^{\eta_3'} \right ) \,
\left ( \zeta ^\dagger (-i \, \sigma ^2) \otimes I_2 \zeta \right )
\no \\
\bar \ep   \Gamma ^a \ep '
& = &
\delta _{\eta _1, \eta _1'} \, \delta _{\eta _2 , \eta _2'} \, \delta _{\eta _3 , \eta _3'}
\, \left ( \zeta ^\dagger \sigma ^1 \otimes  \sigma ^3 \sigma ^a \zeta \right )
\eea
Using the expressions for the metric factors $f_i$ of  (\ref{radii}),
we may convert the bilinears in $\zeta$ in the above expressions
in terms of metric factors $f_i$ and the constants $c_i$.
Also, using the fact that $\zeta ^\dagger \sigma ^1 \otimes  \sigma ^3 \sigma ^a \zeta=0$
for $a=9,\natural$, we obtain,
\bea
\bar \ep  \Gamma ^{a_1} \ep '
& = &
2 c_1 f_1 \, \left ( \bar \chi _1  \gamma ^{a_1} \chi _1' \right ) \,
\delta _{\eta _2, \eta _2'} \, \delta _{\eta _3 , \eta _3'}
\no \\
\bar \ep   \Gamma ^{a_2} \ep '
& = &
2 c_2 f_2 \, \delta _{\eta _1, \eta _1'}  \,
\left ( -i \chi _2  ^\dagger \gamma ^{a_2} \chi _2' \right ) \,
 \delta _{\eta _3 , \eta _3'}
\no \\
\bar \ep   \Gamma ^{a_3} \ep '
& = &
2 c_3 f_3 \,  \delta _{\eta _1, \eta _1'} \,  \delta _{\eta _2 , \eta _2'} \,
\left ( -i \chi _3 ^\dagger  \gamma ^{a_3} \chi _3' \right )
\no \\
\bar \ep   \Gamma ^a \ep '
& = &  0
\eea
It remains to exhibit the relations between the bilinears in $\chi_i$
and the Killing vectors on the unit radius spaces
$AdS_3$, $S_2^3$, and $S_3^3$.

\subsection{Normalization of the unit radius $AdS_3$ Killing vectors}

For $AdS_3$, the Killing spinor equations were derived in \cite{DHoker:2008wc},
and may be expressed as follows in terms of 4-component Dirac spinors $\chi_1$,
\bea
\left ( d + \o ^{(t)}_1  \right ) \chi _1
= \left ( d + {1 \over 4} ( \o_1)  _{a_1 b_1 } \gamma ^{a_1 b_1} \otimes I_2
+ {1 \over 2} (e_1) _{a_1} \gamma ^{a_1} \otimes \sigma ^3 \right ) \chi _1 = 0
\eea
where $a_1,b_1 =0,1,2$, and $\o^{(t)} _1  = U_1  ^{-1} dU_1 $ is parametrized
by $U_1$ in the spinor representation of $ SO(2,2)$. The Killing spinor equations
for 2-component Weyl spinors $\chi^\eta$ \cite{DHoker:2008wc}, may be obtained
from the equations for $\chi_1$  by projection  onto definite chiralities.
The general solution for Dirac spinors $\chi_1$, and for the Weyl spinors
$\chi _1 ^\eta$ are given by
\bea
\chi _1 = U_1 \ep _1 \hskip 0.2in & \hskip 1in &
\no \\
\chi _1 ^\eta  \equiv P_\eta  U_1 \ep _1 & \hskip 1in &
P_\eta \equiv \half ( 1 +  \eta I_2 \otimes \s^3 )
\eea
Here, $\ep_1$ is an arbitrary constant 4-component spinor, which we
conveniently normalize to $\bar \ep _1 \ep_1 =1$, so that
the normalization condition (\ref{spinornorm}) for $\chi_1$ holds. Since
$P_\eta$ commutes with the $SO(2,2)$ generators in
the spinor representation, the actions of the
simple factors $SO(2,1)_+ \times SO(2,1)_-=SO(2,2)$ may be identified
with the generators $P_\pm \gamma ^{a_1 b_1 }$. The Killing spinors
$\chi _1 ^+$ and $\chi _1 ^-$ transform under $SO(2,1)_+\otimes SO(2,1)_-$
respectively as  $({\bf 2},{\bf 1})$ and $({\bf 1}, {\bf 2})$.
It is now immediate to evaluate the Killing vector
combinations $\bar \chi _1 \gamma ^{a_1} \chi _1 '$, and
we have
\bea
\bar \chi _1^\eta  \gamma ^{a_1} \chi _1^{-\eta}
& = & 0
\no \\
(v_1)^{a_1} _\pm \equiv \bar \chi _1 ^\pm  \gamma ^{a_1} \chi _1 ^\pm
& = &
\ep_1^\dagger U_1^\dagger \gamma ^0  \gamma ^{a_1}
U_1 P_\pm \ep_1
\eea
The $(v_1)^{a_1}_+$ and $(v_1)^{a_1} _-$ are respectively the Killing vectors
generating the isometry groups $SO(2,1)_+$ and $SO(2,1)_-$
on $AdS_3$ of unit radius.

\subsection{Normalization of the unit radius $S^3$ Killing vectors}

Similarly, for $S^3_i$, with $i=2,3$, the Killing spinor equations,
expressed in terms of 4-component Dirac spinors,  are given by,
\bea
\left ( d + \o ^{(t)} _i \right ) \chi  _i \left ( d + {1 \over 4} (\o _i) _{a_i b_i } \gamma ^{a_i b_i} \otimes I_2
+ {i \over 2} (e_i) _{a_i} \gamma ^{a_i} \otimes \s^3 \right ) \chi  _i = 0
\eea
where $a_2, b_2=3,4,5$, $a_3,b_3=6,7,8$, and
$\o^{(t)}_i  = U_i ^\dagger dU_i $ is parametrized by $U_i$
in the spinor representation of $ SO(4)_i$.
The general solution $\chi _i $ and the chiral solutions
$\chi _i^\eta$ are given by
\bea
\chi _i = U_i \ep _i
\hskip 1in
\chi _i ^\eta  \equiv P_\eta  U_i  \ep _i
\hskip 1in
P_\eta \equiv \half ( 1 +  \eta I_2 \otimes \s^3 )
\eea
Here, $\ep_i$ are arbitrary constant spinors, which we
conveniently normalize to  $\ep _i ^\dagger \ep_i=1$, so that
the normalization conditions (\ref{spinornorm}) for $\chi_i$ hold. Since
$P_\eta$ commutes with the $SO(4)_i$ generators in
the spinor representation, the actions of the
simple factors $SU(2)_{i+} \times SU(2)_{i -}=SO(4)_i$ may be identified
with the rotations $P_\pm \gamma ^{a_i b_i}$. The Killing spinors
$\chi _i^+$ and $\chi _i ^-$ transform under $SU(2)_{i+}\otimes SU(2)_{i-}$
respectively as  $({\bf 2},{\bf 1})_i$ and $({\bf 1}, {\bf 2})_i$.
It is now immediate to evaluate the Killing vector
combinations $\chi _i^\dagger \gamma ^{a_i} \chi _i '$, and
we have
\bea
(\chi ^\eta_i  )^\dagger \gamma ^{a_i} \chi ^\eta_i
& = & 0 \hskip 2in \eta = \pm
\no \\
(v_i) ^{a_i} _\pm \equiv (\chi _i ^\mp )^\dagger \gamma ^{a_i} \chi _i ^\pm
& = &
\ep_i^\dagger U_i^\dagger \gamma ^{a_i} U_i P_\pm \ep_i
\eea
The $(v_i)^{a_i}_+$ and $(v_i)^{a_i} _-$ are respectively the Killing vectors
generating the isometry groups $SU(2)_{i+}$ and $SU(2)_{i-}$
on the spheres of unit radii $S_i^3$.

\subsection{Summary of results}

Combining all results, we find that the spinor bilinears evaluate
as follows,
\bea
\label{Killing1}
\bar \ep  \Gamma ^{a_1} \ep '
& = &
2 c_1 f_1 (v_1)^{a_1} _\pm \,
\delta _{\eta _2, \eta _2'}  \delta _{\eta _3 , \eta _3'}
\no \\
\bar \ep   \Gamma ^{a_2} \ep '
& = &
- 2 i c_2 f_2  (v_2)^{a_2} _\pm  \,
\delta _{\eta _1, \eta _1'} \delta _{\eta _3 , \eta _3'}
\no \\
\bar \ep   \Gamma ^{a_3} \ep '
& = &
- 2 i c_3 f_3  (v_3)^{a_3} _\pm \,
\delta _{\eta _1, \eta _1'}  \delta _{\eta _2 , \eta _2'}
\no \\
\bar \ep   \Gamma ^a \ep '
& = &  0
\eea
where $(v_i)_\pm ^A$, with $i=1,2,3$, and $A=0,1,2,3,4,5,6,7,8$
are the Killing vectors for the factors $AdS_3$, $S_2^3$, and $S_3^3$
with unit radius respectively, expressed in terms of frame indices $A$.
To obtain the Killing vectors in the customary Einstein indices $M$,
as in (\ref{Kill1}),  it suffices to scale the Killing vectors $(v_i)_\pm ^A$
with frame indices, by the appropriate corresponding
metric factors $f_i$ of (\ref{frame1}). This rescaling precisely absorbs the
prefactors $f_i$ in (\ref{Killing1}), but not the constants $c_i$.

\sm

The presence of the remaining factors $c_i$ reveals that the closure of the
fermionic generators is precisely that of $D(2,1;c) \times D(2,1;c)$
with $c=c_2/c_1$. This may be seen by consulting \cite{Gauntlett:1998kc},
(or \cite{sorba} page  201), for the precise structure relations of $D(2,1;c)$. Given that the
maximal bosonic subalgebra of this superalgebra must be
$SO(2,2) \times SO(4) \times SO(4)$, we find that, more precisely,
the fermionic generators close onto the following real form,
$D(2,1;c;0) \times D(2,1;c;0)$.


\newpage




\begin{thebibliography}{99}

\itemsep 0.03in

{\small



\bibitem{Freund:1980xh}
  P.~G.~O.~Freund and M.~A.~Rubin,
  ``Dynamics Of Dimensional Reduction,''
  Phys.\ Lett.\  B {\bf 97} (1980) 233.

\bibitem{Romans:1984an}
  L.~J.~Romans,
  ``New Compactifications Of Chiral N=2 D = 10 Supergravity,''
  Phys.\ Lett.\  B {\bf 153} (1985) 392.



\bibitem{FigueroaO'Farrill:2002ft}
  J.~Figueroa-O'Farrill and G.~Papadopoulos,
  ``Maximally supersymmetric solutions of ten- and eleven-dimensional
  supergravities,''
  JHEP {\bf 0303} (2003) 048
  [arXiv:hep-th/0211089].


\bibitem{Gran:2007eu}
  U.~Gran, J.~Gutowski, G.~Papadopoulos and D.~Roest,
  ``IIB solutions with $N >28$ Killing spinors are maximally supersymmetric,''
  JHEP {\bf 0712} (2007) 070
  [arXiv:0710.1829 [hep-th]].


\bibitem{Lin:2004nb}
  H.~Lin, O.~Lunin and J.~M.~Maldacena,
  ``Bubbling AdS space and 1/2 BPS geometries,''
  JHEP {\bf 0410} (2004) 025
  [arXiv:hep-th/0409174].

\bibitem{Berenstein:2004kk}
  D.~Berenstein,
  ``A toy model for the AdS/CFT correspondence,''
  JHEP {\bf 0407}, 018 (2004)
  [arXiv:hep-th/0403110].



\bibitem{DHoker:2007xy}
  E.~D'Hoker, J.~Estes and M.~Gutperle,
  ``Exact half-BPS Type IIB interface solutions I: Local solution and
  supersymmetric Janus,''
  JHEP {\bf 0706}, 021 (2007)
  [arXiv:0705.0022 [hep-th]].

\bibitem{DHoker:2007xz}
  E.~D'Hoker, J.~Estes and M.~Gutperle,
  ``Exact half-BPS Type IIB interface solutions. II: Flux solutions and
  multi-janus,''
  JHEP {\bf 0706}, 022 (2007)
  [arXiv:0705.0024 [hep-th]].


\bibitem{DHoker:2007fq}
  E.~D'Hoker, J.~Estes and M.~Gutperle,
  ``Gravity duals of half-BPS Wilson loops,''
  JHEP {\bf 0706} (2007) 063
  [arXiv:0705.1004 [hep-th]].

\bibitem{DHoker:2008wc}
  E.~D'Hoker, J.~Estes, M.~Gutperle and D.~Krym,
  ``Exact Half-BPS Flux Solutions in M-theory I, Local Solutions,''
  arXiv:0806.0605 [hep-th].

\bibitem{Gauntlett:2002sc}
  J.~P.~Gauntlett, D.~Martelli, S.~Pakis and D.~Waldram,
  ``G-structures and wrapped NS5-branes,''
  Commun.\ Math.\ Phys.\  {\bf 247}, 421 (2004)
  [arXiv:hep-th/0205050].

\bibitem{Gauntlett:2002fz}
  J.~P.~Gauntlett and S.~Pakis,
  ``The geometry of D = 11 Killing spinors,''
  JHEP {\bf 0304}, 039 (2003)
  [arXiv:hep-th/0212008].

\bibitem{Gauntlett:2005bn}
  J.~P.~Gauntlett,
  ``Classifying supergravity solutions,''
  Fortsch.\ Phys.\  {\bf 53}, 468 (2005)
  [arXiv:hep-th/0501229].
  
\bibitem{Gauntlett:2004zh}
  J.~P.~Gauntlett, D.~Martelli, J.~Sparks and D.~Waldram,
  ``Supersymmetric AdS(5) solutions of M-theory,''
  Class.\ Quant.\ Grav.\  {\bf 21}, 4335 (2004)
  [arXiv:hep-th/0402153].

\bibitem{Kirsch:2005uy}
  I.~Kirsch and D.~Vaman,
  ``The D3/D7 background and flavor dependence of Regge trajectories,''
  Phys.\ Rev.\  D {\bf 72}, 026007 (2005)
  [arXiv:hep-th/0505164].


\bibitem{Buchbinder:2007ar}
  E.~I.~Buchbinder, J.~Gomis and F.~Passerini,
  ``Holographic Gauge Theories in Background Fields and Surface Operators,''
  JHEP {\bf 0712} (2007) 101
  [arXiv:0710.5170 [hep-th]].


\bibitem{Harvey:2008zz}
  J.~A.~Harvey and A.~B.~Royston,
  ``Gauge/Gravity duality with a chiral N=(0,8) string defect,''
  arXiv:0804.2854 [hep-th].

\bibitem{Aharony:1998xz}
  O.~Aharony, A.~Fayyazuddin and J.~M.~Maldacena,
  ``The large N limit of N = 2,1 field theories from three-branes in
  F-theory,''
  JHEP {\bf 9807} (1998) 013
  [arXiv:hep-th/9806159].

\bibitem{Grana:2001xn}
  M.~Grana and J.~Polchinski,
  ``Gauge / gravity duals with holomorphic dilaton,''
  Phys.\ Rev.\  D {\bf 65} (2002) 126005
  [arXiv:hep-th/0106014].

\bibitem{Boonstra:1998yu}
  H.~J.~Boonstra, B.~Peeters and K.~Skenderis,
  ``Brane intersections, anti-de Sitter spacetimes and dual superconformal
  theories,''
  Nucl.\ Phys.\  B {\bf 533} (1998) 127
  [arXiv:hep-th/9803231].

\bibitem{Gauntlett:1998kc}
  J.~P.~Gauntlett, R.~C.~Myers and P.~K.~Townsend,
  ``Supersymmetry of rotating branes,''
  Phys.\ Rev.\  D {\bf 59} (1999) 025001
  [arXiv:hep-th/9809065].

\bibitem{deBoer:1999rh}
  J.~de Boer, A.~Pasquinucci and K.~Skenderis,
  ``AdS/CFT dualities involving large 2d N = 4 superconformal symmetry,''
  Adv.\ Theor.\ Math.\ Phys.\  {\bf 3} (1999) 577
  [arXiv:hep-th/9904073].

\bibitem{VanProeyen:1986me}
  A.~Van Proeyen,
  ``Superconformal Algebras,''\\
http://www.slac.stanford.edu/spires/find/hep/www?irn=1943812\\
{\it  in  Vancouver 1986, Proceedings, Super Field Theories, 547-555 }


\bibitem{Maldacena:1997re}
  J.~M.~Maldacena,
  ``The large N limit of superconformal field theories and supergravity,''
  Adv.\ Theor.\ Math.\ Phys.\  {\bf 2}, 231 (1998)
  [Int.\ J.\ Theor.\ Phys.\  {\bf 38}, 1113 (1999)]
  [arXiv:hep-th/9711200].

\bibitem{Gubser:1998bc}
  S.~S.~Gubser, I.~R.~Klebanov and A.~M.~Polyakov,
  ``Gauge theory correlators from non-critical string theory,''
  Phys.\ Lett.\  B {\bf 428}, 105 (1998)
  [arXiv:hep-th/9802109].

\bibitem{Witten:1998qj}
  E.~Witten,
  ``Anti-de Sitter space and holography,''
  Adv.\ Theor.\ Math.\ Phys.\  {\bf 2}, 253 (1998)
  [arXiv:hep-th/9802150].

\bibitem{Aharony:1999ti}
  O.~Aharony, S.~S.~Gubser, J.~M.~Maldacena, H.~Ooguri and Y.~Oz,
  ``Large N field theories, string theory and gravity,''
  Phys.\ Rept.\  {\bf 323}, 183 (2000)
  [arXiv:hep-th/9905111].

\bibitem{D'Hoker:2002aw}
  E.~D'Hoker and D.~Z.~Freedman,
  ``Supersymmetric gauge theories and the AdS/CFT correspondence,''
  arXiv:hep-th/0201253.

\bibitem{Drukker:1999zq}
  N.~Drukker, D.~J.~Gross and H.~Ooguri,
  ``Wilson loops and minimal surfaces,''
  Phys.\ Rev.\  D {\bf 60}, 125006 (1999)
  [arXiv:hep-th/9904191].


\bibitem{Bianchi:2002gz}
  M.~Bianchi, M.~B.~Green and S.~Kovacs,
  ``Instanton corrections to circular Wilson loops in N = 4 supersymmetric
  Yang-Mills,''
  JHEP {\bf 0204}, 040 (2002)
  [arXiv:hep-th/0202003].



\bibitem{Gukov:2006jk}
  S.~Gukov and E.~Witten,
  ``Gauge theory, ramification, and the geometric langlands program,''
  arXiv:hep-th/0612073.


\bibitem{Gukov:2008sn}
  S.~Gukov and E.~Witten,
  ``Rigid Surface Operators,''
  arXiv:0804.1561 [hep-th].


\bibitem{Drukker:2008wr}
  N.~Drukker, J.~Gomis and S.~Matsuura,
  ``Probing N=4 SYM With Surface Operators,''
  arXiv:0805.4199 [hep-th].

\bibitem{Gomis:2007fi}
  J.~Gomis and S.~Matsuura,
  ``Bubbling surface operators and S-duality,''
  JHEP {\bf 0706} (2007) 025
  [arXiv:0704.1657 [hep-th]].


\bibitem{DHoker:2006uv}
  E.~D'Hoker, J.~Estes and M.~Gutperle,
  ``Interface Yang-Mills, supersymmetry, and Janus,''
  Nucl.\ Phys.\  B {\bf 753} (2006) 16
  [arXiv:hep-th/0603013].


\bibitem{Gaiotto:2008ak}
  D.~Gaiotto and E.~Witten,
  ``S-Duality of Boundary Conditions In N=4 Super Yang-Mills Theory,''
  arXiv:0807.3720 [hep-th].


\bibitem{Gaiotto:2008sa}
  D.~Gaiotto and E.~Witten,
  ``Supersymmetric Boundary Conditions in N=4 Super Yang-Mills Theory,''
  arXiv:0804.2902 [hep-th].



\bibitem{Gaiotto:2008sd}
  D.~Gaiotto and E.~Witten,
  ``Janus Configurations, Chern-Simons Couplings, And The Theta-Angle in N=4
  Super Yang-Mills Theory,''
  arXiv:0804.2907 [hep-th].


\bibitem{Constable:2002xt}
  N.~R.~Constable, J.~Erdmenger, Z.~Guralnik and I.~Kirsch,
  ``Intersecting D3-branes and holography,''
  Phys.\ Rev.\  D {\bf 68} (2003) 106007
  [arXiv:hep-th/0211222].


\bibitem{Breitenlohner:1982bm}
  P.~Breitenlohner and D.~Z.~Freedman,
  ``Positive Energy In Anti-De Sitter Backgrounds And Gauged Extended
  Supergravity,''
  Phys.\ Lett.\  B {\bf 115} (1982) 197.

\bibitem{Skenderis:2002vf}
  K.~Skenderis and M.~Taylor,
  ``Branes in AdS and pp-wave spacetimes,''
  JHEP {\bf 0206} (2002) 025
  [arXiv:hep-th/0204054].


\bibitem{Karch:2000ct}
  A.~Karch and L.~Randall,
  ``Locally localized gravity,''
  JHEP {\bf 0105}, 008 (2001)
  [arXiv:hep-th/0011156].

\bibitem{Karch:2000gx}
  A.~Karch and L.~Randall,
  ``Open and closed string interpretation of SUSY CFT's on branes with
  boundaries,''
  JHEP {\bf 0106}, 063 (2001)
  [arXiv:hep-th/0105132].

\bibitem{Cederwall:1996ri}
  M.~Cederwall, A.~von Gussich, B.~E.~W.~Nilsson, P.~Sundell and A.~Westerberg,
  ``The Dirichlet super-p-branes in ten-dimensional type IIA and IIB
  supergravity,''
  Nucl.\ Phys.\  B {\bf 490} (1997) 179
  [arXiv:hep-th/9611159].

\bibitem{Aganagic:1996pe}
  M.~Aganagic, C.~Popescu and J.~H.~Schwarz,
  ``D-brane actions with local kappa symmetry,''
  Phys.\ Lett.\  B {\bf 393} (1997) 311
  [arXiv:hep-th/9610249].

\bibitem{Bergshoeff:1996tu}
  E.~Bergshoeff and P.~K.~Townsend,
  ``Super D-branes,''
  Nucl.\ Phys.\  B {\bf 490} (1997) 145
  [arXiv:hep-th/9611173].


\bibitem{Gomis:2006sb}
  J.~Gomis and F.~Passerini,
  ``Holographic Wilson loops,''
  JHEP {\bf 0608} (2006) 074
  [arXiv:hep-th/0604007].



\bibitem{DeWolfe:2001pq}
  O.~DeWolfe, D.~Z.~Freedman and H.~Ooguri,
  ``Holography and defect conformal field theories,''
  Phys.\ Rev.\  D {\bf 66} (2002) 025009
  [arXiv:hep-th/0111135].

\bibitem{Erdmenger:2002ex}F
  J.~Erdmenger, Z.~Guralnik and I.~Kirsch,
  ``Four-dimensional superconformal theories with interacting boundaries or
  defects,''
  Phys.\ Rev.\  D {\bf 66} (2002) 025020
  [arXiv:hep-th/0203020].



\bibitem{Karch:2002sh}
  A.~Karch and E.~Katz,
  ``Adding flavor to AdS/CFT,''
  JHEP {\bf 0206}, 043 (2002)
  [arXiv:hep-th/0205236].


\bibitem{Schwarz:1983qr}
  J.~H.~Schwarz,
  ``Covariant Field Equations Of Chiral N=2 D=10 Supergravity,''
  Nucl.\ Phys.\  B {\bf 226} (1983) 269.

\bibitem{Yamaguchi:2006te}
  S.~Yamaguchi,
  ``Bubbling geometries for half BPS Wilson lines,''
  Int.\ J.\ Mod.\ Phys.\  A {\bf 22}, 1353 (2007)
  [arXiv:hep-th/0601089].


\bibitem{Lunin:2006xr}
  O.~Lunin,
  ``On gravitational description of Wilson lines,''
  JHEP {\bf 0606}, 026 (2006)
  [arXiv:hep-th/0604133].



\bibitem{Gomis:2008qa}
  J.~Gomis, S.~Matsuura, T.~Okuda and D.~Trancanelli,
  ``Wilson loop correlators at strong coupling: from matrices to bubbling
  geometries,''
  arXiv:0807.3330 [hep-th].

\bibitem{Okuda:2008px}
  T.~Okuda and D.~Trancanelli,
  ``Spectral curves, emergent geometry, and bubbling solutions for Wilson
  loops,''
  arXiv:0806.4191 [hep-th].





\bibitem{Gomis:2006cu}
  J.~Gomis and C.~Romelsberger,
  ``Bubbling defect CFT's,''
  JHEP {\bf 0608}, 050 (2006)
  [arXiv:hep-th/0604155].





\bibitem{Bagger:2007vi}
  J.~Bagger and N.~Lambert,
  ``Comments On Multiple M2-branes,''
  JHEP {\bf 0802} (2008) 105
  [arXiv:0712.3738 [hep-th]].

\bibitem{Bagger:2007jr}
  J.~Bagger and N.~Lambert,
  ``Gauge Symmetry and Supersymmetry of Multiple M2-Branes,''
  Phys.\ Rev.\  D {\bf 77} (2008) 065008
  [arXiv:0711.0955 [hep-th]].

\bibitem{Gustavsson:2007vu}
  A.~Gustavsson,
  ``Algebraic structures on parallel M2-branes,''
  arXiv:0709.1260 [hep-th].

\bibitem{Aharony:2008ug}
  O.~Aharony, O.~Bergman, D.~L.~Jafferis and J.~Maldacena,
  ``N=6 superconformal Chern-Simons-matter theories, M2-branes and their
  gravity duals,''
  arXiv:0806.1218 [hep-th].


\bibitem{Aharony:1997an}
  O.~Aharony, M.~Berkooz and N.~Seiberg,
  ``Light-cone description of (2,0) superconformal theories in six
  dimensions,''
  Adv.\ Theor.\ Math.\ Phys.\  {\bf 2} (1998) 119
  [arXiv:hep-th/9712117].


\bibitem{Kim:2002tj}
  N.~Kim and J.~T.~Yee,
  ``Supersymmetry and branes in M-theory plane-waves,''
  Phys.\ Rev.\  D {\bf 67} (2003) 046004
  [arXiv:hep-th/0211029].


\bibitem{Lunin:2007ab}
  O.~Lunin,
  ``1/2-BPS states in M theory and defects in the dual CFTs,''
  JHEP {\bf 0710} (2007) 014
  [arXiv:0704.3442 [hep-th]].

\bibitem{Cremmer:1978km}
  E.~Cremmer, B.~Julia and J.~Scherk,
  ``Supergravity theory in 11 dimensions,''
  Phys.\ Lett.\  B {\bf 76} (1978) 409.

\bibitem{D'Hoker:2008qm}
  E.~D'Hoker, J.~Estes, M.~Gutperle and D.~Krym,
  ``Exact Half-BPS Flux Solutions in M-theory II: Global solutions asymptotic
  to $AdS_7 x S^4$,''
  arXiv:0810.4647 [hep-th].
  

\bibitem{Howe:1997ue}
  P.~S.~Howe, N.~D.~Lambert and P.~C.~West,
  ``The self-dual string soliton,''
  Nucl.\ Phys.\  B {\bf 515} (1998) 203
  [arXiv:hep-th/9709014].


\bibitem{Kac:1977qb}
  V.~G.~Kac,
  ``A Sketch Of Lie Superalgebra Theory,''
  Commun.\ Math.\ Phys.\  {\bf 53} (1977) 31.


\bibitem{Kac:1977em}
  V.~G.~Kac,
  ``Lie Superalgebras,''
  Adv.\ Math.\  {\bf 26} (1977) 8.



\bibitem{Scheunert:1976uf}
  M.~Scheunert, W.~Nahm and V.~Rittenberg,
  ``Classification Of All Simple Graded Lie Algebras Whose Lie Algebra Is
  Reductive. 1,''
  J.\ Math.\ Phys.\  {\bf 17}, 1626 (1976).

\bibitem{Scheunert:1976ug}
  M.~Scheunert, W.~Nahm and V.~Rittenberg,
  ``Classification Of All Simple Graded Lie Algebras Whose Lie Algebra Is
  Reductive. 2. Construction Of The Exceptional Algebras,''
  J.\ Math.\ Phys.\  {\bf 17}, 1640 (1976).

\bibitem{sorba} L. Frappat, A. Sciarrino, and P. Sorba,
 {\sl Dictionary on Lie Algebras and Superalgebras}, Academic Press, 2000.

\bibitem{Parker:1980af}
  M.~Parker,
  ``Classification Of Real Simple Lie Superalgebras Of Classical Type,''
  J.\ Math.\ Phys.\  {\bf 21} (1980) 689.

\bibitem{Bars:1982ep}
  I.~Bars and M.~Gunaydin,
  ``Unitary Representations Of Noncompact Supergroups,''
  Commun.\ Math.\ Phys.\  {\bf 91}, 31 (1983).

\bibitem{Gunaydin:1985tc}
  M.~Gunaydin and N.~P.~Warner,
  ``Unitary Supermultiplets Of Osp(8/4,R) And The Spectrum Of The S(7)
  Compactification Of Eleven-Dimensional Supergravity,''
  Nucl.\ Phys.\  B {\bf 272}, 99 (1986).

\bibitem{Gunaydin:1990ag}
  M.~Gunaydin and R.~J.~Scalise,
  ``Unitaru Lowest Weight Representations Of The
  Noncompact Supergroup OSp(2m*/2n),''
  J.\ Math.\ Phys.\  {\bf 32}, 599 (1991).

\bibitem{Brown:1986nw}
  J.~D.~Brown and M.~Henneaux,
  ``Central Charges in the Canonical Realization of Asymptotic Symmetries: An
  Example from Three-Dimensional Gravity,''
  Commun.\ Math.\ Phys.\  {\bf 104}, 207 (1986).

\bibitem{HennTeit}
M. Henneaux and C. Teitelboim, ``Asymptotically Anti-de Sitter Spaces",
Commun. Math. Phys. {\bf 98} (1985) 391.

\bibitem{Hollands:2005wt}
  S.~Hollands, A.~Ishibashi and D.~Marolf,
  ``Comparison between various notions of conserved charges in  asymptotically
  AdS-spacetimes,''
  Class.\ Quant.\ Grav.\  {\bf 22}, 2881 (2005)
  [arXiv:hep-th/0503045].

\bibitem{Skenderis:2002wp}
  K.~Skenderis,
  ``Lecture notes on holographic renormalization,''
  Class.\ Quant.\ Grav.\  {\bf 19} (2002) 5849
  [arXiv:hep-th/0209067].

\bibitem{Abbott:1981ff}
  L.~F.~Abbott and S.~Deser,
  ``Stability Of Gravity With A Cosmological Constant,''
  Nucl.\ Phys.\  B {\bf 195}, 76 (1982).

\bibitem{helgason}
S. Helgason, {\sl Differential Geometry, Lie groups, and Symmetric Spaces},
Academic Press, 1978

\bibitem{Besse} A. Besse, {\sl Einstein manifolds}, Springer-Verlag (1986).

\bibitem{KN} S. Kobayashi and K. Nomizu, {\sl Foundations of Differential Geometry},
Vol II, John Wiley Interscience (1969).

\bibitem{Lin:2005nh}
  H.~Lin and J.~M.~Maldacena,
  ``Fivebranes from gauge theory,''
  Phys.\ Rev.\  D {\bf 74} (2006) 084014
  [arXiv:hep-th/0509235].

\bibitem{vanAnders:2007ky}
  G.~van Anders,
  ``General Lin-Maldacena solutions and PWMM instantons from supergravity,''
  JHEP {\bf 0703}, 028 (2007)
  [arXiv:hep-th/0701277].

\bibitem{Shieh:2007xn}
  H.~H.~Shieh, G.~van Anders and M.~Van Raamsdonk,
  ``Coarse-Graining the Lin-Maldacena Geometries,''
  JHEP {\bf 0709} (2007) 059
  [arXiv:0705.4308 [hep-th]].

\bibitem{yu} E. D'Hoker and Y. Guo, in preparation.

\bibitem{Gomis:2006im}
  J.~Gomis and F.~Passerini,
  ``Wilson loops as D3-branes,''
  JHEP {\bf 0701}, 097 (2007)
  [arXiv:hep-th/0612022].

\bibitem{Kim:2005ez}
  N.~Kim,
  ``AdS(3) solutions of IIB supergravity from D3-branes,''
  JHEP {\bf 0601}, 094 (2006)
  [arXiv:hep-th/0511029].
  
\bibitem{Gauntlett:2006ns}
  J.~P.~Gauntlett, N.~Kim and D.~Waldram,
  ``Supersymmetric AdS(3), AdS(2) and bubble solutions,''
  JHEP {\bf 0704}, 005 (2007)
  [arXiv:hep-th/0612253].

\bibitem{SSJ} M.M. Sheikh-Jabbari, unpublished.

\bibitem{FigueroaO'Farrill:2007gj}
  J.~M.~Figueroa-O'Farrill,
  ``Deformations of M-theory Killing superalgebras,''
  Class.\ Quant.\ Grav.\  {\bf 24}, 5257 (2007)
  [arXiv:0706.2600 [hep-th]].
  
\bibitem{Kim:2006qu}
  N.~Kim and J.~D.~Park,
  ``Comments on AdS(2) solutions of D = 11 supergravity,''
  JHEP {\bf 0609}, 041 (2006)
  [arXiv:hep-th/0607093].

\bibitem{MacConamhna:2006nb}
  O.~A.~P.~Mac Conamhna and E.~O Colgain,
  ``Supersymmetric wrapped membranes, AdS(2) spaces, and bubbling geometries,''
  JHEP {\bf 0703}, 115 (2007)
  [arXiv:hep-th/0612196].
  

\bibitem{Gauntlett:2006qw}
  J.~P.~Gauntlett, O.~A.~P.~Mac Conamhna, T.~Mateos and D.~Waldram,
  ``New supersymmetric AdS(3) solutions,''
  Phys.\ Rev.\  D {\bf 74}, 106007 (2006)
  [arXiv:hep-th/0608055].

\bibitem{Figueras:2007cn}
  P.~Figueras, O.~A.~P.~Mac Conamhna and E.~O Colgain,
  ``Global geometry of the supersymmetric AdS(3)/CFT(2) correspondence in
  Phys.\ Rev.\  D {\bf 76}, 046007 (2007)
  [arXiv:hep-th/0703275].
 
  
\bibitem{Nilsson:1984bj}
  B.~E.~W.~Nilsson and C.~N.~Pope,
  ``Hopf Fibration Of Eleven-Dimensional Supergravity,''
  Class.\ Quant.\ Grav.\  {\bf 1}, 499 (1984).

\bibitem{Klebanov:2008vq}
  I.~Klebanov, T.~Klose and A.~Murugan,
  ``$AdS_4/CFT_3$ -- Squashed, Stretched and Warped,''
  arXiv:0809.3773 [hep-th].


\bibitem{sorba2}
L. Frappat, A. Sciarrino and P. Sorba,  ``Structure of Basic Lie
Superalgebras and of their Affine Extensions",
Commun. Math. Phys. {\bf 121} (1989) 457.

  }

\end{thebibliography}
\end{document}